\RequirePackage{fix-cm}
\documentclass[twocolumn,epjc3]{svjour3}  
\RequirePackage{graphicx}
\RequirePackage{latexsym}
\RequirePackage{amsmath}
\usepackage{amssymb}
\usepackage{yfonts}
\usepackage{bbold}
\usepackage[normalem]{ulem}
\usepackage{cite}
\usepackage{tcolorbox}

\RequirePackage[colorlinks,citecolor=blue,urlcolor=blue,linkcolor=blue]{hyperref}

\newcommand{\ket}[1]{|#1 \rangle}
\newcommand{\bra}[1]{\langle #1 |}

\newcommand{\eg}{{\it e.g., }}
\newcommand{\ie}{{\it i.e., }}
\newcommand{\del}{\partial}
\newcommand{\mC}{\mathcal{C}}

\journalname{Eur. Phys. J. C}
\begin{document}

\title{Quantum Computational Complexity %\thanksref{t1}
}
\subtitle{From Quantum Information to Black Holes and Back}


\author{Shira Chapman\thanksref{e1,addr1}
        \and
        Giuseppe Policastro\thanksref{e2,addr2} %etc.
}

\thankstext{e1}{e-mail: schapman@bgu.ac.il}
\thankstext{e2}{e-mail: giuseppe.policastro@phys.ens.fr }


\institute{Department of Physics, Ben-Gurion University of the Negev, Beer Sheva 84105, Israel \label{addr1}
       \and
           Laboratoire de Physique de l'\'Ecole normale sup\'erieure, ENS, Universit\'e PSL, CNRS, Sorbonne
Universit\'e, Universit\'e de Paris, F-75005 Paris, France \label{addr2}
           %\and
           %\emph{Present Address:} if needed\label{addr3}
}

\date{Received: date / Accepted: date}

\maketitle

\begin{abstract}
\sloppy
Quantum computational complexity estimates the difficulty of constructing  quantum states from elementary operations, a problem of prime importance for quantum computation. Surprisingly, this quantity can also serve to study a completely different physical problem - that of information processing inside black holes. Quantum computational complexity was suggested  as a new entry in the holographic dictionary, which extends the connection between geometry and information and resolves the puzzle of why black hole interiors keep growing for a very long time. In this pedagogical review, we present the geometric approach to complexity advocated by Nielsen and show how it can be used to define complexity for generic quantum systems; in particular, we focus on Gaussian states in QFT, both pure and mixed, and on certain classes of CFT states. We then present the conjectured relation to gravitational quantities within the holographic correspondence and discuss several examples in which different versions of the conjectures have been tested. We highlight the relation between complexity, chaos and scrambling in chaotic systems. We conclude with a discussion of open problems and future directions. \emph{This article was written for the special issue of EPJ-C Frontiers in Holographic Duality}.

\hspace{5pt}

%Include keywords, PACS and mathematical subject classification numbers as needed. 
\end{abstract}

\tableofcontents

\section{Introduction}
\label{intro}
\sloppy
Surprising connections between geometry and information have an honorary place in current research in theoretical physics.
These ideas date back to the Bekenstein-Hawking formula  \cite{Bekenstein:1973ur,Hawking:1975vcx} relating the entropy and area of a black hole. The discovery of the AdS/CFT correspondence 
-- the observation that certain gauge theories are equivalent (or ``dual'') to gravitational theories in one higher dimension (see \eg  \cite{Aharony:1999ti,ammon2015gauge}) -- enabled putting the relation between gravity and information on firm ground.
Specifically, it permitted Ryu and Takayanagi (RT) to formulate a proposal \cite{Nishioka:2009un} (later proven by \cite{Lewkowycz:2013nqa}) that relates the entanglement entropy -- a quantity characterizing quantum correlations between two regions in conformal field theory (CFT) -- and areas of minimal surfaces in asymptotically anti-de Sitter (AdS) spaces. 

The RT proposal was the starting point for many interesting developments. It was used to study entanglement in strongly correlated systems and as a consequence improved our understanding of critical points and topological phases, chaos and thermalization, and RG flows (see \cite{rangamani2017holographic} for a review). Furthermore, it provides an interpretation of spacetime as emergent from  quantum entanglement. Specifically, it can be used to understand the way in which the boundary information is encoded in the bulk, and vice versa, in the AdS/CFT correspondence.

However, black holes pose a barrier for our understanding of spacetime in terms of entanglement. The reason is that the space behind the horizon of black holes is only partially accessible via the minimal surfaces in the RT proposal and therefore a lot of the geometry remains uninterpreted in terms of quantum information. This is not a technicality but rather it has been suggested that it is not possible to fully reconstruct the geometry behind the horizon using the boundary data and this topic is still being debated (see, \eg \cite{RevModPhys.88.015002}). Furthermore, despite recent progress in reconstructing the Page curve of black hole evaporation \cite{Penington:2019npb,Almheiri:2019psf} we still lack a full understanding of how black holes process and store information about objects which are thrown into them.

One aspect of these problems is that the volume behind the horizon of black holes keeps growing for a very long  time while the entanglement of a subsystem saturates at times of the order of the subsystem size \cite{Hartman:2013qma}. In fact, it is non-trivial to identify dual field theory quantities which have a similar long-term growth behavior.

To begin addressing this difficulty, Susskind et al. proposed that the volume behind a black hole horizon should be dual to a quantity from quantum information theory known as quantum computational complexity  \cite{Susskind:2014rva,Stanford:2014jda,Brown1,Brown2}. Quantum computational complexity tries to estimate how hard it is to construct a given quantum ``target state'', starting with a simple (usually unentangled) ``reference state'' using a set of simple universal  ``gates'' \cite{watrous2008quantum,Aaronson:2016vto}. For example, if we start with a quantum system consisting of a large number  of spins initiated to be all aligned, we could ask, what is the minimal number of one and two-spin unitary operations taken from a given set required to get to a given target state.

As we will explain in this review, in chaotic systems the complexity grows linearly as time evolves and reacts to perturbations in a distinctive way. All these behaviors have a counterpart in the behavior of the volume behind the horizon. The duality between complexity and certain geometric quantities -- specifically the volume and gravitational action -- was conjectured based on these similar features. We will refer to these  conjectures as the ``holographic complexity proposals''. 

At first, the holographic complexity proposals suffered from lack of rigor due to the absence of a proper definition of complexity outside the traditional spin-chain formulation, in particular for quantum field theory (QFT) states. However, this difficulty was circumvented, first for Gaussian states in free and weakly interacting field theories \cite{QFT1,QFT2,Hackl:2018ptj,Khan:2018rzm,Bhattacharyya:2018bbv} and later for strongly interacting conformal field theories using different approaches  \cite{Caputa:2017yrh,Caputa:2018kdj,Chagnet:2021uvi,Flory:2020dja,Flory:2020eot,Erdmenger:2020sup}. In fact, the study of complexity in field theory is interesting in its own right,   apart from the relation to black holes. Quantum Computational Complexity is expected to have purely condensed matter applications for the detection of phase transitions \cite{Liu:2019aji,Camilo:2020gdf} and in the study of thermalization and chaos \cite{Bao:2020caj,Balasubramanian:2019wgd} as a natural extension of entanglement entropy.

With the surge in literature on complexity in field theory and holography, and with many people coming into this field from different disciplines, we thought it would be good to have an introductory text. This review was written to be comprehensible but by no means comprehensive. We only review those ingredients which are strictly necessary to enter the field with the hope of getting the reader to a point where it is easy to read relevant research articles in the field.

This article was written for the special issue of EPJ-C \emph{Frontiers in Holographic Duality}. Other aspects of the relation between holography and quantum information are reviewed in \cite{ReviewArnab,ReviewAyan,ReviewLata}, submitted as a part of the same issue.

This review is organized as follows. In \S
\ref{sec:Primer} we begin with an overview of quantum computation. Then, in \S
\ref{sec:ComQub} we define  Quantum Computational Complexity and discuss its properties in spin chains with fast scrambling dynamics and how it relates with scrambling and chaos. In \S
\ref{sec:ContComp} we present a continuous definition of complexity due to Nielsen. In 
\S \ref{sec:CompSHO} we discuss the complexity of systems of coupled simple harmonic oscillators in preparation of our study of complexity in free and weakly interacting QFTs. In \S \ref{sec:Complexity-free-QFT} we review the complexity of Gaussian and coherent states in free and weakly interacting QFTs, both pure and mixed, and discuss complexity in strongly interacting CFTs. In \S \ref{sec:Holo1}- \S \ref{sec:Holo2} we discuss the holographic complexity conjectures and the relevant evidence.  
We conclude in \S \ref{sec:SumDisc} with a summary and outline of open questions.

\section{A Quantum Computation Primer}
\label{sec:Primer}
Quantum computers can famously achieve exponential speed-up of computation compared to classical ones, at least for some problems. They can do this by taking advantage of the possibility of putting a quantum system in a superposition of states; performing operations on a superposition is, roughly speaking, like performing the computation in parallel on all the states in the superposition. Of course this is not precisely true, since in order to read the result one has to perform a measurement, which will cause the collapse of the state of the system on an eigenstate of the measured observable. One might then expect that each input requires a different measurement and the advantage of having the superposition is lost. But this is not the case: by a judicious choice of the algorithm and the initial state one can extract the information in an efficient way.

It is very instructive to see how these ideas work in practice on a  simple example: the Deutsch's algorithm (we follow the presentation given in \cite{NielsenChuang}). Suppose we have the task of computing a function $f(x): \{0,1\} \to \{0,1\}$. One can build a circuit that implements the 2-qubits unitary operator $U_f: \ket{x,y} \to \ket{x,y+f(x)}$ where the addition is understood to be mod 2. We could read  out the value of $f(x)$ by applying the operator on $\ket{x,0}$ and reading the second qubit, and we assume that this operation can be done with the same efficiency as in the classical case. Now let us consider an initial state in a superposition. Let us define $\ket{\pm} = \frac{\ket{0}\pm\ket{1}}{\sqrt{2}}$. First observe that 
$U_f \ket{x,-} = (-)^{f(x)} \ket{x,-}$. Then one can compute   
\begin{align}
\begin{split}
    U_f \ket{+,-} = & \,\frac12  ((-)^{f(0)}+(-)^{f(1)}) \ket{+,-} \\ + &  \, \frac12 ((-)^{f(0)}-(-)^{f(1)}) \ket{-,-} \, .
\end{split}
\end{align}
If we project the first qubit on the $\ket{\pm}$ basis, we can read off whether $f(0)=f(1)$ or $f(0)\neq f(1)$ (we could equivalently say that we computed $f(0)+f(1)$ mod $2$). The point of the example is that there is no way of doing this classically without computing separately $f(0)$ and $f(1)$, whereas quantum mechanically we get the result with a single computation. Not only the computations proceed in parallel, but they can be recombined by using interference of different states. This simple example is not very impressive, but it can be generalized to an analogous problem involving a function on $n$ qubits; the Deutsch-Jozsa algorithm solves the problem with one computation instead of the $2^{n-1}+1$ required classically (see \cite{NielsenChuang}).

Another important point illustrated by the example is that an efficient computation will typically require a particular initial state. We started from $\ket{\psi_0} = \ket{+-}$, but supposing that our computer starts in a canonical state $\ket{00}$, we will need to apply some operations to prepare $\ket{\psi_0}$. Analogously, in the final step we need to measure the state in the $\ket{\pm}$ basis, but if we can only measure in the computational basis (i.e., the  $\ket{0},\ket{1}$ basis), we have to use another operator to move between the two bases . 

We can then formalize a quantum computation as a series of operations on a set of qubits, and the number of operations required to go from the initial to the final state is a measure of the difficulty of the task. This is the notion of {\it quantum computational complexity}. In the next section we will give a more precise definition.

We should point out that the notion of computational complexity is related to the question of the resources needed to solve a problem. We are typically interested in finding the fastest algorithm for a given problem. Assuming that each quantum operation (gate) requires a fixed amount of time, the number of operations is a measure of the total time required for the computation. The real physical time will of course depend on the physical implementation of the gates, but there are some  unavoidable limits imposed by quantum mechanics; the Margolus-Levitin  \cite{Margolus:1997ih} and the 
Aharonov-Anandan-Bohm 
\cite{anandan1990geometry,aharonov1961time,mandelstam1991uncertainty} bounds give the minimum time required for evolving a given state into an orthogonal state\footnote{Note however that it may not always be necessary to use orthogonal states to distinguish the outcome, see \cite{Sinitsyn}.} $t_{min} = \frac{\pi \hbar}{2 E}$, where $E = \langle H - E_0 \rangle$ is the expectation value of the energy above the ground state or the variance of the energy in the state $(\langle E^2\rangle - \langle E \rangle ^2)^{1/2}$, respectively.

Alternative notions of complexity exist, related to the optimization of different resources. For example one could take into account the number of qubits  used in a quantum algorithm similarly to storage in classical complexity. A different notion is the {\it Kolmogorov complexity}. In the classical setup, this is the length of the minimal program that can produce a given string; so it is a measure of the amount of information contained in the string, or how much it can be compressed without losing information. Quantum versions of Kolmogorov complexity have also been proposed \cite{BERTHIAUME2001201}. One can of course also combine the requirements of limitation on time,  storage space and algorithmic complexity all together.

In this review, we will focus only on one notion of quantum computational complexity, related to the number of operations. The reason is that this notion has been found (or rather, conjectured) to play an interesting role  
in the holographic duality, in connection with the properties of black hole interiors, and as a consequence it has been developed in the last few years from a point of view slightly different from that of quantum computing. We cannot rule out that other notions will also become relevant as we understand more and more of the relation between geometry and information (see for example \cite{Brown:2017jil} for a discussion of the Kolmogorov complexity in the context of holography).

\section{Complexity in Qubit Systems}
\label{sec:ComQub}

\subsection{Quantum Computational Complexity}
\label{sec:QuantumComputationalComplexity}

We have explained that a quantum computation can be formalized as the problem
of producing a certain state, from an initial state, through a series of
unitary operations. In practice we can only build a quantum circuit using a
discrete set of gates, each one \ implementing a simple operation, typically
acting only on one or two qubits at the time. Two questions arise naturally:
first, is it possible to construct an arbitrary unitary operator using a
finite predetermined set of gates? Second, if a unitary can be constructed, how
many gates are needed?

For the first question, it is obvious that the set of all \ finite circuits
built out of a finite set of gates can only reproduce a discrete subset of the
unitary group. However if we allow for a margin of error, \ie if we only ask
that for any operator $U$ we can find a circuit that gives an operator $V$
such that $\| U - V \| < \varepsilon$,  in the operator norm,\footnote{The operator norm is defined as the maximal eigenvalue, \ie $\| U \| = \max_{|\psi\rangle} |\langle \psi |U |
\psi \rangle|$ where the maximization is over all normalized states $|\psi\rangle$.} then the answer is positive: \ there
exist sets of {\it universal gates}, using which any unitary
can be constructed with arbitrary precision. The full argument can be found
in \cite{NielsenChuang}. Here, we only give an outline of the proof. Let us consider
first operators acting on a single qubit, \ie elements of  $SU(2)$.
A generic element can be written as a rotation of an angle $\theta$ around the
axis $\vec n$, \ $R_{\vec n} (\theta)\equiv e^{-i \theta \vec n \vec \sigma/(2|\vec n|)}$, where $\vec \sigma$ is the vector of Pauli matrices. We can use two gates:
the Hadamard gate (denoted by $H$) and the T gate (sometimes referred to as the $\pi/8$ phase gate)
\begin{equation}
\begin{split}
& H = \frac{1}{\sqrt{2}} \left(\begin{array}{cc}
     1 &  1\\
     1 & - 1
   \end{array}\right) = \frac{1}{\sqrt{2}}  (\sigma_x + \sigma_z),\\
& T = \left(\begin{array}{cc}
     e^{-i \pi/8} & 0 \\
     0 & e^{i \pi / 8} 
   \end{array}\right) \,. 
\end{split}
\end{equation}
One can check that $H T H = R_{\hat x} (\pi / 4)$, and $T H T H = R_{\vec n} (\theta)$,
where $\vec n = (\cos \pi / 8, \sin \pi / 8, \cos \pi / 8 )$ and $\cos(\theta/2) = \cos^2(\pi/8)$. Note that the angle  $\theta$ is an irrational multiple of $2 \pi$. This implies that we can approximate any angle of rotation by taking powers of $R_{\vec n} (\theta) $. Furthermore one
can see that $H R_{\vec n} (\theta) H = R_{\vec m} (\theta) $ with $\vec m = (\cos \pi / 8, - \sin
\pi / 8, \cos \pi / 8 )$. Since $\vec m$ and $\vec n$ are not parallel, one can find a
parametrization of an arbitrary rotation  as 
\begin{equation}
    U = e^{i \phi} R_{\vec n} (\alpha) R_{\vec m}
(\beta) R_{\vec n} (\gamma)\,.
\end{equation}
These would be the Euler angles in the case where $\vec m \perp \vec n$.  This shows that the gates $H, T$ are universal for a single qubit.

For the case of more than one qubit, an arbitrary unitary cannot be approximated using only the $H$ and $T$ gates since those do not generate quantum correlations between multiple qubits. However, it turns out that adding one kind of two-qubit operation is enough to generate a universal gate set on any number of qubits. An example of such a gate is the CNOT gate: 
\begin{equation}
    CNOT =\frac{1}{2} (1 + \sigma_z^{(1)}) \otimes
\mathbb{1^{(2)} } + \frac{1}{2}(1 - \sigma_z^{(1)}) \otimes \sigma_x^{(2)}.
\end{equation}
One can easily see that in the computational basis,\footnote{The computational basis is the basis of states in which each qubit is in an eigenstate of $\sigma_z$.} this gate flips the second qubit only if the first qubit is in the state 1. With the CNOT gate in hand the proof of universality amounts to a linear algebra theorem and it proceeds as follows. 
First, we can show that any unitary operator can be decomposed as a product of two-level operators, which act non-trivially only
on a subspace spanned by two computational basis vectors. 
Then, essentially one has to map any two-dimensional
subspace to a single qubit;
this can be achieved by acting with the CNOT gate.\footnote{As a simple illustration, let us consider a two-dimensional subspace spanned by the two vectors $(0,1,\ldots),(1,0,\ldots)$ that differ in the first two qubits. Acting with the CNOT gate on these qubits turns the states into $(0,1,\ldots),(1,1,\ldots)$ and the states now differ only in the first qubit.} 
This proves that every unitary operation can be decomposed as a product of $H$ and $T$ gates acting on the different qubits and CNOT gates acting on all pairs of qubits.

\begin{figure}[htbp]
\centerline{\includegraphics[scale=.5]{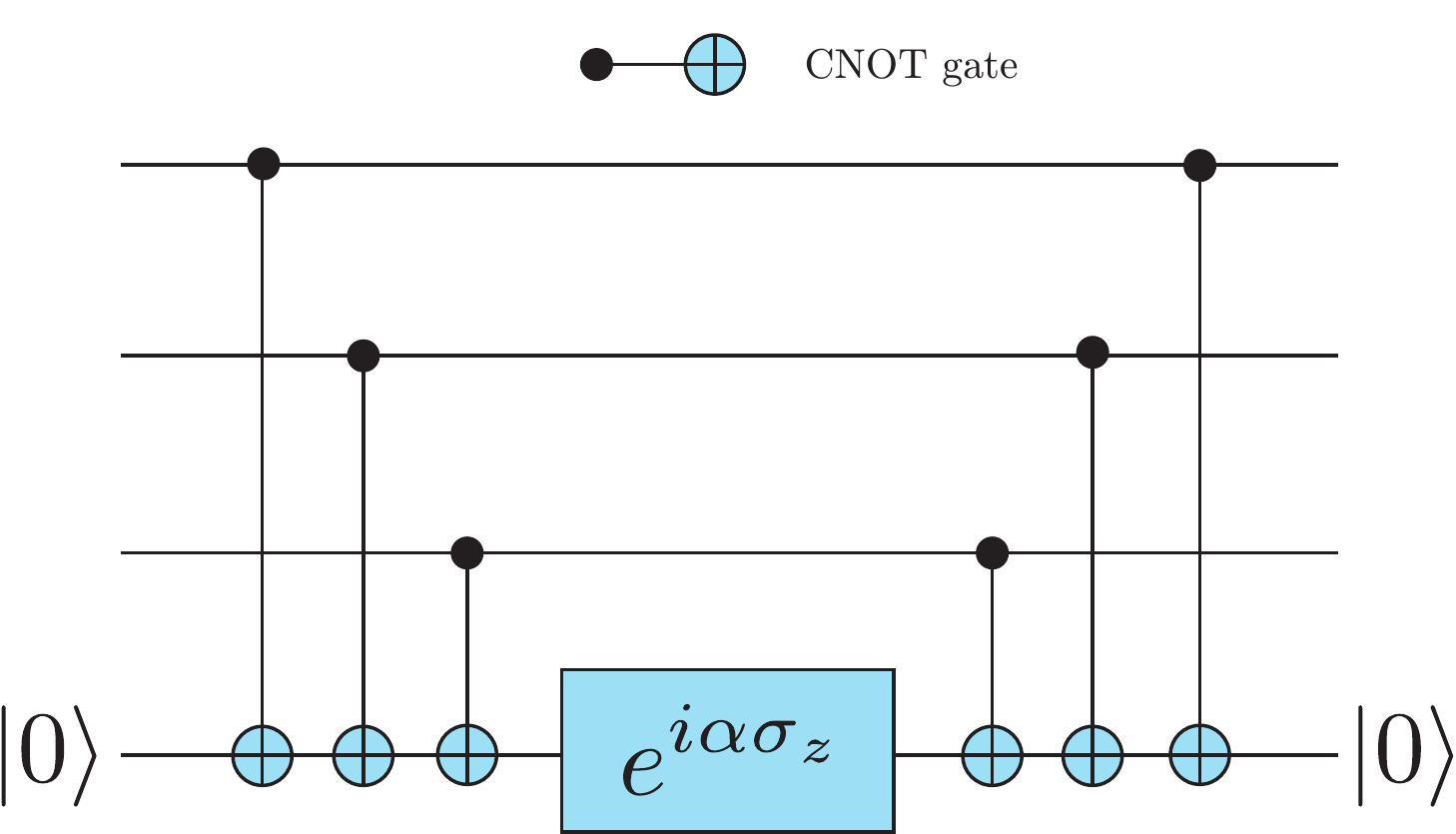}}
\caption{Illustration of a circuit implementing the unitary transformation $\exp{\left(i \alpha \prod_i \sigma_z^i\right)}$.}
\label{fig:circuit}
\end{figure}

An alternative proof can be given, which is perhaps more suggestive and closer to a physicist's mindset. 
We can write a generic unitary operator as 
\begin{equation}
     U = \exp \left( i \sum y_a h_a \right) 
\end{equation}
where the sum is over all operators of the form $h_a = \prod_i \sigma_{k_i}^{(i)}$. 
This can be approximated as $U = (\prod_a e^{i \frac{y^a}{n} h_a})^n + {\cal O}(4^N/n)$, where $N$ is the number of qubits. Note that we assume that $n\gg 4^N$ so that the correction is much smaller than the leading term.
Using single-qubit operations, one can convert any $h_a$ into $h=\prod_i \sigma_z^{(i)}$. The operator $e^{i \alpha h}$ can be implemented using a single-qubit operator and the CNOT gates as follows: we apply successively the CNOT to the $j$-th qubit and an ancillary qubit. The effect is to encode the product of all bits on the extra qubit, and then one can act on it with $e^{i \alpha \sigma_z}$, and reverse the series of CNOTs. The circuit is represented in Fig. \ref{fig:circuit}. In this way, we have demonstrated that using only one and two-qubit operations any unitary can be constructed to arbitrary precision. The arbitrary precision is achieved by tuning $n$ to be as large as we wish.

Notice that this logic could be applied also at the level of one qubit: any element of $SU(2)$ can be written as $e^{a_x \sigma_x +a_y \sigma_y +a_z \sigma_z}$ and can be approximated 
using the three gates $e^{i \epsilon \sigma_x}, e^{i \epsilon \sigma_y}$, $e^{i \epsilon\sigma_z}$. In fact, the third gate $e^{i \epsilon\sigma_z}$ can be replaced by further combinations of the first two gates using the group commutation relations. We have again a set of two universal gates on one qubit. However, now the gates have to be adjusted according to the required precision; moreover, for $\epsilon$ very small, the gates are very close to the identity and a circuit built with them would be very susceptible to noise, although such considerations are outside our purview.

Having established the possibility of approximating an arbitrary unitary
operator, we can address the second question: how efficiently can we simulate
a given operator? This question leads us finally to the notion of complexity.  Let us start with a definition.

\vspace{7pt}
\centerline{\begin{tcolorbox}[width=0.95\linewidth,colback=blue!10,colframe=blue]
{\bf  \emph{Quantum computational complexity} $\mathcal{C}(U)$ of an operator $U$ is the minimal number $n$ such that $\|U - \prod_{i=1}^n U_i\| < \epsilon$, where $U_i$ belong to a set of allowed gates.}
\end{tcolorbox}}

The answer should depend on the allowed error $\epsilon$ (also known as the tolerance), on the allowed set of gates,  and
on the size of the system, that is on the number of qubits $N$. At the single
qubit level, the Solovay-Kitaev theorem \cite{Kitaev97} states that any operator can be built
with $O \left( \log^c  \frac{1}{\epsilon} \right)$ gates, where $c \approx 2$. 
For a system of $N$ qubits, we can give an
estimate by computing how many balls of radius $\epsilon$ are needed to cover
the unitary group $U (K\equiv2^N)$. This group has dimension $K^2$, and its volume (see \eg \cite{lando2004graphs} Corollary 3.5.2) is given by\footnote{Here we work with the group of unitary transformations $U(K)$ but since overall phases are not important in physical applications, a similar estimate is often done for the special unitary group $SU(K)$, see \eg \cite{Susskind:2018pmk}.} 
\begin{equation}
 \text{Vol} \, (U (K)) = \frac{(2 \pi)^{(K^2 + K) / 2}}{2!3! \ldots (K - 1) !}
   \, . 
\end{equation}
The volume of an $\epsilon$-ball of the same dimension is\footnote{Since we consider a small ball, we can use the result for the volume in flat space. The exact result, and the large-$K$ asymptotics, for the volume of a ball of any radius in $U(K)$ can be found in \cite{Wei17}. Interestingly, as discussed in this reference, the result is related to a number of information-theoretical properties.}
\begin{equation}
\text{Vol} (B_{\epsilon}) = \frac{(\sqrt{\pi} \epsilon)^{K^2}}{ (K^2 / 2) !}
\end{equation}
and the ratio of the two volumes gives an estimate of the required number of balls. For large $N$ one finds, using the Stirling's
formula, 
\begin{equation}\label{volvol}
    \log \left( \frac{\text{Vol} (U (2^N))}{\text{Vol} (B_{\epsilon})} \right)
   \sim 2^{2N}\left( \frac{N}{2} \log 2 +  \log \frac{1}{\epsilon} \right)\,.
\end{equation}
The main thing to notice is that the dependence on the error is only
logarithmic, just as in the case of one qubit, but the dependence on the size
of the system is exponential. Given a set of $p$ gates, the number of circuits
with $m$ elements is bounded by $p^m$. Therefore, the number of unitaries with complexity less than or equal to $m$ is bounded by $p^m$. Together with equation \eqref{volvol}, this implies  that most unitary transformations are exponentially complex. In other words, 
simulating a unitary operator is generically exponentially hard. Enlarging the set of gates cannot improve the situation: one can show that if a circuit can be built with $m$ gates, then it can be build with $\mathcal{O}(m \log^c(\frac{m}{\epsilon}))$ gates from a different universal set \cite{cleve2000introduction}. Combining the  estimate in equation \eqref{volvol} with the Solovay-Kitaev theorem, one can show that a unitary over $N$ qubits may be approximated with tolerance $\epsilon$ using at most $O(N^2 2^{2N}\log^c(N^2 2^{2N}/\epsilon)$ gates \cite{NielsenChuang}.

In this section we have considered the operator complexity; the question
of the complexity of a state is related but not identical, because many
unitary operators can produce the same state.

\vspace{7pt}
\centerline{\begin{tcolorbox}[width=0.98\linewidth,colback=blue!10,colframe=blue]
{\bf Quantum computational complexity of a state is defined by the minimal operator complexity over all operators which produce a given target state $|\psi_T\rangle$ starting with a simple reference state $|\psi_R\rangle$, \ie
\begin{equation}
   \mathcal{C}(|\psi_T\rangle) = \min_{U|\psi_R\rangle = |\psi_T\rangle} \mathcal{C}(U).
\end{equation}}
\end{tcolorbox}}
We will dwell more on the difference between the two later on; for now we can just notice that a similar counting
argument shows that the state complexity has the same qualitative behavior as the operator complexity in that the discretized number of states in $\mathbb{CP}^{K-1}$ is exponential in $N$ and logarithmic $\epsilon$.

\subsection{Complexity in Fast Scramblers}
\label{sec:fast-scramblers}

In the previous section we have considered the complexity from the point of view of computation, \ie we focused on the complexity of  a unitary  operation designed to perform a certain task. From a physics perspective, unitaries arise as operators that describe the evolution in time of a system. It is natural then to consider the question of how complexity changes with time. Under some assumptions, the result will follow from the volume counting of last section. We follow here the presentation given in \cite{Stanford:2014jda,Susskind:2018pmk}.

\begin{figure}[htbp]
\centerline{\includegraphics[scale=.5]{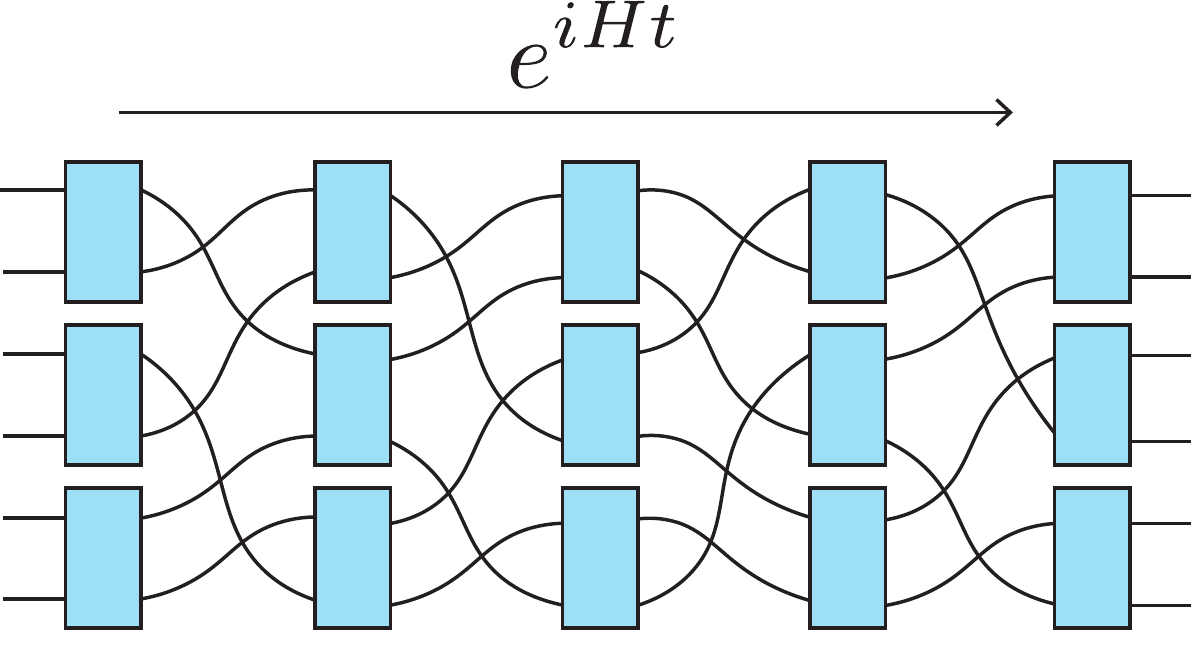}}
\caption{Illustration of a circuit representing time evolution according to a k-local (in this case 2-local) Hamiltonian.}
\label{fig:k-circuit}
\end{figure}

We model the evolution of a Hamiltonian system with a discrete circuit of the form shown in Fig. \ref{fig:k-circuit}. 

\begin{figure}[htbp]
\centerline{\includegraphics[scale=.7]{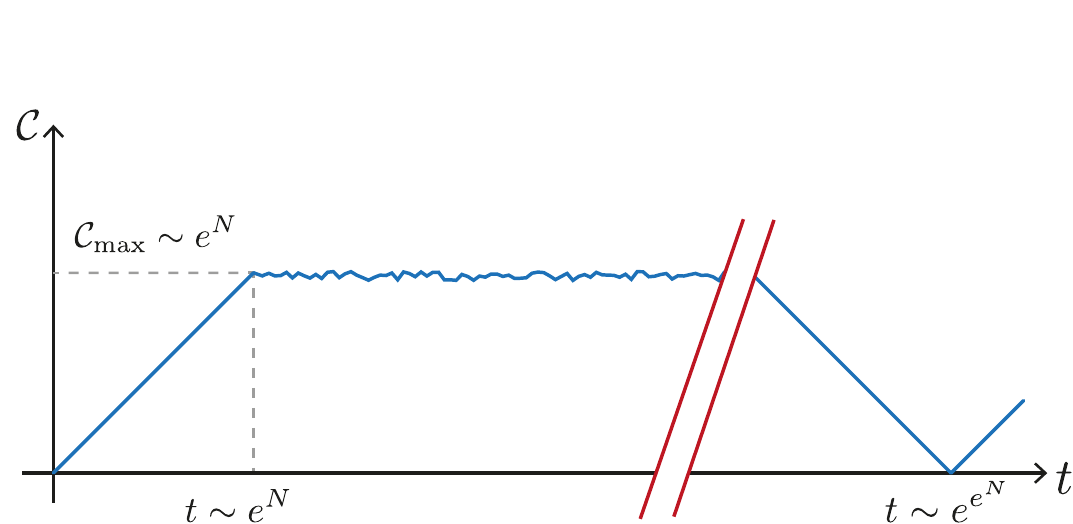}}
\caption{Illustration of the time dependence of complexity during chaotic Hamiltonian evolution. The complexity grows linearly until it reaches its maximal value which is exponential in the number of degrees of freedom, and is expected to decrease significantly around the quantum recurrence time which is doubly exponential in the number of degrees of freedom in the system, once the full unitary group has been explored.}
\label{fig:TimeDep1}
\end{figure}

We assume that the circuit contain only \emph{$k$-local} gates, \ie gates that act on $k\ll N$ qubits at the time.
The evolution happens in discrete steps, at each step the qubits are divided in groups of $k$ and 
acted on by the gates; however the partition changes at every step, so the qubits are all interacting with each other. This is a feature of systems that have the property of \emph{fast scrambling}, namely, the information contained in a part of the system is quickly distributed over the whole system \cite{Sekino:2008he}. 
After $n$ steps of evolution, the number of unitaries that could be generated is  
\begin{equation}
    \left( \frac{N!}{(N/k)!\ (k!)^{N/k}} \right)^n \sim \exp \left( n \frac{k-1}{k} N \log N \right) \,.
\end{equation}
This is much smaller than  the total number of unitaries in \eqref{volvol}, unless $n$ is exponentially large. We can often assume that all these unitaries are different from each other, and that there is no other circuit that generates them more efficiently; under these assumptions, the complexity is
\begin{equation}\label{circuit-growth}
    \mathcal{C}=n N/k \,,
\end{equation}
so it grows linearly with the number of steps and with the size of the system. The linear growth is expected to continue until most of the group has been explored, which happens for $n = \mathcal{O}(2^{2N})$, and then the complexity saturates and oscillates close to its maximal value. Eventually quantum recurrence will make it return to small values but on a  doubly-exponential time scale, see Fig. \ref{fig:TimeDep1}.

\begin{figure}[htbp]
\centerline{\includegraphics[scale=.7]{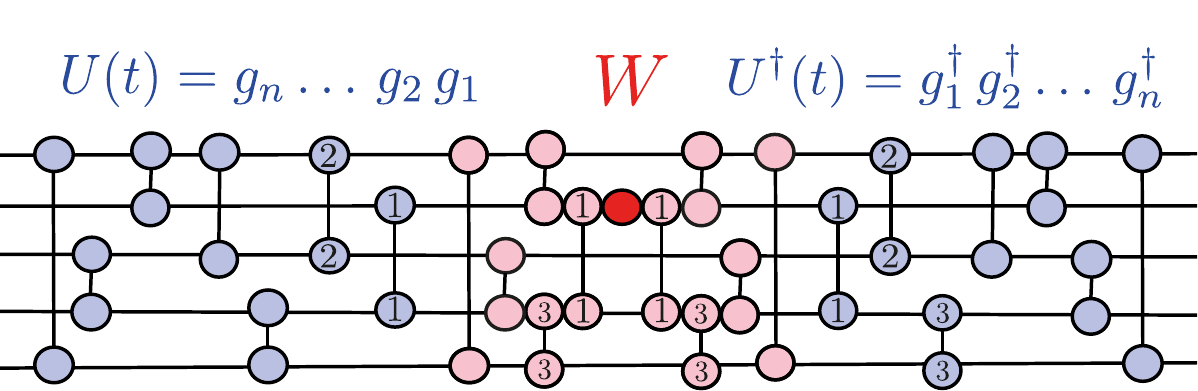}}
\caption{Illustration of the switchback effect. The perturbation $W$ is acted on by $U(t)$ on the left and $U^\dagger(t)$ on the right to create the precursor operator. Two qubit gates participating in the most efficient preparation of $U(t)$ are labeled $g_i$ and they appear as light-purple circles before applying them to $W$ (and as light-red circles after being applied). Estimating the complexity of the precursor operator at different times depend on delicate cancellations which can be seen after applying the gates. For example, the gate $g_2$ commutes with the perturbation and the previously applied gates and therefore does not contribute to the complexity.}
\label{fig:switchbacker}
\end{figure}

Another natural question that one can ask is: how does the complexity grow when the system is subject to a perturbation? We can consider an operator $W$ that is simple, \eg it acts on a single qubit, and let it evolve, so we need to find the complexity of the so-called \emph{precursor}
\begin{equation}
    W(t) = U(t) W U(-t) \,.
\end{equation}
A precursor is defined \cite{Susskind:2013lpa} as any non-local operator which acts at one time, to simulate the effect of a local operator acting at a different time (later or earlier). For the present purposes, we can just think of the forward or backward time evolution of a local operator. 
It is clear that this is a very different question from finding the complexity of $U(t)$ itself; for instance, when $W$ is the identity operator, $W(t)$ is also the identity operator for any $t$, so its complexity does not grow. The circuit model explains why \cite{Susskind:2014jwa,Susskind:2021esx}: a discretized version of the circuit that represents $W(t)$ can be drawn like in Fig. \ref{fig:switchbacker}, with a layer in the middle representing $W$, and series of layers on the left and the right representing $U(t),U(-t)$. In fact, we have discretized time here into a series of discrete time steps which we will label $n$. The gates on the right are the inverse of the corresponding ones on the left. But this is not the optimal circuit for $W(t)$, because gates on the two sides that act on qubits that are not affected by $W$ will have no effect and can be canceled out. At the second layer, the cancellation is obstructed not only by the qubit acted on by $W$ but also by those qubits that have interacted with it. 

Let us define $s(n)$ to be the number of qubits that have been infected after the action of $n$ layers of the circuit, and $p(n)=s(n)/N$ the fraction of infected qubits. When another layer is applied, the probability that a qubit is infected is the probability that it was already infected plus the probability that it was not, multiplied by the probability that one of the $k-1$ qubits that it interacts with is infected.\footnote{Recall, that at every step the qubits are divided randomly in groups of $k$ on which the gates act.} It is easier to write it in terms of $q(n)=1-p(n)$. The evolution of the infection is described by 
\begin{equation}
    q(n+1) = q(n)^{k} \,.
\end{equation}
This can be easily solved and we find the number of infected qubits:
\begin{equation}
    s(n) = N \left( 1 - \left(1-\frac{s_0}{N}\right)^{k^n}\right),
\end{equation}
where $s_0$ is the initial number of infected qubits. 
When the initial operator is small, we can approximate this expression for small $n$ with $s(n) \sim s_0 k^n$.
The complexity is given by the sum of the infected sites at different steps. 
We cannot perform the sum analytically, however 
we can see that because of the exponential behavior, $(s(n+1)-s(n))/s(n)$ becomes small after a few steps. We can then replace the difference equation by a differential equation 
\begin{equation}
    \frac{ds}{dn} = (N-s)\left( 1-\left(1-\frac{s}{N}\right)^{k-1} \right) \,.
\end{equation}
The solution can be given explicitly for the inverse function $n(s)$: 
\begin{equation}\label{episol}
    n  = \left. \frac{1}{k-1} \log \left( \frac{1-(1-\frac{s}{N})^{k-1}}{(1-\frac{s}{N})^{k-1}} \right) \right|_{s_0}^s.
\end{equation}
This expression can be inverted as follows
\begin{equation}\label{episolinv}
\begin{split}
     \frac{s}{N} & =  1-\left(1+\mathfrak{c}\, e^{(k-1)n} \right)^{-\frac{1}{k-1}}, \\
   \mathfrak{c} & =  \left(1-\frac{s_0}{N}\right)^{-(k-1)} -1,
\end{split}
\end{equation}
from which we can extract the early time behavior: $s(n) \sim s_0 e^{(k-1) n}$, and the late time behavior: $s(n) \sim N (1 - \mathfrak{c}^{-\frac{1}{k-1}} e^{-n})$, where for these limits we have assumed that $s_0\ll N$ and therefore $\mathfrak{c}\sim\frac{s_0(k-1)}{N}$.
We can also see that the time it takes for a small perturbation to spread to a finite fraction of the system (the {\it scrambling time}) is of order $n_* \sim \frac{1}{k-1} \log \left(\frac{N}{s_0(k-1)}\right)$.\footnote{Here we are using the term time for the number of steps in anticipation of it becoming the physical time of some Hamiltonian evolution later on.} 

In the case of a 2-local circuit, $k=2$, the solution \eqref{episolinv} takes the form
\begin{equation}
    s(n) = \frac{N s_0 e^n}{N+s_0(e^n-1)}\,.
\end{equation}
We can then compute the complexity which is obtained by summing over the  number of infected qubits  at different times:
\begin{equation}\label{epidemic-comp}
    \mathcal{C}(n) = \int_0^n s(n')d n' =N \log\left(1+e^{(n-n_*)}\right) \,,
\end{equation}
where here again, we have assumed $s_0 \ll N$ and defined $n_*=\log \frac{N}{s_0}$.

\begin{figure}[htbp]
\centerline{\includegraphics[scale=.4]{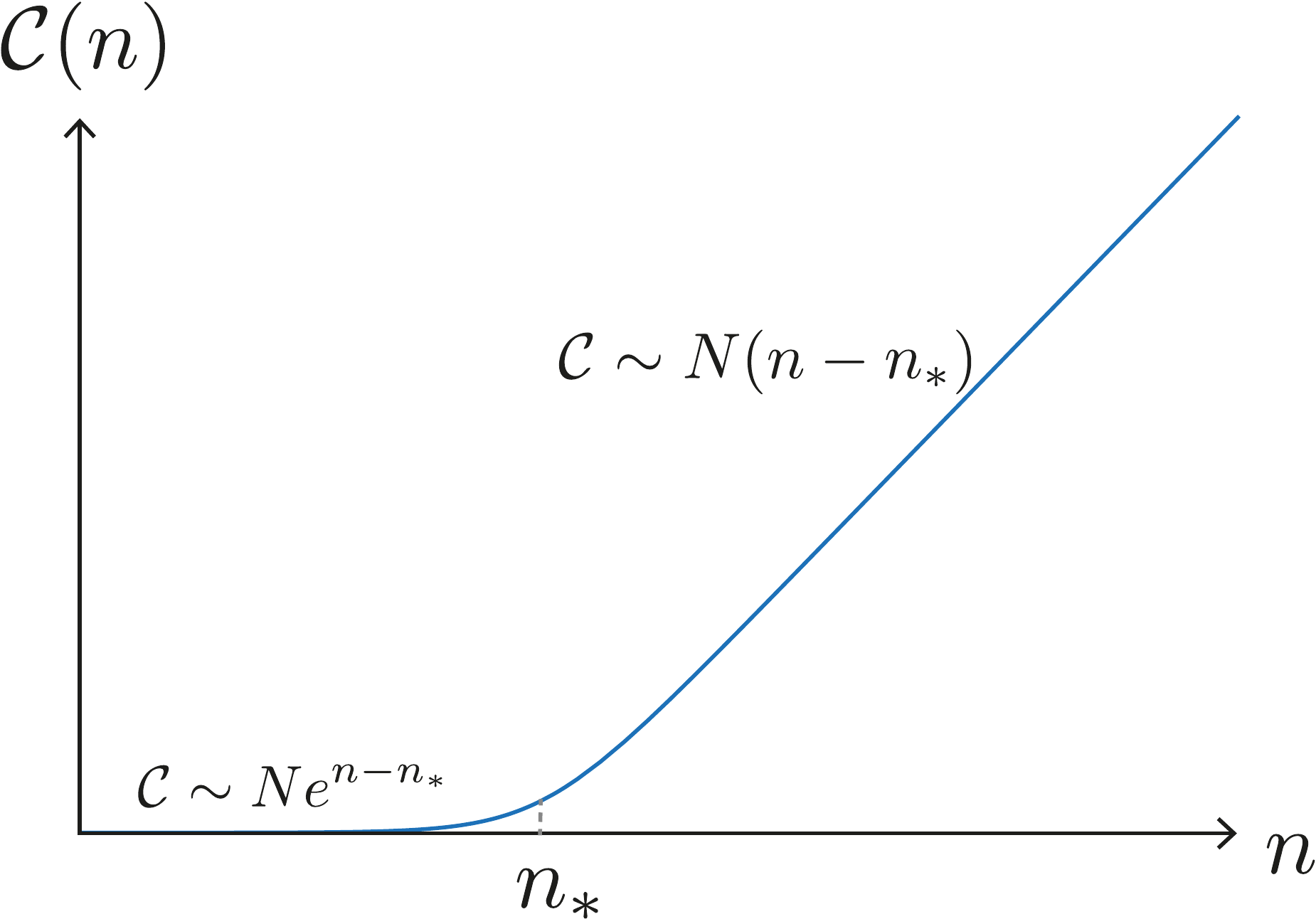}}
\caption{Illustration of the time dependence of complexity of the precursor. An initial exponential  regime is followed by linear growth starting at the scrambling time $n_*$.}
\label{fig:switchbacker2}
\end{figure}

There are two notable features of this result. 
1) It grows linearly for times larger than the scrambling time; the delay in the onset of the linear growth is called the \emph{switchback effect} \cite{Stanford:2014jda}; 
just as for the unperturbed evolution, the linear growth will eventually come to an end and the complexity will saturate on exponentially long time scales. This linear growth behavior is very important; it is one of the motivations for the holographic conjectures that we will present later in section \ref{sec:Complexity-conjectures}. We will comment further on this in the discussion section.
2) The early-time behavior is exponentially growing, but with a small prefactor that is suppressed as $1/N$. It can be argued that this  behavior is related to the Lyapunov growth of the out-of-time-order correlators \cite{Maldacena:2015waa} which is a signature of quantum chaos. Under the assumption of maximal chaos, this yields the identification $ (k-1) n = 2 \pi T t$.  The number of qubits corresponds to the entropy of the system. Up to prefactors, we find that the rate of growth is expected to be proportional to $TS$. This expectation is borne out by the two holographic complexity proposals CV and CA applied to black holes which we will discuss later in section \ref{sec:Holo1}. The time dependence of the complexity of the precursor is illustrated in Fig. \ref{fig:switchbacker2}.

\section{Continuous Complexity}
\label{sec:ContComp}

\subsection{Nielsen's Approach}

We have estimated the number of gates needed to reproduce a given unitary,
but how can one go about finding the actual optimal circuit that does the job?
This appears to be a very difficult problem. 

An approach to this question, proposed by Nielsen \cite{Nielsen05,Nielsen06,Nielsen07} turns the question into a
geometric problem, and as such provides a universally applicable strategy. The
idea is suggested in the proof of universality given in the previous section:
if the universal gates are chosen to be $e^{i \epsilon h}$, then a circuit
will explore the unitary group by small steps, and in the limit $\epsilon
\to 0$ will give a continuous path, which can be constructed by means of a
time-dependent Hamiltonian,
\begin{equation}\label{controlprob1}
U(t) = \overleftarrow{\mathcal{P}} \exp \left( \int_0^{t} H (s) ds \right) . 
\end{equation}
The Hamiltonian can be expanded in a basis of operators
\begin{equation}\label{controlprob2}
 H (t) = \sum_I Y^I (t)  \mathcal{O}_I 
\end{equation}
and the complexity is defined by the minimization of a suitable {\it cost
functional} $F [Y^I]$ as
\begin{equation}
    \mathcal{C }_F [U] = \min_{\{ Y \}} \int dt \, F [Y^I (t)] \,,
\end{equation}
with the constraint that the desired operator is reached at some fixed time $t_f$.\footnote{When the cost function is homogeneous of degree 1, we can take $t_f=1$ without loss of generality.} 
In this way the problem is translated into a Hamiltonian control problem.
The cost function, if it satisfies the appropriate requirements, defines a notion of distance on the space of unitaries, and optimal circuits correspond to minimal length geodesics. 
However the complexity is not uniquely defined, as it depends on the choice of
the cost function. For instance, a quite general family of cost functions,
that we will use in the following, is given by
\begin{equation}\label{NielsenYNorms}
    F_{k,\{p\}} [Y^I] = \left(\sum_I p_I | Y^I |^k\right)^{\frac{1}{k}},
\end{equation}
where the positive \emph{penalty factors} $p_I>0$ 
account for the relative difficulty of implementing different gates.\footnote{For the notation, in the following the penalty factors should be understood to be absent, \ie all set to one, unless explicitly indicated; so $F_k$ will refer to the unpenalized cost, and $\mathcal{C}_k$ to the corresponding complexity.} In the
case $k = 2$ the cost function is the distance induced by a
Riemannian metric on the space of unitaries. This metric is always
right-invariant, as it is defined in terms of $H (t) =  \partial_t
U (t) U^{-1} (t)$, but in general it is not left-invariant.\footnote{The cost function could in principle depend both on the position $U(t)$ and the velocity $Y(t)$ along the path. This would give rise to inhomogeneous metrics on the group, but we will not consider such cases.}

Notice that the complexity thus defined will depend on the choice of the
basis of operators used and in general it is not invariant under a change
of basis. One can obtain a basis-independent notion using the Schatten norm:
\begin{equation}\label{NielsenHNorms}
S_k [H] \quad = \left( \textrm{tr} (H^{\dagger} H)^{\frac{k}{2}}
   \right)^{\frac{1}{k}} . 
\end{equation}
If the operators of the basis are chosen so that $\frac{1}{2}\textrm{tr} (\mathcal{O}_I 
\mathcal{O}_{J }^{\dagger}) = \delta_{IJ}$, then $F_{2 k} [H] = (1/\sqrt{2}) S_k
[H]$. In this case $F_2$ corresponds to the left- and right-invariant metric, and is invariant under an \emph{orthogonal} change of basis.  

One may wonder whether the ``continuous'' complexity defined in this section
can be related precisely to the discrete notion defined by the number of
gates. The argument given in \cite{Nielsen06}  shows that this is the case, and at the same
time it illustrates the role of the penalty factors. They consider a
Hamiltonian of the form
\begin{equation}
H = \sum_a Y^a \sigma_a + \sum_i  \widetilde{Y^i} \sigma_i 
\end{equation}
where $\sigma_a $ are one- or two-qubit gates, and $\sigma_i$ are
three or higher qubit gates, taken to be tensor products of Pauli-matrices. Note that these generators are not normalized as before but rather 
$\text{tr}(\sigma_A\sigma_B) = 2^N \delta_{AB}$. With this choice, the relation between the cost functions \eqref{NielsenYNorms}-\eqref{NielsenHNorms} is rescaled accordingly. We will keep this normalization until the end of the section to match with the reviewed literature. 
The cost function is chosen as $F = \left(
\sum_a (Y^a)^2 + p \sum_i  (\widetilde{Y^i})^2\right)^{1/2}$ .
When the penalty factor $p$ is taken to be very large, one can expect that the optimal path will use only the ``easy'' gates. This can be formalized using the
projector $P \sigma_a = \sigma_a, P \sigma_i = 0$. First, one can show that if $U = \exp \int H (t)$, $U_P = \exp\int P H (t)$ , then 
\begin{equation}
\| U - U_P \| \leq \frac{2^N}{\sqrt{p}}  \mathcal{C}_F [U]\,.
\end{equation}
This shows that, by penalizing enough the higher order gates, 
the operator can be approximated with arbitrary precision using only one and two-qubit gates. For instance, choosing $\sqrt{p} > 4^N$, we obtain $\| U - U_P \| \leq   \mathcal{C}_F [U]/2^N$.

Then, replacing the functions $Y^a (t)$ with step-wise constant functions,
one can effectively discretize the integral, and exhibit a circuit built with one and two-qubit gates that approximates $U$. The discrete complexity $\mathcal{C}_{\text{d}}(U,\epsilon)$, defined as the number of gates in the optimal circuit that builds $U$ with a tolerance $\epsilon$, is then related to the continuous one as 
\begin{equation}
\mathcal{C}_\text{d}(U,\epsilon) \leq c \frac{N^6 \mathcal{C}_F[U]^3}{\epsilon^2} \,
\end{equation}
for some constant $c$. 
Moreover, as proven in \cite{Nielsen05}, the complexity gives also a lower bound on
the number of gates, provided the cost function satisfies certain conditions: given an exactly universal set of gates $\mathcal{G} =\{ e^{i X_i} \}$, which allows us to reach the target unitary exactly, and a cost function that satisfies $F[X_i]<1 \, \forall i$,\footnote{By $F[X_i]$ here, we mean the $F$ cost function defined with respect to the Hamiltonian $H=X_i$ and a choice of basis generators $\mathcal{O}_I$ from which the control functions $Y^I$ can be extracted.} then for any, unitary it holds that $\mathcal{C}_F[U] \leq \mathcal{C}_{\cal{G}}[U]$, where the latter is the {\it exact} discrete complexity of $U$ with respect to the gate set.  
This shows that the notions of discrete and continuous complexity are polynomially 
related to each other. It is not known what cost function gives the tightest bound; notably, $F_2$ is not optimal, since for all operators $F_2(U)\leq \pi$. 

\subsection{Complexity of One Qubit}

In order to get a better understanding of the complexity geometry, it is useful to consider the simplest possible case: a system of a single qubit. We follow mainly the presentation in 
\cite{Brown:2019whu}. 

As explained in the previous section, the choice of a cost function of the type $F_2$ is equivalent to the choice of a right-invariant metric on $SU(2)$. As is well-known, there is a unique (up to rescaling) right-and-left invariant metric; when equipped with this metric, the group is isometric to the round sphere $S^3$. The general right-invariant metric can be written using the right-invariant 1-forms $\omega^a$ defined by $dg \, g^{-1} = \omega^a i \sigma_a$: 
\begin{equation}
     ds^2 =  I_{ab} \, \omega^a \omega^b \,. 
\end{equation}
The maximally symmetric round-sphere is obtained when $I_{ab} = I \delta_{ab}$. If we choose, for instance, a diagonal matrix\footnote{For the basis-independent cost functions, we can always choose a basis that diagonalises the matrix.} but with different entries: $I_{xx}=I_{yy}=1, I_{zz} = p$, then the geometry is that of a squashed 3-sphere. Let us consider the following parametrization of $SU(2)$: 
\begin{equation}
    g = \begin{pmatrix} z_1 & z_2\\-\bar z_2 & \bar z_1 \\ \end{pmatrix},
\end{equation}
with $(z_1,z_2) \in \mathbb{C^2} \,, |z_1|^2+|z_2|^2=1$. 
In these coordinates the metric with the penalty factor $p$ is the pullback on $S^3$ of the following metric on $\mathbb{C^2}$:
\begin{equation}\label{C2penaltymetric} 
\begin{split}
    ds^2 =   & ~~~~~~dz_1 d \bar z_1 + dz_2 d \bar z_2 \\    & - \frac{p-1}{4} (z_1 d\bar z_1 - \bar z_1 dz_1 + z_2 d\bar z_2 - \bar z_2 dz_2)^2 \,.
\end{split}
\end{equation} 
The geodesics can be described explicitly  as follows \cite{Podobryaev}: the geodesic starting from the identity with tangent vector $v$ is given by 
\begin{equation}
    g(t) = R_{J}(t |J|) \, R_{\hat z}(t \gamma J_3) 
\end{equation}
where we used the same notation for the rotations as in section \ref{sec:QuantumComputationalComplexity}, $\gamma = \frac{1}{p}-1$ and $J$ is the angular momentum, related to the angular velocity as $J_a=I_{ab} v^b$.  Clearly for $\gamma=0$ we recover the usual geodesics on the sphere. \\
In coordinates, the geodesic trajectories are
\begin{equation}\label{S3geodesics}
\begin{split}
    z_1(t) = & e^{-i \gamma J_3 t/2} \left(\cos \frac{|J| t}{2} - i \hat J_3 \sin \frac{|J| t}{2} \right) \,, \\
     z_2(t) = & e^{-i \gamma J_3 t/2} (\hat J_1+i \hat J_2)  \sin \frac{|J| t}{2}  \,, \quad \hat J = J/|J| \,.
\end{split}
\end{equation}
It is instructive to consider the behavior of neighboring geodesics $g_J(t), g_{J+\delta J}(t)$; their difference gives the Jacobi vector field, whose length tells us whether geodesics converge or diverge; more precisely one has  \cite{do1992riemannian}
\begin{equation}
||\delta_w g_v(t)||^2 = t^2 - \frac{1}{3}K_{v,w} t^4 + o(t^4)
\end{equation}
where $v=\dot g_v(0)$, $w$ is a unit vector orthogonal to $v$, and $K_{v,w}$ is the sectional curvature of the plane spanned by $v,w$.  The calculation gives 
\begin{equation}
\begin{split}
    K_{1,3}=K_{2,3} & \propto p \,, \\
    K_{1,2} & \propto 4-3p \,.
\end{split}
     \end{equation}
We see that for $p=1$ all the sectional curvatures are equal, as the metric is isotropic. For $p>4/3$ the sectional curvature becomes negative in the plane $1,2$ spanned by the easy generators. This is a general feature, which can be understood as follows: since the commutator of two easy gates gives a hard one, it may be more efficient, in order to go from $\sigma_x$ to $\sigma_y$, to travel along the two axis rather than the hypotenuse. This appearance of hyperbolic geometry is a striking feature of complexity geometry, and can illustrate one important aspect, namely the fact that the distance in complexity can be much larger than the distance in the operator norm. In fact, there always exists a small ball around each point, inside which the direct geodesics are the shortest paths. Then for sufficiently small $\mathcal{C}_2(U)=\epsilon$, one has  $\epsilon \leq \mathcal{C}_{2,p}(U) \leq \sqrt{p} \, \epsilon$. For $p$ large the two distances can be very different. even though they go to zero together, so the complexity is still a continuous function of the distance. The difference becomes more significant when we consider systems with more degrees of freedom: in that case, as we have already seen, the complexity can increase exponentially in the number of qubits while the Hilbert space distance cannot. 

\begin{figure}[htbp]
\centerline{\includegraphics[scale=.6]{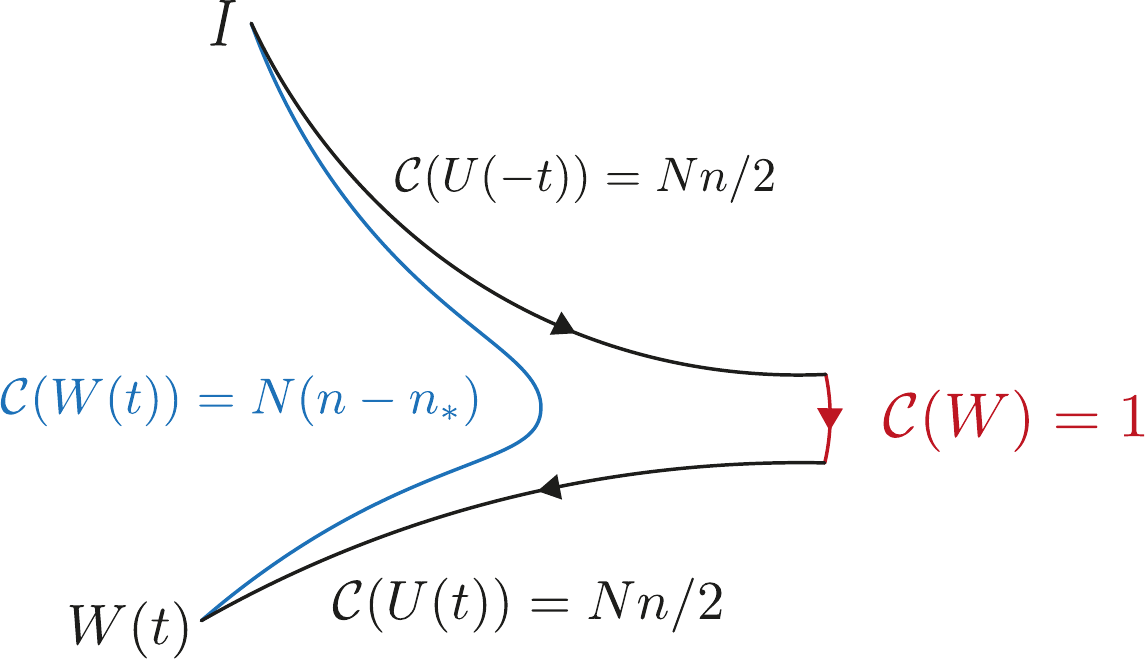}}
\caption{Illustration of the evolution of the complexity of the precursor as a geodesic deviation in negatively curved space.}
\label{fig:switchbacker34}
\end{figure}

As pointed out in \cite{Susskind:2014jwa}, a hyperbolic geometry similar to what we saw above but for a larger number of qubits accounts for the switchback effect discussed in section \ref{sec:fast-scramblers}. 
An initially small operator can be represented as a short segment in the space of unitaries. The precursor is obtained evolving in time the two ends of the segment. Connecting the ends with geodesics sweeps out a two-dimensional surface; if we assume a constant negative curvature on this surface, then one can show that the geodesic distance grows in time with the same features described by the switchback, \ie initially exponential and later linear with a time offset. This behavior is illustrated in Fig. \ref{fig:switchbacker34}.   

Finally, we can analyze in detail in this example the difference between operator complexity and state complexity. For the latter, we want to find the shortest path in operator space requiring that we reach a certain target state, so we define 
\begin{equation}\label{state-comp}
    \mathcal{C}(\ket{\psi_T},\ket{\psi_R}) = \min_U 
     \mathcal{C}(U), ~~~\text{s.t.}~~~ U\ket{\psi_R}=\ket{\psi_T} \,.
\end{equation}

The space of states of a qubit is $\mathbb{CP}^1 \approx S^2$. It can be identified with the coset $ SU(2)/H$ where $H$ is the stabilizer group of the action of $SU(2)$ on the states. Explicitly we can parametrize the group as 
\begin{equation}
    (z_1,z_2) =  \left(\frac{x}{\sqrt{1+x \bar x}}e^{i \alpha},\frac{1}{\sqrt{1+x \bar x}}e^{-i \alpha} \right) \, 
\end{equation}
and identify $x$ with the local coordinate on $\mathbb{CP^1}$. The minimization over the stabilizer in \eqref{state-comp} means that locally we have to choose a direction along the fiber that minimizes the length. When we write the metric \eqref{C2penaltymetric} in these coordinates, we find that one can extract a term $(d\alpha + \ldots)^2$. Setting this term to zero minimizes the length, and one is left with a metric which is best written in angle coordinates using the stereographic projection  $x=\cot(\frac{\theta}{2})e^{i \phi}$ : \begin{equation}\label{State-distance-p}
    ds^2 =\frac{1}{4} \left( d\theta^2 + \frac{p \sin^2 \theta}{\sin^2\theta + p \cos^2 \theta} d\phi^2 \right) \,.
\end{equation}
It is clear from the definition \eqref{state-comp} that the state complexity is in general not left-invariant, since the operator complexity is not: $\mathcal{C}(g\ket{\psi_T},g\ket{\psi_R}) \neq \mathcal{C}(\ket{\psi_T},\ket{\psi_R})$, and indeed the metric \eqref{State-distance-p} is not homogeneous. For large $p$ it has negative curvature everywhere except in a small region around the equator. 

So far we have considered only the geometry corresponding to the penalized $F_2$ cost. We could ask what is the distance for other costs, for instance $F_1$. Unfortunately, it is quite complicated to compute the geodesics, even in this simple setup of a single qubit. Looking at the definition \eqref{NielsenYNorms}, it is clear  that there is a simple case in which $\mathcal{C}_1$ and $\mathcal{C}_2$ coincide: when there is only one non-vanishing $Y^I$. In this case the geodesic can be written as the exponential of a single gate, and we should assume that the gate is contained in the basis. However the inspection of the geodesics \eqref{S3geodesics} shows that they do not have this simple form, except for the unpenalized case $\gamma=0$, or for the special geodesics with $J_3=0$.

\section{Complexity of Harmonic Oscillators}
\label{sec:CompSHO}

So far we have discussed the complexity of states over spin chains. Those states live in a finite dimensional Hilbert space. We can also study the complexity in infinite-dimensional Hilbert spaces as long as we focus on a specific sub-manifold of states generated by a closed algebra of operators. One example is that of Gaussian states of bosonic or fermionic systems. We will develop some technology to deal with this example which will come in handy later when studying complexity in free scalar quantum field theory.

\subsection{Complexity of Gaussian States}\label{sec:SHO}

Gaussian states can be fully characterized by their one- and two-point functions. To make use of this fact we will define the Gaussian states in terms of their {\it covariance matrix} and {\it displacement vector}, see \eg \cite{adesso2014continuous,eisert2003introduction,weedbrook2012gaussian} 
\begin{equation}\label{GOmega}
    \text{Tr}(\hat \rho \,\hat\xi^a \hat\xi^b) = \frac{1}{2}(G^{(ab)}+i \Omega^{[ab]}),\qquad \text{Tr}(\hat \rho \,\hat\xi^a)=w^a,
\end{equation}
where $\hat \rho$ is the density matrix representing the Gaussian state and $\hat\xi^a = (\hat q_1,\ldots, \hat q_N,\hat p_1,\ldots,\hat p_N)$ are $2N$ degrees of freedom on the quantum phase space consisting of position and momentum operators  which can be either fermionic or bosonic. In the case of a pure state \eqref{GOmega} simply becomes
\begin{equation}\label{GOmegaPure}
    \langle \psi | \hat \xi^a \hat \xi^b |\psi\rangle = \frac{1}{2}(G^{(ab)}+i \Omega^{[ab]}),\quad 
    \langle \psi |\hat \xi^a |\psi\rangle=w^a\,.
\end{equation}
In equations \eqref{GOmega}-\eqref{GOmegaPure},  $G^{(ab)}$ encodes the symmetric part of the correlation function and $\Omega^{[ab]}$ encodes its anti-symmetric part. 
To begin with, we take the simplifying assumption that the states have vanishing one-point functions $w^a=0$ in equations \eqref{GOmega}-\eqref{GOmegaPure}. The case of non-vanishing displacement will be treated later in section \ref{sec:Coherent-states}. 

We will focus mostly on the bosonic case below, but a lot of this machinery has also been adapted for studying fermionic states, see \eg \cite{eisert2010colloquium,Hackl:2020ken,Ge:2019mjt}.   
For a bosonic system $\Omega^{[ab]}$ is trivially fixed by the canonical commutation relations of the phase space operators
\begin{equation}
\Omega = \begin{pmatrix}0&\mathbb{1}_{n\times n}\\-\mathbb{1}_{n\times n}&0\\\end{pmatrix},
\end{equation}
and the only non-trivial information is in $G^{(ab)}$. Hence, from now on we will refer to $G^{(ab)}$ as the covariance matrix of the state $\hat \rho$.

For our complexity study we will focus on quantum circuits which move entirely within the space of Gaussian states with vanishing displacement and will therefore be parametrized using covariance matrices. Such circuits are generated by exponentiating quadratic generators as follows
\begin{equation}\label{kKrel1}
    \hat \rho(\sigma) = \widehat U(\sigma) \hat \rho(0) \widehat U^\dagger(\sigma), \quad 
    \widehat U (\sigma) =e^{-\frac{i}{2} \hat\xi^a k_{(ab)}(\sigma) \hat\xi^b}
\end{equation}
where $\widehat U(\sigma)$ is a unitary transformation parametrized by a symmetric matrix $k_{(ab)}(\sigma)$ and  $\hat\rho(\sigma)$ is the instantaneous density metric along the circuit with $\sigma\in[0,1]$ a path-parameter along the circuit.\footnote{Here, $\sigma$ plays the role of the time in the Hamiltonian control problem of section \ref{sec:ContComp}. We have changed the name here to distinguish it from the physical time of our systems which we will also be using in some of the calculations below.}
Then, with some algebra one can easily demonstrate that (see \eg \cite{TFD,ChargedTFD})
\begin{equation}
\begin{split}\label{kKrel2}
    & \widehat U^{\dagger}(\sigma) \, \hat\xi^a \, \widehat U(\sigma) = S(\sigma)^a{}_b(\sigma) \hat\xi^b, \\
    & G(\sigma) = S(\sigma) \cdot G(0)  \cdot S^T(\sigma),\\
    & S^a{}_b(\sigma) =  \left(e^{K(\sigma)}\right)^a{}_b, ~~~~ K^a{}_b = (\Omega \cdot k(\sigma))^a{}_b,
\end{split}
\end{equation}
where $G(\sigma)$ is the covariance matrix of the state $\hat \rho(\sigma)$ along the circuit. 
Note that $S(\sigma)$ in the last equation  belongs to the symplectic group $Sp(2N,\mathbb{R})$ by virtue of satisfying
\begin{equation}
    S(\sigma) \cdot \Omega \cdot S^T(\sigma) = \Omega.
\end{equation}

To make connection with the complexity functionals of equation \eqref{NielsenYNorms}, we should decompose the symplectic transformation using a fixed basis of generators $K_I$ of the symplectic group $Sp(2N,\mathbb{R})$
\begin{equation}\label{selgatebelow}
    S(\sigma) = \overleftarrow{\mathcal{P}} \exp \int_0^\sigma d\sigma' \, Y^I(\sigma') K_I
\end{equation}
and extract the control functions $Y_I$. 

The complexity depends on this choice of basis.  
One option is to fix the basis of generators $K_I$ in terms of our choice $\hat\xi^a$ of the operators on the quantum phase space. That is, we select  
\begin{align}
    &(K_{I=(a',b')})^a{}_b=(\Omega \cdot k_{I=(a',b')})^a{}_b,\quad a',b'\in{1,\ldots,2N},\nonumber
    \\
    &k^I_{(ab)} = \frac{1}{\sqrt{1+\delta_{a'b'}}} (\delta_a^{a'} \delta_b^{b'}+\delta_b^{a'} \delta_a^{b'}),\label{eq:basischoice}
\end{align}
which represent the generator $\exp{\left[-i\frac{ \hat\xi_{a'}\hat\xi_{b'}+\hat\xi_{b'}\hat\xi_{a'}}{2\sqrt{1+\delta_{a'b'}}}\right]}$, see  equations \eqref{kKrel1}-\eqref{kKrel2}.
The proportionality factor is fixed such that the different generators are orthonormal, \ie  $\frac{1}{2}\text{Tr}(K_I K_J^T) = \delta_{IJ}$. 
With this choice of basis we can extract the control functions
\begin{equation}
    Y^I = \frac{1}{2}\text{Tr}(\del_{\sigma}S S^{-1} K_I^T)\,.
\end{equation}
The norm \eqref{NielsenYNorms} with $p_I=1$ and $k=2$, which we refer to as the unpenalized $F_2=\sqrt{\sum_I |Y^I|^2}$ norm, can be expressed directly from the matrices $S(\sigma)$ along the circuit as follows
\begin{equation}\label{ds2Gmetricold}
    ds^2 = \frac{1}{2} \text{Tr} \left(dS \, S^{-1} \,  \, (dS \, S^{-1})^T \,  \right).
\end{equation}
This expression is written covariantly and does not require a particular choice of basis to be evaluated. However, note that to prove its equivalence with the unpenalized $F_2$ norm, we had to assume that the generators of the circuit are chosen to be orthonormal.

A natural generalization of the $F_2$ norm in equation \eqref{NielsenYNorms} is defined in terms of a given covariance matrix $G_{\text{metric}}$
\begin{equation}\label{ds2Gmetric}
    ds^2 = \frac{1}{2} \text{Tr} \left(dS \, S^{-1} \, G_{\text{metric}} \, (dS \, S^{-1})^T \, G_{\text{metric}}^{-1} \right).
\end{equation}
In effect, the choice of $G_{\text{metric}}$ introduces some penalty factors into the definition of the $F_2$ norm. When the generators of the symplectic group satisfy
\begin{equation}\label{orthorth}
    \frac{1}{2} \text{Tr} \left(K_I \, G_{\text{metric}} \, K_J^T G_{\text{metric}}^{-1} \right) = \delta_{IJ},
\end{equation}
we recover the unpenalized $F_2$ norm. 
More generally, we have 
\begin{equation}
    \frac{1}{2} \text{Tr} \left(K_I \, G_{\text{metric}} \, K_J^T G_{\text{metric}}^{-1} \right) = \gamma_{IJ}
\end{equation}
and $F_2 = \sqrt{\gamma_{IJ}Y^I Y^J}$ where $\gamma_{IJ}$ function as penalty factors. 
We would like to emphasize that the unpenalized $F_2$ norm is basis dependent. While remaining unmodified under orthogonal transformations which mix the positions among themselves (accompanied by the same orthogonal transformation on momenta), the  unpenalized $F_2$ norm in fact changes under more general symplectic transformations which modify the orthogonality condition \eqref{orthorth}, even with $G_{\text{metric}}=1$.

The complexity problem, \ie finding the optimal trajectory (or circuit) between a reference state $G_R$ and a  target state $G_T$ within the complexity geometry \eqref{ds2Gmetric}, can now be formulated explicitly as a geodesic problem, namely
\begin{equation}
\begin{split}
    ~~~~~\mathcal{C}_2 = \min_{S(\sigma)} \int_0^1 d\sigma \left(\frac{ds}{d\sigma}\right), \qquad \text{such that}
    \\
    ~~~~S(\sigma=1) G_R S^T (\sigma=1) = G_T\,.
\end{split}
\end{equation}
It was proven \cite{Hackl:2018ptj,TFD}, that when the matrix $G_{\text{metric}}$ used to define the geometry \eqref{ds2Gmetric} coincides with the covariance matrix $G_R$ of the reference state,  the geodesics from the reference state to the target state take a particularly simple form of ``straight lines'', \ie
\begin{equation}\label{STT}
    S(\sigma) = \exp\left[\frac{\sigma}{2} \log \Delta \right], \qquad \Delta \equiv G_T G_R^{-1},
\end{equation}
where $\Delta$ is the relative covariance matrix between the reference and the target state.

With the choice $G_{\text{metric}}=G_R$, and for generators satisfying the condition \eqref{orthorth}, the unpenalized $\mathcal{C}_2$ complexity, associated with the unpenalized $F_2$ cost function reads
\begin{equation}
    \mathcal{C}_2 (G_R,G_T) = \frac{1}{2\sqrt{2}}\sqrt{\text{Tr}[(\log \Delta)^2]}\,.
\end{equation}

While the trajectory \eqref{STT} does not necessarily minimize the unpenalized $F_1$ norm given by equation \eqref{NielsenYNorms} with $k=1$ and $p_I=1$, we could still evaluate its cost to obtain an upper bound on the unpenalized $\mathcal{C}_1$ complexity
\begin{equation}\label{C1UB}
    \mathcal{C}_1 \leq \mathcal{C}_1^{UB} = \sum_I |Y^I| = \frac{1}{4} \sum_I |\text{Tr}(\log\Delta \cdot K_{I}^T)|\,.
\end{equation}

\subsection{Single Harmonic Oscillator}\label{sec:SHO2}

As a specific example, let us focus on the bosonic case of a simple Harmonic oscillator described by the following Hamiltonian\footnote{In this section we will omit the hats from operators to simplify the notation. It should be clear from the context if we are considering an operator or an expectation value.}
\begin{equation}\label{SHOHamil}
    H = \frac{1}{2M}P^2+\frac{1}{2}M\omega^2 Q^2
\end{equation}
with $M$ and $\omega$ the mass and frequency of the oscillator, respectively, and $Q$ and $P$ are its position and momentum. In what follows it will be more convenient to work in terms of dimensionless position and space coordinates and hence we rescale
\begin{equation}\label{pqomg}
    p \equiv P/\omega_g, \qquad q\equiv\omega_g Q.
\end{equation}
(In the case of several positions and momentum operators we rescale all of them).
Later on, the scale $\omega_g$ will participate in defining a {\it gate scale} when discussing complexity. More precisely it will play a role in rendering the control functions $Y^I$ dimensionless. With the rescaled variables, the Hamiltonian takes the form
\begin{equation}\label{omegagH}
    H = \frac{\omega_g^2}{M} \,
    \left(\frac{1}{2}p^2 + \frac{1}{2} \lambda^2 q^2\right), \qquad \lambda\equiv\frac{M\omega}{\omega_g^2}.
\end{equation}

A general Gaussian wavefunction takes the form
\begin{equation}\label{SHOwavefunctionGeneral}
    \psi(q) = \langle q | \psi \rangle = \left(\frac{a}{\pi}\right)^{1/4} \exp \left[-\frac{1}{2} (a+ib)q^2\right]
\end{equation}
where $a$ and $b$ are real numbers and $a$ has to be positive in order for the wavefunction to be normalizable. For the special case of the vacuum state of the Hamiltonian \eqref{omegagH} we have $a=\lambda$ and $b=0$.

Explicitly evaluating the covariance matrix for the wavefunction \eqref{SHOwavefunctionGeneral} we obtain
\begin{equation}
    G = \begin{pmatrix} \frac{1}{a} &~ -\frac{b}{a} \\ -\frac{b}{a} &~~ \frac{a^2+b^2}{a} \end{pmatrix}
\end{equation}
and in particular for the vacuum state
\begin{equation}
    G_{\text{vac}} = \begin{pmatrix} \frac{1}{\lambda} &~ 0 \\ 0 &~~ \lambda \end{pmatrix}.
\end{equation}
As we will motivate later when discussing complexity in QFT, the reference state is often taken to be the ground state of another Hamiltonian with a different frequency $\omega=\mu$ and hence its covariance matrix is 
\begin{equation}\label{lambdarcovmat}
    G_R = \begin{pmatrix} \frac{1}{\lambda_R} &~ 0 \\ 0 &~~ \lambda_R \end{pmatrix}, \qquad \lambda_R = \frac{M\mu}{\omega_g^2}.
\end{equation}
The relative covariance matrix between the reference state and the vacuum reads
\begin{equation}\label{DLL}
    \Delta = \begin{pmatrix} \frac{\lambda_R}{\lambda} &~ 0 \\ 0 &~~ \frac{\lambda}{\lambda_R} \end{pmatrix}
\end{equation}
and so the unpenalized $\mathcal{C}_2$ complexity is simply
\begin{equation}\label{C2complexity1SHO}
    \mathcal{C}_2 (G_R,G_T) =
    \frac{1}{2} \left| \log \left(\frac{\lambda }{\lambda_R}\right)\right| = \frac{1}{2} \left| \log \left(\frac{\omega }{\mu}\right)\right|.
\end{equation} 
Note that in this expression the gate scale $\omega_g$ has canceled.

To obtain the bound \eqref{C1UB} on the  unpenalized $\mathcal{C}_2=\mathcal{C}_1^{UB}$ complexity  we should first select a basis. As described around equation \eqref{eq:basischoice}, we could consider circuits associated with the generators
\begin{equation}
    \widehat K_1 = \frac{1}{2}\left(pq+qp\right), \quad 
    \widehat K_2 = \frac{q^2}{\sqrt{2}},\quad
    \widehat K_3 = \frac{p^2}{\sqrt{2}}.
\end{equation}
Using the relations \eqref{kKrel1}-\eqref{kKrel2} we may read the relevant matrices $k_{(ab)}$ 
\begin{equation}
    \begin{split}
        k_{(ab)}^1 = \begin{pmatrix}0&1\\1&0\end{pmatrix},~
        k_{(ab)}^2 = \begin{pmatrix}\sqrt{2}&0\\0&0\end{pmatrix},~
        k_{(ab)}^3 = \begin{pmatrix}0&0\\0&\sqrt{2}\end{pmatrix},
    \end{split}
\end{equation}   
and the corresponding $Sp(2,\mathbb{R})$ generators:
\begin{equation}
    \begin{split}
        K_1 = \begin{pmatrix}1&0\\0&-1\end{pmatrix},~
        K_2 = \begin{pmatrix}0&0\\-\sqrt{2}&0\end{pmatrix},~
        K_3 = \begin{pmatrix}0&\sqrt{2}\\0&0\end{pmatrix}.
    \end{split}
\end{equation}
This leads to
\begin{equation}\label{C1complexity1SHO}
   \mathcal{C}_1^{UB} =
\frac{1}{2} \left| \log \left(\frac{\lambda}{\lambda_R }\right)\right| = \frac{1}{2} \left| \log \left(\frac{\omega }{\mu}\right)\right|.
\end{equation}

Note that in this very special case we have obtained the same result for the two cost functions. Generally this will not be the case. If we consider for example a system of many decoupled harmonic oscillators, each with Hamiltonian of the form \eqref{SHOHamil} but with  different frequencies $\omega_i$, the complexities will simply be given by
\begin{equation}\label{cmanyosc}
    \mathcal{C}_1^{UB} = \frac{1}{2} \sum_{i}  \left|\log \frac{\omega_i}{\mu}\right|;
\quad
    \mathcal{C}_2 =\frac{1}{2} \sqrt{\sum_{i} \left(\log \frac{\omega_i}{\mu}\right)^2}.
\end{equation}

\subsection{Complexity of Coherent states} 
\label{sec:Coherent-states}

We can extend the discussion of sections \ref{sec:SHO} to the case of Gaussian states with  non-zero displacement (cf. \eqref{GOmega}-\eqref{GOmegaPure}), \ie coherent states. We follow mostly the treatment of  \cite{Guo:2018kzl}, with some modifications (see also \cite{Yang:2017nfn} for a different approach). 
For simplicity, we focus on wavefunctions of the form 
\begin{equation}\label{coherent-wf}
    \psi(q_i) = {\cal N} \textrm{exp} \left[ - \frac{1}{2} A_{ij} (q_i -a_i)(q_j - a_j) \right] \,,
\end{equation}
with $A_{ij}$ and $a_i$ for $i\in \{1,\ldots,N\}$ real parameters. As a consequence, the displacement 
vector in \eqref{GOmegaPure} is non-vanishing only in the coordinates directions and is zero in the momenta, $\langle q_i \rangle = a_i,$ $\langle p_i \rangle =0$. Clearly this restricts the choice of symplectic transformations, as we can only allow transformations that do not mix coordinates and momenta.\footnote{We could, of course, use general symplectic transformations along the path and only impose the restriction on the final state, but for simplicity, we will not consider this possibility. Instead, we will restrict the gates along the entire circuit.} The transformations we consider take the form
\begin{equation}\label{coheretransform}
    q_i \to m_{ij} (q_j + b_j),
\end{equation}
where $m_{ij}$ is a general real matrix. These transformations keep us within the class of real wavefunctions  \eqref{coherent-wf}, in addition to keeping the vanishing expectation value of the momentum. The transformations \eqref{coheretransform} form the group $GL(N,\mathbb{R}) \ltimes \mathbb{R}^N$.

We could generalize the discussion of section \ref{sec:SHO} by introducing new gates that move within the space of coherent states. We will follow a different route which allows us to borrow the previous results directly. We observe that a coherent state wavefunction can be interpreted as a Gaussian wavefunction in a space with one more coordinate.  We rewrite \eqref{coherent-wf} as 
\begin{equation}\label{coherent-wf-ext}
    \psi(q_I) = {\cal N} \textrm{exp} \left[ - \frac{1}{2} \tilde A_{IJ} q_I q_J \right] \,,
\end{equation}
with $q_I = (q_0,q_i)$. At $q_0=1$, this reduces to \eqref{coherent-wf} if $\tilde A_{ij} = A_{ij}$, $\tilde A_{i0}=- A_{ij} a_j$, whereas the value of $\tilde A_{00}$ can be reabsorbed in the normalization factor and so is irrelevant. 

The  transformations \eqref{coheretransform} can be embedded into the group of linear transformations $GL(N+1,\mathbb{R})$ on the operators\footnote{Here we mean the transformation of $\hat q^i$ as in equation \eqref{kKrel2}. If the operator $U$ is such that $U^\dagger \hat q U = M \hat q$, the wavefunction will transform as $U \psi(q) = \psi(M^{-1}q)$.}
of the extended space as follows: 
\begin{equation}\label{matrixM}
   M = \begin{pmatrix}
    1 &  0  \\ m {\bf b} & m
    \end{pmatrix} \,, \quad m \in GL(N, \mathbb{R}) \,.
\end{equation}
The action on the wavefunction induced by $\hat q \to M \hat q$ is given by $q\rightarrow M^{-1} q$ or equivalently $\tilde A \to M^{-1}{}^T \tilde A \, M^{-1}$. Notice that the value of $q_0$ does not change under the action of $M$. 

In order to apply the formulas of section \ref{sec:SHO} we need the covariance matrix of the state and the symplectic transformations that act on it. They have a block-diagonal form: 
\begin{equation}\label{coherentGStrans}
    G = \begin{pmatrix}
    \tilde A^{-1} & 0 \\ 0 & \tilde A 
    \end{pmatrix} \,, \quad
    S = \begin{pmatrix}
    M & 0 \\ 0 & M^{-1}{}^T 
    \end{pmatrix} \,.
\end{equation}

With these ingredients at hand, we can use the formula \eqref{ds2Gmetric} for the metric.  
Choosing as before 
$G_{metric} =G_R= \begin{pmatrix}
\frac{1}{\lambda_R} \mathbb{1} & 0 \\ 0 &  \lambda_R \mathbb{1}
\end{pmatrix}$, this gives 
\begin{equation} \label{coherent-metric}
\begin{split}
    ds^2 &=  \textrm{tr} \left( dM \, M^{-1} (dM\, M^{-1})^T   \right) \\  
    & =  \textrm{tr}\left( dm \, m^{-1} (dm \, m^{-1})^T   \right) +  d {\bf b}^T m^T m d{\bf b}\,.
\end{split}
\end{equation}
We find that the $\mathbb{R}^N$ factor has a flat metric, but it is non-trivially fibered over the $GL(N)$ factor. 

In order to give a more explicit description of the geometry we restrict now to the case $N=2$. We can use the following parametrization of a $GL(2)$ matrix: 
\begin{equation}\label{mparam}
    m = \begin{pmatrix}
    \cos \alpha & -\sin \alpha \\  \sin \alpha & \cos \alpha \end{pmatrix} \begin{pmatrix}
    e^{-y_1} & 0 \\ 0 & e^{-y_{2}}
    \end{pmatrix} \begin{pmatrix}
    \cos \beta & -\sin \beta \\  \sin \beta & \cos \beta \end{pmatrix} .
\end{equation}
In these coordinates the metric \eqref{coherent-metric} reads 
\begin{equation}
    \begin{split}\label{metriccoherent}
        ds^2 =&  dy_1^2 + dy_2^2 + 2 d\alpha^2 \\ +&  4 \cosh(y_1-y_2) d\alpha d\beta +2 \cosh(2y_1-2y_2) d\beta^2 + \\
        +& e^{-2 y_1} (\cos \beta \, db_1 - \sin \beta \, db_2)^2 \\
        + & e^{-2 y_2} (\sin \beta \, db_1 + \cos \beta \, db_2)^2 \,.
        \end{split}
\end{equation}
The equations for the geodesics in this geometry cannot be solved analytically. An interesting property of this geometry, as was shown in \cite{Guo:2018kzl}, is that if we want to start from the reference state $A_R = \lambda_R \mathbb{1}$, ${\bf a}_R =0$ and arrive at the target state $A =\begin{pmatrix}
\lambda_1 & 0 \\ 0 & \lambda_2 
\end{pmatrix}$ with $\lambda_1 \neq \lambda_2$, and with $a_1,a_2$ both non-vanishing, then the corresponding geodesic will pass through states in which the two oscillators are entangled, even though in both the initial and final states the two oscillators are unentangled. 

If instead we turn on only one component of the displacement vector, it is possible to find simple geodesics analytically.   
One can show that the geodesics satisfying $\alpha=\beta=n\pi$, $b
_2=0$ can be obtained from the induced metric on this slice : 
\begin{equation}\label{H2R}
    ds^2 = dy_1^2+dy_2^2+e^{-2 y_1} db_1^2 \,.
\end{equation}
This geometry is $\mathbb{H}^2\times \mathbb{R}$, and we see the hyperbolic space in the coordinates $y_1$, $b_1$ arising from the fibration. 
The target states corresponding to this submanifold have $\langle q_2 \rangle =0$ and are unentangled in the given coordinates. It is easy to evaluate the complexity of a target state with $A_T=\text{diag}(\lambda_1,\lambda_2 )$, ${\bf a}=(a_1,0)$ and obtain
\begin{equation}
    \mathcal{C}_2  = \sqrt{ \frac{1}{4} \log^2 \frac{\lambda_2}{\lambda_R} + \textrm{arccosh}^2\left( \frac{\lambda_R+\lambda_1+ \lambda_1 a_1^2}{2 \sqrt{\lambda_R \lambda_1}} \right)}\,.
\end{equation}
The geometry \eqref{H2R} is simple enough that in this case we can compute explicitly the complexity also for the $F_1$ cost function, rather than just giving an upper bound.\footnote{Recall that the $F_1$ cost function depends on a choice of basis. Here we use the basis described in equation \eqref{eq:basischoice}, \ie we construct our gates with respect to the coordinates of the two oscillators and the new fictitious coordinate $\hat x^0$. This is explained in detail in \cite{Guo:2018kzl}.} Since the $y_2$ direction is decoupled, we can consider trajectories in the $y_1,b_1$ direction; for simplicity we rename them as $y,b$. The cost function is
\begin{equation}
    F_1 = \int ds \left( |\dot y | + e^{-y} |\dot b| \right) \,.
\end{equation}
This is a singular functional, so we cannot find solutions from the equations of motion. Let us consider a trajectory from $(y_i,b_i)$ to $(y_f,b_f)$ and assume for simplicity that $y_f> y_i, b_f > b_i$. If we assume that $\dot y(s)>0$, the first term is independent of the trajectory, and the second term is minimized by making $y$ as large as possible. The minimal trajectory will move in a straight line first along the $y$ axis, and then along the $b$ axis at $y=y_f$. The cost of this path is $ \Delta y + e^{- y_f} \Delta b$. 
But it can be more convenient to minimize the second term by moving along $b$ at a larger value of $y$, say $\tilde y$, paying the price of backtracking in the $y$ direction. The minimum length is obtained for $e^{\tilde y} = \Delta b/2$, and is $2+2 \log \frac{\Delta b}{2} - y_i - y_f$. This path has shorter length when $\tilde y>y_f$, or $e^{-y_f}\Delta b >2$. In terms of the parameters of the wavefunction, moving from the reference state to the target state $\lambda$ , $a$ (with $\lambda>\lambda_R$) and using the relations $y_f=\frac{1}{2}\log\frac{\lambda}{\lambda_R}$, $b_f=\sqrt{\frac{\lambda}{\lambda_R}}\, a$ and $b_i=y_i=0$,\footnote{
To obtain these relations, use the target and reference state matrices defined below equation \eqref{metriccoherent} and relate them to 
$\tilde A$ using the relations below equation \eqref{coherent-wf-ext}. The values of $b$ and $y$ at the end of the trajectory can then be fixed in terms of the wavefunction transformation below equation \eqref{matrixM} (see also equation \eqref{mparam} for the parametrization of $m$).}
 we find a cost 
\begin{equation}
    \begin{split}
    \mathcal{C}_1 & = \frac{1}{2} \log \frac{\lambda}{\lambda_R} + |a| \,, \quad &  |a| < 2\,, \\
    \mathcal{C}_1 & = \frac{1}{2} \log \frac{\lambda}{\lambda_R} + 2 + 2 \log \frac{ |a|}{2} \,, \quad &  |a| > 2 \,.
     \end{split}
\end{equation}
Similar results can be obtained for $\lambda < \lambda_R$. Notice that the contribution from the displacement is frequency-independent. 
The dependence on $a$ is linear for small $a$, whereas it is quadratic for the $\mathcal{C}_2$ case. For large $a$ the leading behavior is $\log (a^2)$ in both cases, but the subleading terms are different and are frequency-dependent for $\mathcal{C}_2$. The path that minimizes $\mathcal{C}_1$ is not the same that minimizes $\mathcal{C}_2$, so the upper bound $\mathcal{C}_1^{UB}$ from the previous sections is not saturated.

\subsection{Complexity of the Thermofield Double State}\label{sec:Complexity-TFD-SHO}

A particularly interesting example of a Gaussian state of vanishing displacement whose complexity can be studied using the techniques of section \ref{sec:SHO} is  the thermofield double (TFD) state of a single harmonic oscillator. The complexity of this state was studied in \cite{TFD} (see also \cite{ChargedTFD}). The TFD state is defined with respect to two identical copies of a given system as follows 
\begin{equation}\label{TFDt}
\ket{TFD(t)} = \mathcal{N}_{\text{TFD}}  \sum_n e^{-\frac{\beta E_n}{2}-iE_n t} |E_n\rangle_L |E_n\rangle_R
\end{equation}
where the two copies have been labeled left and right ($L/R$), $E_n$ are the energy eigenstates, $t$ is the time, $\beta$ is the inverse temperature and $\mathcal{N}_{\text{TFD}}$ is a normalization constant. The TFD state is a pure state which evolves non-trivially under time evolution.\footnote{Although it is invariant under the action of $H_L-H_R$.} It is also a particularly symmetric purification of the thermal state, \ie when considering the reduced density matrix and tracing out the right subsystem we are left with a mixed thermal state on the left subsystem -- more on that in the next section.

If we focus on the example of the single harmonic oscillator from section \ref{sec:SHO}, we will have energy eigenstates defined according to the Hamiltonian \eqref{SHOHamil}
\begin{equation}
    H |n\rangle = \omega \left(n+\frac{1}{2}\right) \ket{n}.
\end{equation}
Of course, since we are working with two copies of the system, we will have both left and right energy eigenstates $|n\rangle_L$ and $|n\rangle_R$. In terms of these eigenstates the TFD state reads
\begin{equation}\label{TFDt2}
\begin{split}
& \ket{TFD(t,\omega)} = \mathcal{N}_{\text{TFD}} \sum_{n=0}^\infty e^{-n\beta\omega/2-i(n+\frac12)\omega t} \ket{n}_L \ket{n}_R\\ 
& =
e^{-i \omega t/2} \mathcal{N}_{\text{TFD}}  \exp{\left[e^{-\beta\omega/2-i\omega t} a_L^\dagger a_R^\dagger\right]} \ket{0}_L \ket{0}_R.
\end{split}
\end{equation}
The second line shows that this state is Gaussian since it is produced from the vacuum state using a quadratic operator. 
It will be convenient to combine the position and momentum operators for the left and right copies as follows
\begin{equation}
    Q_{\pm} = \frac{1}{\sqrt{2}}(Q_L\pm Q_R), \quad P_{\pm} = \frac{1}{\sqrt{2}}(P_L\pm P_R),
\end{equation}
and define their dimensionless versions according to equation \eqref{pqomg}. In these $\pm$ coordinates, the 4x4 covariance matrix is block diagonal. The blocks have the form
\begin{equation}
\begin{split}\label{covTFDt}
   &G_{\text{TFD}}^{\pm}(t) =
   \\&{\footnotesize
   \begin{bmatrix}
    \frac{(\cosh(2\alpha)\pm\sinh(2\alpha)\cos(\omega t))}{\lambda} &  \mp\sinh(2\alpha)\sin(\omega t)  \\ \mp\sinh(2\alpha)\sin(\omega t) &  \lambda (\cosh(2\alpha)\mp\sinh(2\alpha)\cos(\omega t))
    \end{bmatrix}} \,, 
\end{split}
\end{equation}
where we have defined
\begin{equation}
    \alpha = \frac{1}{2}\log \left[\frac{1+e^{-\beta\omega/2}}{1-e^{-\beta\omega/2}}\right],
\end{equation}
and $\lambda$ has been defined in equation~\eqref{omegagH}. The reference state for each of the blocks is taken as in equation \eqref{lambdarcovmat} and selecting $G_{metric}=G_R$ as described above equation \eqref{STT}, we can evaluate the $\mathcal{C}_2$ complexity as before. At $t=0$, we obtain
\begin{equation}
    \mathcal{C}_2 = \sqrt{\frac{1}{2} \log^2 \frac{\omega}{\mu} + 2\alpha^2}.
\end{equation}
Note that the gate scale $\omega_g$ canceled from this expression.
Evaluating the length of the $\mathcal{C}_2$
optimal circuit with the $F_1$ cost function yields at $t=0$ in the basis defined with respect to the $Q^{\pm}$ and $P^{\pm}$ coordinates
\begin{equation}
    \mathcal{C}_1^{(\pm),UB} = \left|\frac{1}{2}\log \frac{\omega}{\mu} + \alpha \right| +
    \left|\frac{1}{2}\log \frac{\omega}{\mu} - \alpha \right|.
\end{equation}
When considering a basis which acts naturally on the physical $L$ and $R$ degrees of freedom rather than the $\pm$ modes, we obtain the following complexity at $t=0$
\begin{equation}
    \mathcal{C}_1^{(LR),UB} = |\log (\omega/\mu)\,| + 2|\alpha|.
\end{equation}
We will see later that the results of the measure $\mathcal{C}_1^{(LR)}$ match best with holography.

It is interesting to compare the complexity of the TFD state at $t=0$ to that of two copies of the vacuum state, see equations \eqref{C2complexity1SHO} and \eqref{C1complexity1SHO}. We refer to this difference in complexities as the \emph{complexity of formation} of the thermal state \cite{Formation}
\begin{equation}\label{formationdef}
    \Delta \mathcal{C} \equiv \mathcal{C} (|TFD(t=0)\rangle) - 2\mathcal{C} (|0\rangle)\,.
\end{equation}
This yields for the various cost functions
\begin{equation}
\begin{split}\label{formationfinal1SHO}
    &\Delta \mathcal{C}_2 = \sqrt{\frac{1}{2} \log^2 \frac{\omega}{\mu} + 2\alpha^2} - \frac{1}{\sqrt{2}} \left|\log \frac{\omega}{\mu} \right|\,,
    \\
    &\Delta\mathcal{C}_1^{(\pm),UB} = \left|\frac{1}{2}\log \frac{\omega}{\mu} + \alpha \right| +
    \left|\frac{1}{2}\log \frac{\omega}{\mu} - \alpha \right|- \left|\log \frac{\omega}{\mu} \right|\,,
    \\
    &\Delta\mathcal{C}_1^{(LR),UB} = 2|\alpha|\,.
\end{split}
\end{equation}
We can also evaluate the complexity at a different time $t\neq 0$, but the expressions are slightly more cumbersome and we will not write them here. We refer the reader to section 4.4 of 
\cite{TFD}. In general at $t\neq 0$ the gate scale $\omega_g$ dependence will not cancel out. However, simplified expressions can be obtained when choosing it such that $\lambda_R=1$. We will make this choice from now on. Let us further remark that due to the periodic time dependence in the covariance matrix \eqref{covTFDt}, it is clear that the complexity will oscillate in time with frequency $\omega$. The contribution of these oscillations to the complexity can be shown to be exponentially suppressed at large $\beta \omega$ (\ie $\Delta\mathcal{C} \sim e^{-\#\beta \omega}$).

\subsection{Complexity of Mixed States}\label{sec:Complexity-Mixed-States}

So far we have focused on the complexity of pure states. However, it is of interest to try and define complexity for mixed states too.  
In this section we will focus on one such definition - the \emph{complexity of purification}, \ie the lowest value of the circuit complexity optimized over the possible purifications of the mixed state we are interested in.

\begin{figure}[htbp]
\centerline{\includegraphics[scale=0.95]{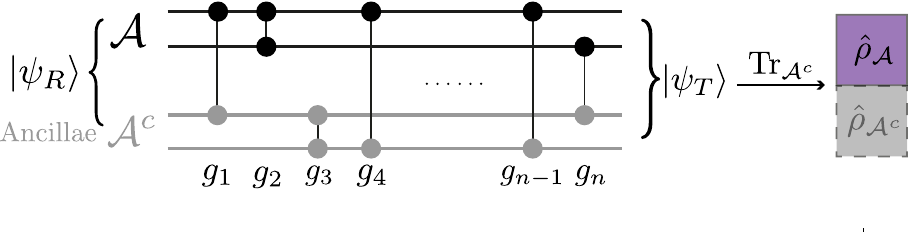}}
\caption{Illustration of the definition of complexity of purification. We purify the reduced density matrix $\rho_\mathcal{A}$ in terms of ancilla degrees of freedom on a system $\mathcal{A}^c$ and optimize the preparation of the state of the combined system.}
\label{fig:purification}
\end{figure}

More precisely, imagine that we start with a  mixed state of a system $\mathcal{A}$ described by the density matrix $\hat\rho_{\mathcal{A}}$. To purify the mixed state we supplement the degrees of freedom in $\mathcal{A}$ with ancillary degrees of freedom in a complementary system $\mathcal{A}^c$. We consider purifications of the state $\hat \rho_{\mathcal{A}}$, \ie pure states on the combined system
$|\psi_{\mathcal{A} \mathcal{A}^c}\rangle$ such that $\hat \rho_{\mathcal{A}} = Tr_{\mathcal{A}^c} |\psi_{\mathcal{A} \mathcal{A}^c}\rangle \langle\psi_{\mathcal{A} \mathcal{A}^c}|$.
The complexity of purification is simply defined as the minimal pure state complexity among all such possible purifications and all possible ancillary system sizes 
$\mathcal{C}(\hat\rho_\mathcal{A}) = \min \mathcal{C}(|\psi_{\mathcal{A} \mathcal{A}^c}\rangle)$ starting with a completely unentangled reference state on the combined $\mathcal{A} \mathcal{A}^c$ system. Fig. \ref{fig:purification} illustrates this process. 

% figure* makes it in the two columns

\begin{figure}[htbp]
\centerline{\includegraphics[scale=.42]{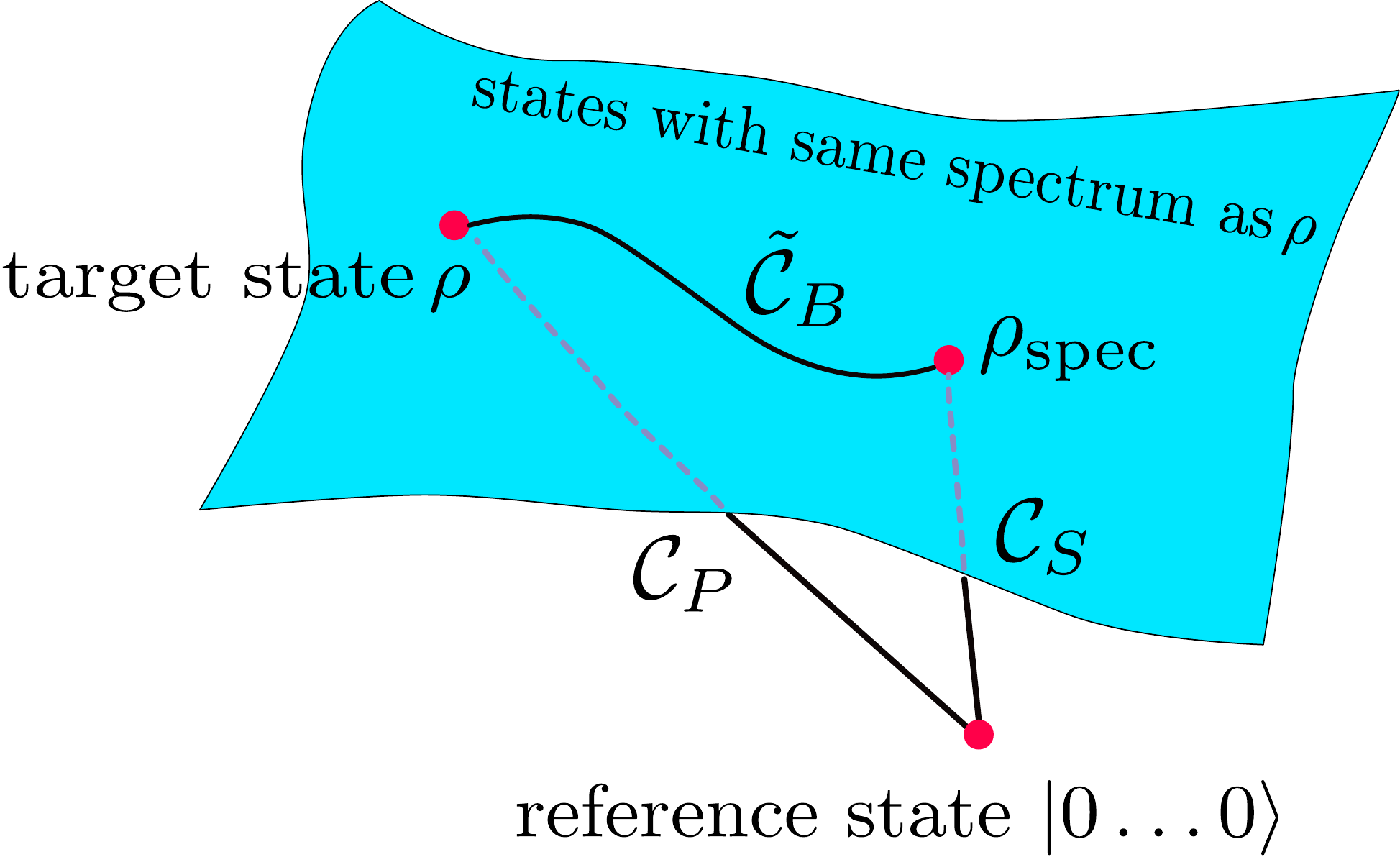}}
\caption{Illustration of spectrum and basis complexity for mixed states.}
\label{fig:spectrum}
\end{figure}

Several alternative definitions for mixed state complexity  have been proposed. For example, we can consider an approach based on the spectrum of eigenvalues $p_i$ of the density matrix $\hat\rho = \sum_i p_i |\phi_i\rangle \langle \phi_i |$, see, \eg \cite{Agon:2018zso}. In this approach, one breaks the process of constructing the state $\hat\rho$ into two separate parts. 
First, we define the \emph{spectrum complexity} $\mathcal{C}_S$ of the state $\hat\rho$ as the minimal complexity of purification among all states with the same spectrum as $\hat \rho$. We will denote the state for which this minimum is achieved by $\hat\rho_{\text{spec}}$. 
Second, we turn the state $\hat\rho_{\text{spec}}$ into our state of interest by using unitary operations with minimal complexity. This is always possible since the two states have the same spectrum. We call this part the basis complexity $\widetilde{\mathcal{C}}_B$. In any case, the 
complexity of purification $\mathcal{C}_P$ is always smaller than $\mathcal{C}_S+\widetilde {\mathcal C}_B$, because reaching the mixed state via $\hat\rho_{\text{spec}}$ is one possible circuit. The spectrum approach to mixed state complexity is illustrated in Fig.  \ref{fig:spectrum}.

Another approach to mixed state complexity is the \emph{ensemble complexity}, see, \eg \cite{Agon:2018zso}. Here, as before, we decompose the mixed state $\hat\rho$ as an ensemble of pure states $\rho = \sum_i p_i |\psi_i\rangle \langle \psi_i|$ and define the ensemble complexity as the weighted average over the complexities of the pure states in this ensemble, minimized over all possible ensembles, \ie  $\mathcal{C}_E(\rho) = \min_{\text{ensemble}} \sum_i p_i \mathcal{C}(|\psi_i\rangle)$.

Yet another approach to mixed state complexity is based on using an information metric metric adapted to trajectories between mixed states directly, without purifying them first. For example, 
\cite{DiGiulio:2020hlz,Ruan:2020vze} considered the Bures metric or Fisher-Rao information metric.

A more detailed discussion of mixed state circuits and complexity can be found in, \eg  \cite{Caceres:2019pgf,Agon:2018zso,Aharonov:1998zf,nielsen2002quantum,DiGiulio:2020hlz,Ruan:2020vze}. However, as we said before, here we will focus on the complexity of purification.

As before, when restricting to Gaussian states we are able to make considerable progress in studying the complexity \cite{Caceres:2019pgf} (see also \cite{Camargo:2018eof}). 
Let us start again with the example of a simple Harmonic oscillator and consider the most general mixed state with real parameters\footnote{The choice of real parameters was made to keep the derivation as simple as possible. A discussion which incorporates  complex wavefunction parameters can be found in appendix C of \cite{Camargo:2018eof}.} 
\begin{equation}\label{mixed1}
    \rho(x,x') \equiv \langle x |\hat \rho| x' \rangle  \propto  e^{-\frac{1}{2} (a x^2+a x'^2 -2 bx x')}
\end{equation}
where the density matrix is Hermitian $\rho(x,x') = \rho^*(x',x)$ as it should be,  and $a$ and $b$ are real parameters satisfying $a>b$ and $b\geq 0$, such that the density matrix is normalizable and positive semidefinite. The normalization constant is fixed by requiring $\text{Tr}(\rho) = \int \rho(x,x)=1$.
The most general purification with two degrees of freedom and real parameters reads
\begin{equation}\label{purification1}
    \psi_{12}(x,y) \equiv \langle x,y |\psi\rangle \propto  e^{-\frac{1}{2}(\omega_1 x^2+\omega_2 y^2 +2 \omega_3 xy)},
\end{equation}
where in order to indeed be a purification of the state \eqref{purification1} should satisfy
\begin{equation}
    \int dy  \, \psi_{12} (x,y) \psi^*_{12} (x',y) = \rho(x,x').
\end{equation}
Explicitly this yields
\begin{equation}
    \omega_1 = a+b, \qquad \omega_2 = \frac{\omega_3^2}{2b},
\end{equation}
where $\omega_3$ remains a free parameter. 
We can easily diagonalize the wavefunction \eqref{purification1} and bring it to the form
\begin{equation}
    \psi_{12} (x_+,x_-) \propto \, e^{-\frac{1}{2}(\omega_+ x_+^2 + \omega_- x_-^2)}
\end{equation}
where $\omega_{\pm}$ are the eigenvalues of the matrix
\footnotesize
$
\left(
\begin{array}{cc}
 a+b & \omega_3 \\
 \omega_3 & \frac{\omega_3^2}{2b} \\
\end{array}
\right)
$.
\normalsize
In this form, the two oscillators decouple and we can use equation \eqref{cmanyosc} to evaluate the complexity. We focus on the $\mathcal{C}_1$ complexity since it will be most closely related to holography as we will see later on. 
We obtain the upper bound 
\begin{equation}
\mathcal{C}_1^{\text{diag},UB} = \min_{\omega_3} \frac{1}{2} \left|\log \frac{\omega_+}{\mu}\right| +\frac{1}{2} \left| \log\frac{\omega_-}{\mu}\right|
\end{equation}
where $\mu$ is the reference state scale and the final answer is obtained by minimizing over the purification free parameter $\omega_3$.
The diag superscript indicates that we evaluate the $\mathcal{C}_1$ complexity in the \emph{diagonal basis}, whose generators are defined with respect to the coordinates $x_{\pm}$ according to the prescription described in equation \eqref{eq:basischoice}. It is also possible to explore the complexity in the \emph{physical basis} which distinguishes naturally the physical and ancillary degrees of freedom  \cite{Caceres:2019pgf} but we will not pursue this possibility here.\footnote{This choice was merely done in order to allow us to present some of the following expressions analytically, but the behaviors obtained  when studying the complexity of mixed states in the diagonal basis and in the physical basis are qualitatively similar.}

In the above example, we purified a mixed state of a single harmonic oscillator  using one additional harmonic oscillator. It is always the case that doubling the number of degrees of freedom in the system is enough to purify it.\footnote{Similarly to the TFD state, we can purify a density matrix $\hat\rho = \sum_i p_i \ket{i}\bra{i}$ with $\ket{\Psi}= \sum_i \sqrt{p_i} \ket{i}_1 \ket{i}_2$.} However, one might wonder if purifications with more degrees of freedom are more efficient from the complexity point of view. Testing the above with purifications of a single oscillator using  two ancillary oscillators, one concludes  that at least for such small systems
optimal purifications are \emph{essential purifications} – which use the smallest number of degrees of freedom necessary for the purification.

We can use the above results to answer the question - is the thermofield double state  of two harmonic oscillators of frequency $\omega$ at $t=0$ (cf. equation \eqref{TFDt2})
\begin{equation}\label{TFDstate}
    |TFD\rangle_{12} = \mathcal{N}_{TFD} \sum_{n=0}^\infty e^{- \beta  \omega n/2} |n\rangle_1 |n\rangle_2
\end{equation}
the optimal purification of the thermal state
\begin{equation}\label{thermalState}
    \hat\rho_{th} = \mathcal{N}_{th} \sum_{n=0}^\infty e^{-\beta  \omega n} |n\rangle \langle n|,
\end{equation}
where $|n\rangle$ are the energy eigenstates of our oscillator and $\beta$ is the inverse temperature.
Using Mehler's formula for summation over Hermite polynomials we can show that the thermal state is Gaussian of the form \eqref{mixed1} with the following  parameters
\begin{equation}
    a=\omega \coth(\beta \omega), \quad b=\frac{\omega}{\sinh(\beta \omega)},
\end{equation}
while the thermofield double state is also Gaussian of the form \eqref{purification1} with parameters 
\begin{equation}
    \omega_1 =\omega_2 = \omega \coth(\frac{\beta \omega}{2}), \quad \omega_3=-\frac{\omega}{\sinh(\frac{\beta\omega}{2})}.
\end{equation}
Minimizing over all possible purifications of the thermal state encloses a larger family of purifications than just the TFD state. Performing the minimization yields the following complexity of purification
\footnotesize
\begin{equation}\label{cth}
\begin{split}
    &\mathcal{C}_1^{UB,\text{diag}}(\hat\rho_{th}) = \\
    &  \begin{cases}
    \frac{1}{2}\log \frac{\mu}{\omega} + \frac{1}{2} \log \left(\frac{\mu\coth\left(\frac{\beta\omega}{2}\right)-\omega }{\mu-\omega \coth\left(\frac{\beta\omega}{2}\right)}\right) 
    & 
    \beta\omega\coth\left(\frac{\beta\omega}{4}\right) \leq \beta \mu 
    \\
    \log \coth \frac{\beta \omega}{4} 
    &
    {\beta\omega\tanh\left(\frac{\beta\omega}{4}\right) \leq \atop \beta \mu \leq \beta\omega\coth\left(\frac{\beta\omega}{4}\right)}
    \\
    \frac{1}{2}\log \frac{\omega}{\mu} + \frac{1}{2} \log \left(\frac{\omega\coth\left(\frac{\beta\omega}{2}\right)-\mu}{\omega-\mu \coth\left(\frac{\beta\omega}{2}\right)}\right) 
    & 
    \beta \mu \leq \beta\omega\tanh\left(\frac{\beta\omega}{4}\right) 
    \end{cases}.
\end{split}
\end{equation}
\normalsize

\begin{figure}[htbp]
\centerline{\includegraphics[scale=.4]{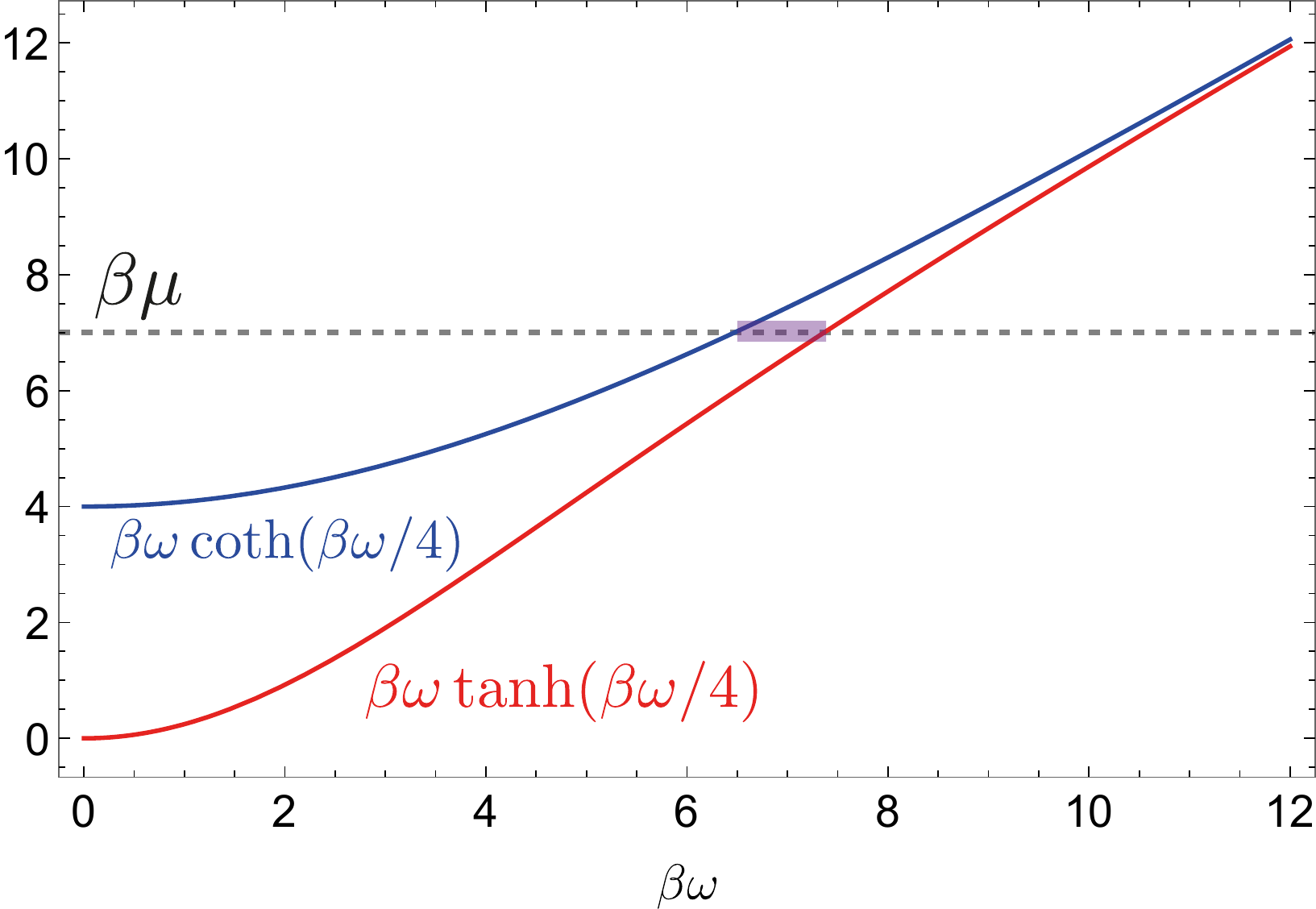}}
\caption{Plot illustrating the three regimes of the complexity of the thermofield double state and the complexity of purification of the thermal state. The gray dashed curve indicates our choice of the reference state scale (in this case we have chosen $\beta \mu=7$) and depending on whether it is higher or lower than the red and blue curves (the functions $\beta\omega \tanh(\beta\omega/4)$ and $\beta\omega \coth(\beta\omega/4)$) tells us which is the relevant complexity regime in equations \eqref{cth}-\eqref{CTFD}. Only in the small range of frequency indicated in purple will the thermofield double be the optimal purification of the thermal state. }
\label{fig:regimesThermal}
\end{figure}

Comparing this to the complexity of the thermofield double, \ie without the additional minimization over purifications, we obtain
\begin{equation}\label{CTFD}
\begin{split}
    &\mathcal{C}_1^{UB,\text{diag}}(|TFD\rangle_{12})  =\\
    &  \begin{cases}
    \log \frac{\mu}{\omega} &  \beta\omega\coth\left(\frac{\beta\omega}{4}\right) \leq \beta \mu 
    \\
    \log \coth \frac{\beta \omega}{4} 
    &
    {\beta\omega\tanh\left(\frac{\beta\omega}{4}\right) \leq \atop \beta \mu \leq \beta\omega\coth\left(\frac{\beta\omega}{4}\right)}
    \\
    \log \frac{\omega}{\mu}  
    & 
    \beta \mu \leq \beta\omega\tanh\left(\frac{\beta\omega}{4}\right) 
    \end{cases}.
\end{split}
\end{equation}
From the comparison of the two above results we see that the thermofield double state is the optimal purification of the thermal state only in the middle regime (which may be quite narrow), see Fig. \ref{fig:regimesThermal}.

\section{Complexity in QFT }
\label{sec:Complexity-free-QFT}
After having extensively studied the complexity of a small number of harmonic oscillators, we are now ready to use those results to study the complexity of states within Quantum Field Theory (QFT) -- the framework studying many body physics with changing particle number. 
We will consider the complexity of the vacuum state, the thermofield double state and several interesting examples of mixed states of free (or nearly free) bosonic field theories. Just like many other quantities in QFT, we will see that also the complexity diverges due to contributions from short distance correlations in the system. We will explain how to regulate those divergences.  We will conclude this section with a discussion of complexity in strongly interacting conformal field theories.

\subsection{Free Scalar QFT}

Here we describe the pioneering works \cite{QFT1,QFT2} which were the first to study complexity in a simple QFT. These works studied the complexity of the vacuum state of a free bosonic QFT in $d$ spacetime dimensions described by the following Hamiltonian
\begin{equation}\label{FreeBosonH}
    H = \frac{1}{2} \int d^{d-1}x \left[\pi(x)^2+\vec\nabla\phi(x)^2+m^2 \phi(x)^2\right].
\end{equation}

Naively, we expect the vacuum state to be simple and therefore to have low complexity. However, the complexity is defined with respect to a reference state. While there is no canonical choice of a state in a Hilbert space,  we will argue below that there is a natural choice of the reference state in the context of studying quantum computational complexity, which is a completely unentangled state. With this choice, it turns out that the complexity of the vacuum state in QFT is highly divergent. This is because the vacuum state has correlations down to arbitrarily short length scales which are absent in the reference state. 
For readers familiar with the notion of entanglement entropy this should not come as a surprise since a similar divergence appears there. 
One way to regularize the divergences is by placing the theory on a spatial lattice. Alternatively, we could use a sharp momentum cutoff. 
Both the entanglement entropy and complexity diverge when the lattice spacing is sent to zero.

\begin{figure}[htbp]
\centerline{\includegraphics[scale=.5]{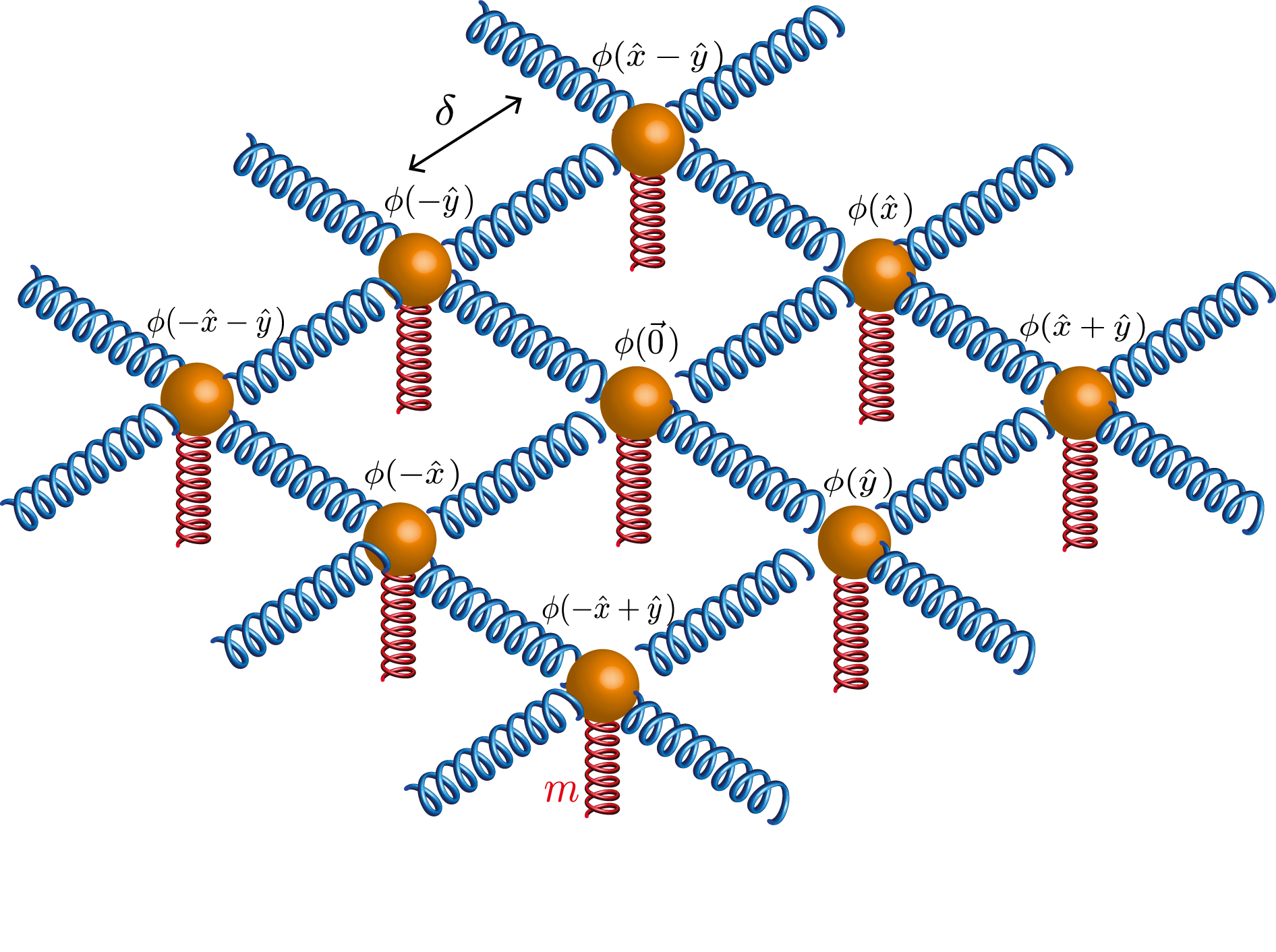}}
\caption{Illustration of a system of coupled harmonic oscillators obtained by discretizing the scalar field theory on a lattice with spacing $\delta$. Red springs represent contributions from the mass $m$ of the scalar field while blue springs introduce couplings between the different oscillators originating from the derivative term in the Hamiltonian.}
\label{fig:lattice}
\end{figure}

As in \cite{QFT1}, we will regularize the divergences by placing the theory on a $d-1$ dimensional periodic lattice with lattice spacing $\delta$ and length $L$ in all directions, see Fig. \ref{fig:lattice}. In this way, the theory becomes that of $N^{d-1} = (L/\delta)^{d-1}$ coupled harmonic oscillators and the complexity is a natural extension of the results of section \ref{sec:SHO2}. 
We will  label the different lattice sites in terms of a $d-1$ dimensional vector $\vec a$ where each component $0\leq a_i \leq N-1$ is an integer. The discretized version of \eqref{FreeBosonH} reads
\begin{equation}\label{DiscPosH}
    H = \sum_{\vec a} \delta \tilde\pi_{\vec a}^2 + m^2 \delta^{-1} \tilde \phi_{\vec a}^2 + \delta^{-3} \sum_j (\tilde \phi_{\vec a+\vec e_j}-\tilde \phi_{\vec a})^2
\end{equation}
where we have defined $\tilde \phi_{\vec a} = \delta^{d/2}\,\phi(\delta \cdot \vec a)$, $\tilde \pi_{\vec a} = \delta^{d/2-1}\pi(\delta \cdot \vec a)$ and $\vec e_j$ denotes the unit vector in the $j$-th direction. Periodicity implies $\tilde\phi_{\vec a+N \vec e_j}=\tilde\phi_{\vec a}$ and $\tilde\pi_{\vec a+N \vec e_j}=\tilde\pi_{\vec a}$ for all $\vec a$-s and $j$-s. The above coordinate and momentum operators satisfy the commutation relations $[\tilde \phi_{\vec a},\tilde \pi_{\vec{b}}] = i \delta_{\vec a \vec b}$. To decouple the different oscillators in \eqref{DiscPosH} we employ a discrete Fourier transform
\begin{equation}
\begin{split}\label{FourierModesFS}
&~~~\phi_{\vec n}  = N^{-\frac{d-1}{2}} \sum_{\vec a} e^{-\frac{2\pi i \vec n \vec a}{N}} \tilde \phi_{\vec{a}},\\
&~~~\pi_{\vec n}  = N^{-\frac{d-1}{2}} \sum_{\vec a} e^{\frac{2\pi i \vec n \vec a}{N}} \tilde \pi_{\vec{a}},
\end{split}
\end{equation}
where $\vec{n}$ is again a $d-1$ dimensional vector of integers running between $0$ and $N-1$. The position and momentum operators in momentum space also satisfy the commutation relations $[\phi_{\vec n},\pi_{\vec{k}}] = i \delta_{\vec n \vec k}$. Using the above transformations, we obtain the diagonalized Hamiltonian in momentum space 
\begin{equation}\label{HamilMom}
    H=\frac{1}{2M}\sum_{k_i=0}^{N-1}\left[|\pi_{\vec k}|^2 +M^2 \omega_{k}^2 |\phi_{\vec k}|^2\right]
\end{equation}
with
\begin{equation}\label{FourierFreq}
    \omega_k^2 = m^2+ \frac{4}{\delta^2}\sum_{i=1}^{d-1} \sin^2\left(\frac{\pi k_i}{N}\right), \qquad M=\frac{1}{\delta}.
\end{equation}
In terms of the momentum space coordinates, the ground-state wave-function reads
\begin{equation}\label{GSwavefunctionQFT}
    \langle \phi_{\vec k} | 0 \rangle = \mathcal{N_{\text{vac}}} \exp{\left[-\sum_{\vec k} \frac{M \omega_k |\phi_{\vec k}|^2}{2}\right]}
\end{equation}
where the normalization constant is given by $\mathcal{N}_{\text{vac}} = \prod_{\vec{k}} \left(\frac{M\omega_{\vec k}}{\pi}\right)^{1/4}$. This wave-function is Gaussian and so we can use our techniques from section \ref{sec:CompSHO} to evaluate its complexity.

As mentioned earlier, the vacuum state is in fact very complex -- its complexity diverges with the lattice spacing. The underlying reason for this divergence is the derivative term in \eqref{FreeBosonH}. This term is the one responsible for entangling the different lattice sites. Without this term, the Hamiltonian would factorize in position space and the quantum state of the different lattice sites would not be correlated.

When we pick a reference state, we want it to satisfy quite the opposite property. We would like the different oscillators to be completely unentangled. Therefore, a natural choice for the reference state is the ground state of an ultra-local Hamiltonian 
\begin{equation}
    H = \frac{1}{2} \int d^{d-1}x \left[\pi(x)^2+\mu^2 \phi(x)^2\right],
\end{equation}
where comparing to equation \eqref{FreeBosonH} we notice that the derivative term has been turned off. The discretized Hamiltonian in momentum space takes the form \eqref{HamilMom} with $\omega_{\vec k}=\mu$ and the relevant wavefunction for the reference state reads
\begin{equation}\label{RSwavefunctionQFT}
    \langle \phi_{\vec k} | \mu \rangle = \mathcal{N_{\mu}} \exp{\left[-\sum_{\vec k} \frac{M \mu \, |\phi_{\vec k}|^2}{2}\right]},
\end{equation}
where again $\mathcal{N_{\mu}}$ is a  normalization constant. Notice that this state is again Gaussian and has a fixed frequency for all momenta.

As in the last section, we will focus on trajectories moving entirely in the space of Gaussian states. The motion between Gaussian states can be studied in terms of symplectic transformations of the corresponding covariance matrices induced by quadratic gates in position and momentum variables.
The optimal trajectory takes the form \eqref{STT} for each momentum mode separately where the relative covariance metric
\eqref{DLL} is replaced with 
\begin{equation}
    \Delta_{\vec k} = \begin{pmatrix} \frac{\mu}{\omega_{ k}} &~ 0 \\ 0 &~~ \frac{\omega_{ k}}{\mu}. \end{pmatrix}
\end{equation}
for each momentum mode. 
The upper bound $\mathcal{C}_1^{UB}$ and the complexity $\mathcal{C}_2$ 
are given by equation \eqref{cmanyosc} summed over the different momentum modes
\begin{equation}\label{c12QFT}
    \mathcal{C}_1^{UB} = \frac{1}{2} \sum_{\vec k}  \left|\log \frac{\omega_k}{\mu}\right|;
\quad
    \mathcal{C}_2 =\frac{1}{2} \sqrt{\sum_{\vec k} \left(\log \frac{\omega_k}{\mu}\right)^2}.
\end{equation}

To improve our intuitive understanding of the optimal circuit constructing the ground state, let us write it explicitly in terms of the relevant unitary transformation in equation \eqref{kKrel1} (see also equations \eqref{kKrel2} and  \eqref{STT}):
\begin{equation}
\begin{split}\label{circuit}
    |\psi(\sigma)\rangle & = \exp \left[-\frac{i}{4} \sigma \sum_{\vec{k}} \log\left(\frac{\mu}{\omega_k}\right) (\phi_{\vec{k}} \pi_{\vec{k}} +\pi_{\vec{k}} \phi_{\vec{k}}) \right]|\mu\rangle,
\end{split}
\end{equation}
with the path parameter $\sigma \in [0,1]$ as before. In this way, we see that the optimal circuit consists of ``squeezing" the wavefunction for each momentum mode separately. Of course, since we have discretized our theory on the lattice, the state obtained at $\sigma=1$ is not exactly the ground state of the original continuum Hamiltonian \eqref{FreeBosonH} but it approximates it on distances larger than the lattice spacing.

Evaluating the result for the complexity \eqref{c12QFT} yields at the leading order in the small lattice spacing 
\begin{equation}
\begin{split}\label{divervac}
    &\mathcal{C}_1^{UB} \simeq \frac{\text{Vol}}{2\delta^{d-1}} |\log \mu\delta\,|+\dots
    \\
    &\mathcal{C}_2 \simeq \frac{1}{2} \left(\frac{\text{Vol}}{\delta^{d-1}}\right)^{1/2} |\log \mu\delta\,| +\cdots
\end{split}
\end{equation}
where $\text{Vol}=L^{d-1}$ is the spatial volume of the system. 
As we will see later, the behavior of $\mathcal{C}_1^{UB}$ matches much better with the results obtained from holography which hints that this cost function is better suited to be identified with the dual of complexity in holography. Note that the free field theory and the strongly coupled holographic theories are very different from each other. However, just as for the entanglement entropy, the structure of divergences is expected to follow a similar pattern. For the above reason, in what follows we will mostly focus on the $\mathcal{C}_1^{UB}$ complexity. 

Our results for the complexity are expressed in terms of $\mu$ -- the characteristic scale of the reference state. How are we to think about this scale? We can obtain a hint from the divergence structure in equation \eqref{divervac}. 
Divergent QFT quantities do not usually mix  logarithmic and polynomial divergences. The appearance of this  divergence in the complexity can be however remedied by choosing the scale of the reference state to depend on the cutoff, \ie $\mu\delta = e^{-\tilde \mu}$, where $\tilde\mu$ is an order one constant. In this case 
\begin{equation}\label{c1mutilde}
    \mathcal{C}_1^{UB} \simeq \frac{\text{Vol}}{2\delta^{d-1}} | \,\tilde \mu\,|+\dots.
\end{equation}
This choice is also natural from a physical point of view -- since we are introducing correlations at all scales down to the lattice scale $\delta$ it is natural to start with a state whose typical frequency is also of the order of the (inverse) lattice spacing. 
The result \eqref{c1mutilde} has a \emph{volume law} divergence. This can be  contrasted with the typical area law divergence of the entanglement entropy.\footnote{A different notion of area law often appears in the condensed matter literature studying entanglement entropy on the lattice in the large volume limit with a fixed lattice spacing. Here instead, we consider the fixed volume and small lattice spacing limit.} We will later see that this behavior is reproduced in holography. 
The complexity of the ground state of fermionic systems has been treated using similar methods and there as well one obtains a volume law \cite{Hackl:2018ptj,Khan:2018rzm}.
The above result is an upper bound on the complexity, however, a simple counting argument shows that the complexity following from exact optimization $\mathcal{C}_1$ will have the same scaling with the cutoff and volume of the system.

Finally, let us make a comment about the scheme of regularization. Above, we have regularized the complexity by placing our theory on a periodic lattice with lattice spacing $\delta$ as in  \cite{QFT1}. Let us now comment on a different scheme of regularization used in \cite{QFT2}. In this case, we work with a continuous momentum variable
\begin{equation}
    \vec{k}_{c} = \frac{2\pi \vec k}{L}
\end{equation}
and replace all the above sums $\sum_{\vec k}$ by integrals $\text{Vol} \int \frac{d^{d-1}k_c}{(2\pi)^{d-1}}$. The momentum integrals are regulated by a sharp momentum cutoff, \ie we cut our momentum integrals at a sharp value $|\vec{k}_c|=\Lambda$.
The results in this regularization scheme can be obtained from the former lattice regularization by initially placing the momentum cutoff significantly below the lattice scale  $\Lambda\ll \frac{2\pi}{\delta}$ and later sending the lattice spacing $\delta\rightarrow 0$ such that the
result remains finite and regulated by the new cutoff $\Lambda$. In that case, we may approximate the frequency in equation \eqref{FourierFreq} by $\omega_{k_c} = \sqrt{k_c^2+m^2}$. As before, the state $|\psi(\sigma=1)\rangle$ constructed by the continuous version of the circuit \eqref{circuit}
\begin{equation}
\begin{split}\label{circuit2}
    |\psi(\sigma)\rangle & = \exp \left[-\frac{i}{4} \sigma \int_{|k_c|<\Lambda} \frac{d^{d-1} k_c}{(2\pi)^{d-1}} \log\left(\frac{\mu}{\omega_k}\right) K(\vec{k}_c) \right]|\mu\rangle,
\\
K(\vec{k}_c) & \equiv \phi(\vec{k}) \pi(\vec{k}_c) +\pi(\vec{k}_c) \phi(\vec{k}_c), \qquad \sigma \in [0,1]
\end{split}
\end{equation}
is not actually the ground state of the Hamiltonian \eqref{FreeBosonH} but it approximates it for momenta below the cutoff momentum. With this regularization scheme, the complexity reads \eqref{c12QFT}
\begin{equation}
\begin{split}
    \mathcal{C}_1^{UB} = \frac{1}{2} \text{Vol} \int \frac{d^{d-1}k_c}{(2\pi)^{d-1}}  \left|\log \frac{\omega_k}{\mu}\right|,
\\
    \mathcal{C}_2 =\frac{1}{2}  \sqrt{ \text{Vol}\int \frac{d^{d-1}k_c}{(2\pi)^{d-1}} \left(\log \frac{\omega_k}{\mu}\right)^2},
\end{split}
\end{equation}
and the leading divergences are as in \eqref{divervac} with the replacement $\delta \rightarrow 1/\Lambda$.

\subsection{Weakly Interacting QFT}
It is clearly of great interest to understand how the analysis of the previous section can be extended to the case of interacting field theories, and study the dependence of the complexity on the couplings. Unfortunately this is a difficult task, and at the time of writing this review  only partial results are available.

The authors of \cite{Bhattacharyya:2018bbv} generalized the previous study by considering the complexity of nearly Gaussian states building on the idea of \emph{quantum circuit perturbation theory} 
\cite{Cotler:2016dha,Cotler:2018ufx,Cotler:2018ehb}. They studied the complexity of the ground state of a $\lambda \phi^4$ theory described by the following Hamiltonian
\begin{equation}
    H = \frac{1}{2}\int d^{d-1} x\left[\pi(x)^2 +(\nabla \phi(x))^2 +m^2 \phi^2 + \frac{\lambda}{12} \phi(x)^4\right]
\end{equation}
with the coefficient $\lambda$ treated perturbatively. 
The authors used perturbation theory in quantum mechanics to express the ground state of this theory as an exponentiated polynomial of order four (rather than two in the Gaussian case). They were then able to enlarge the set of gates used to manipulate Gaussian states up to order six in position and momentum to manipulate these states. This led to a well defined notion of Nielsen-type complexity. However, they found that within this approach the reference state could not be taken to be Gaussian but had to contain some non-quadratic terms. As a consequence, the cost functional also had to be made dependent on the coupling in order to have a smooth zero-coupling limit. As an aside, the authors proposed an alternative mean field theory approximation where one simply includes perturbative corrections to the mass in the Gaussian wavefunction. In this approximation the authors were able to show that at the Wilson-Fisher fixed point around four dimensions the interaction has slightly increased the complexity compared to the Gaussian fixed point.

\subsection{Complexity of the Thermofield Double State}\label{sec:TFDcompandtdep}
Another interesting example of a Gaussian state in free bosonic QFT is the thermofield double state \cite{TFD}. For the case of a single harmonic oscillator this state was studied in section \ref{sec:Complexity-TFD-SHO}. In the full bosonic QFT \eqref{HamilMom}, the TFD is simply the product of the different TFD states for each of the momentum modes, \ie
\begin{equation}
    |TFD(t)\rangle = \bigotimes_{
    \vec k} |TFD(t,\omega_k)\rangle ,
\end{equation}
where we defined the TFD for each mode in \eqref{TFDt2}. We will take the assumption that the optimal trajectory does not mix the different momentum modes. This assumption is natural because if we introduce entanglement between the different modes, this entanglement will have to be removed in the final state and that will increase the length of the circuit. However, recall that  we have seen  the case of coherent states which behaved counterintuitively in this regard in section \ref{sec:Coherent-states}. 

\begin{figure}[htbp]
\centerline{\includegraphics[scale=.45]{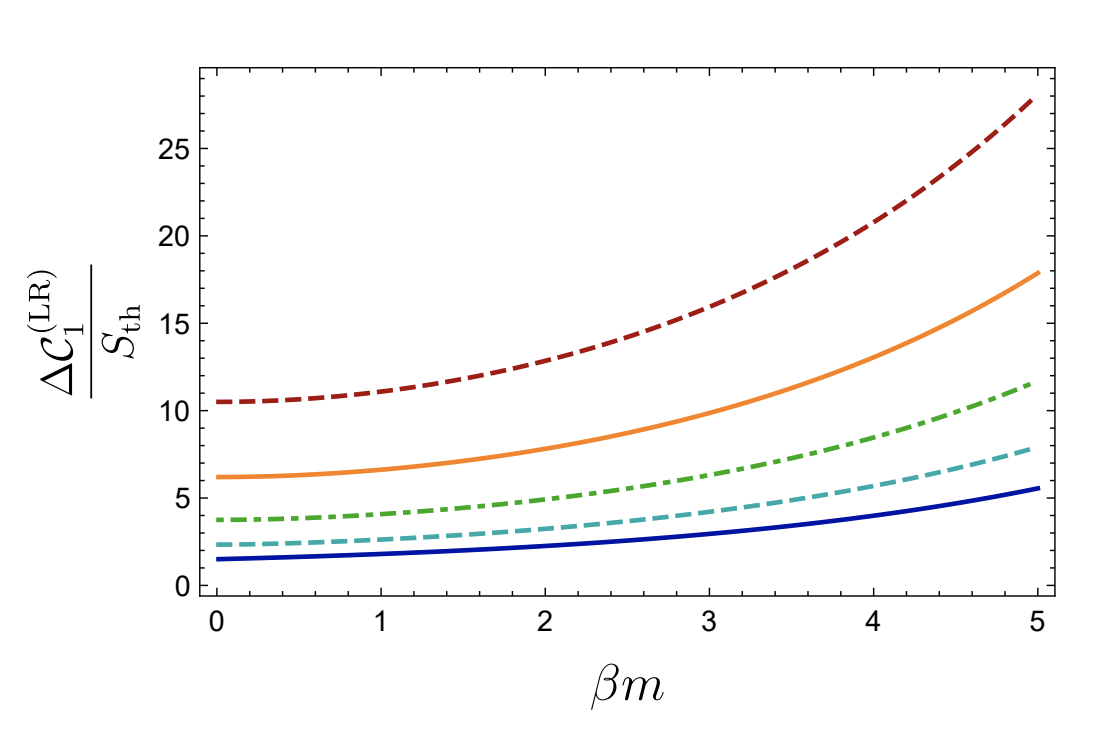}}
\caption{Complexity of formation as a function of the mass in various dimensions from $d=2$ (bottom curve) to $d=6$ (top curve). Figure taken from \cite{TFD}.}
\label{fig:formationmdep}
\end{figure}

Under the no-mode-mixing assumption, the complexity is simply the sum of complexities for each of the momentum modes. We will be particularly interested in the complexity of formation -- the difference in complexities between the TFD state at $t=0$ and two copies of the vacuum state -- cf. equation \eqref{formationfinal1SHO}, which is given by
\begin{equation}
    \Delta\mathcal{C}(|TFD(t=0)\rangle) = \sum_{\vec k} \Delta\mathcal{C}(|TFD(t=0,\omega_k)\rangle),
\end{equation}
where the expressions for the complexity of formation of the individual modes can be found in equation
\eqref{formationfinal1SHO}.
For reasons that we explain below, here we will focus on the $\mathcal{C}_1^{(LR),UB}$ complexity
\begin{equation}
\begin{split}
    &\Delta \mathcal{C}_1^{(LR),UB} = \text{Vol} \int_{k\leq \Lambda} \frac{d^{d-1}k}{(2\pi)^{d-1}} 2|\alpha_k|,\\
    &\alpha_k = \frac{1}{2}\log\left[\frac{1+e^{-\beta\omega_k/2}}{1-e^{-\beta\omega_k/2}}\right], \quad \omega_k = \sqrt{k^2+\omega^2}\,.
\end{split}
\end{equation}
This integral is finite due to the exponential suppression coming from the $\alpha_k$ at large frequency. Therefore we may remove the cutoff $\Lambda$ and simply integrate all the way to infinity.  
The result obtained by integrating this expression in the limit of vanishing mass is simply proportional to the thermal entropy of the system 
\begin{equation}
    S_{\text{th}} = \text{Vol} \int \frac{d^{d-1}k}{(2\pi)^{d-1}}
    \left[\frac{\beta \omega_k}{e^{\beta \omega_k}-1}-\log(1-e^{-\beta\omega_k})\right]
\end{equation}
with proportionality factor
\begin{equation}\label{formationfinalfree}
    \left.\frac{\Delta \mathcal{C}_1^{(LR),UB}}{S_{\text{th}}}\right|_{\beta m=0} = \frac{2^d-1}{d}.
\end{equation}
The proportionality of the complexity of formation and the thermal entropy is a property of complexity which is reproduced in holographic calculations \cite{Formation}. For finite mass the results are shown in Fig. \ref{fig:formationmdep}. 
The complexity of formation in the diagonal basis $\mathcal{C}_1^{(\pm),UB}$ and the $\mathcal{C}_2$ complexity vanish for temperatures much lower than the cutoff scale $T\ll \Lambda$, which is the physical regime. Therefore, we regard them as less useful measures of complexity of the state.

\begin{figure}[htbp]
\centerline{\includegraphics[scale=.3]{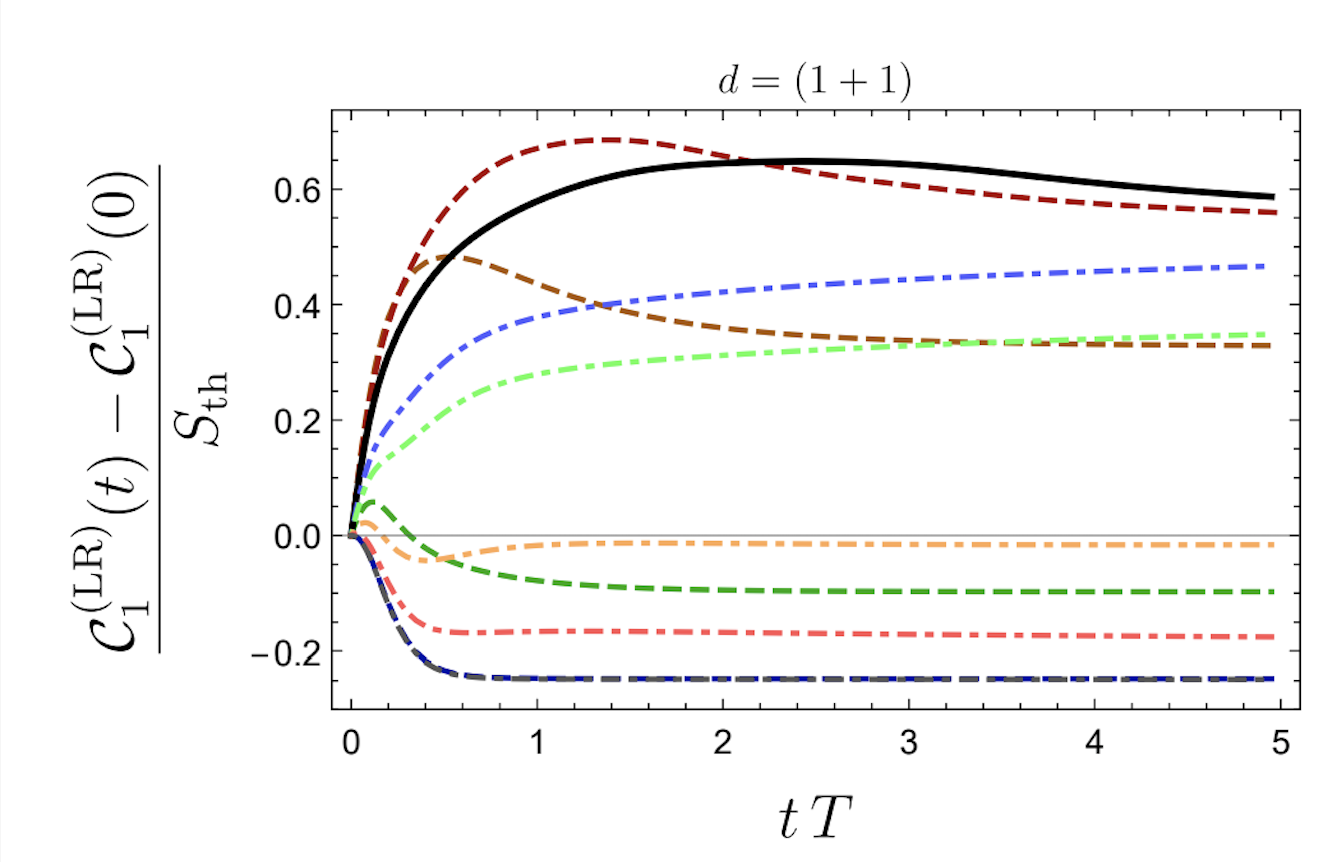}
\includegraphics[scale=.24]{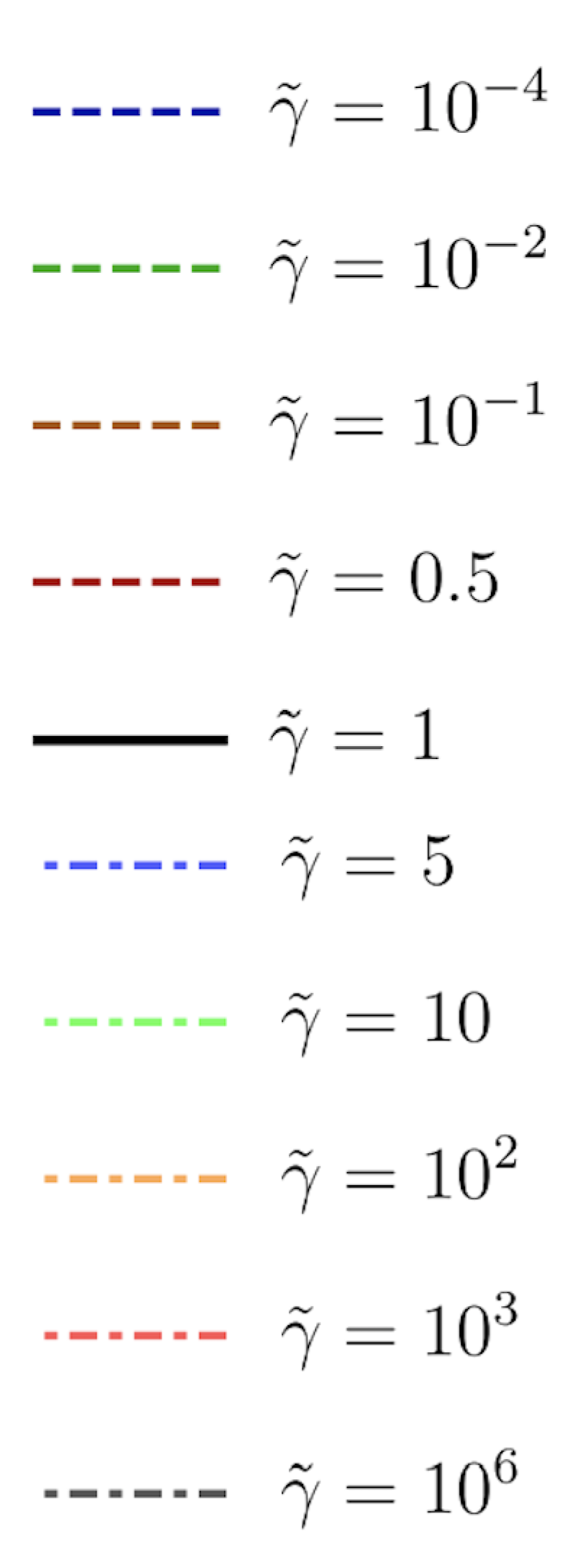}}
\caption{Time dependence of the complexity of the thermofield double state $\tilde \gamma \equiv(\beta \mu)^{-1}$. Figure taken from \cite{TFD}.}
\label{fig:formationtdep}
\end{figure}

While we did not write explicit expressions for the time dependence of the complexity of the TFD state at $t\neq 0$, such expressions follow directly from its covariance matrix in equation \eqref{covTFDt} and the time dependence can then be evaluated by summing the complexity of the different momentum modes. A plot of the time dependence of the complexity of the TFD state can be found in Fig. \ref{fig:formationtdep}. In this figure, taken from \cite{TFD}, the complexity evolves in time (either increases or decreases) and saturates after a time of the order of the inverse temperature. This is natural since each mode oscillates and and the oscillations are aligned at $t=0$ but the different modes become dephased at later times and so the contributions from the different normal modes averages out. Because of the exponential suppression of the oscillations mentioned at the end of section \ref{sec:Complexity-TFD-SHO} with large $\beta \omega$, modes with frequency higher than $1/\beta$ hardly contribute to the complexity and so the saturation is dominated by modes with $\omega\lesssim 1/\beta$ and happens at times $t\sim \beta$. 

We see, that in the free bosonic QFT, the complexity of the TFD saturates rather fast and this is because of the free nature of the system. In holography describing chaotic systems we will see a very different behavior. This highlights a general lesson to be learned about which properties are expected to be similar in free QFT and holography and which are not. In general, static quantities will have common properties while dynamical quantities will differ.

\subsection{Mixed State Complexity in QFT}\label{sec:mixedqft}
In section \ref{sec:Complexity-Mixed-States} we discussed the complexity of mixed states via the \emph{complexity of purification}. These results can be used to evaluate the complexity of various interesting mixed states of free quantum field theory. 

For example, let us start by considering the complexity of thermal states. The thermal state in free QFT can be decomposed as follows
\begin{equation}
    \hat \rho (\beta) = \otimes \hat \rho_{th}(\beta,\omega_k), \quad \omega_k = \sqrt{k^2+m^2},
\end{equation}
where $\hat \rho_{th}(\beta,\omega_k)$ is the thermal state of a single oscillator defined in equation \eqref{thermalState}. 
Hence the complexity is simply\footnote{Here we focus on the $\mathcal{C}_1$ complexity in the diagonal basis from equations \eqref{cth}-\eqref{CTFD} since those had nice analytic expressions. However, all the qualitative results which we describe below hold equally well in the physical basis, see \cite{Caceres:2019pgf}.}
\begin{equation}
    \mathcal{C}_{1,th}^{UB,\text{diag}}(\beta) = \sum_k  \mathcal{C}_{1,th}^{UB,\text{diag}}(\beta,\omega_k),
\end{equation}
where the complexity for each momentum mode can be found in equation \eqref{cth}. 
Note that the divergences in complexity come from integrating the $\log \left| \frac{\mu}{\omega_k}\right|$ contributions in equations \eqref{cth}-\eqref{CTFD}. Hence, we see that the complexity of the thermofield double state is twice as divergent as that of the thermal state. This reflects a general property that the purification which preserves the most symmetry between the ancillary degrees of freedom and the physical ones is not always the most efficient one. In the case of the thermofield double state for example, we work very hard to establish short distance correlations between the ancillary degrees of freedom themselves, which would then be removed upon tracing out this part of the system anyway and so that is useless work.

\begin{figure}[htbp]
\centerline{\includegraphics[scale=.5]{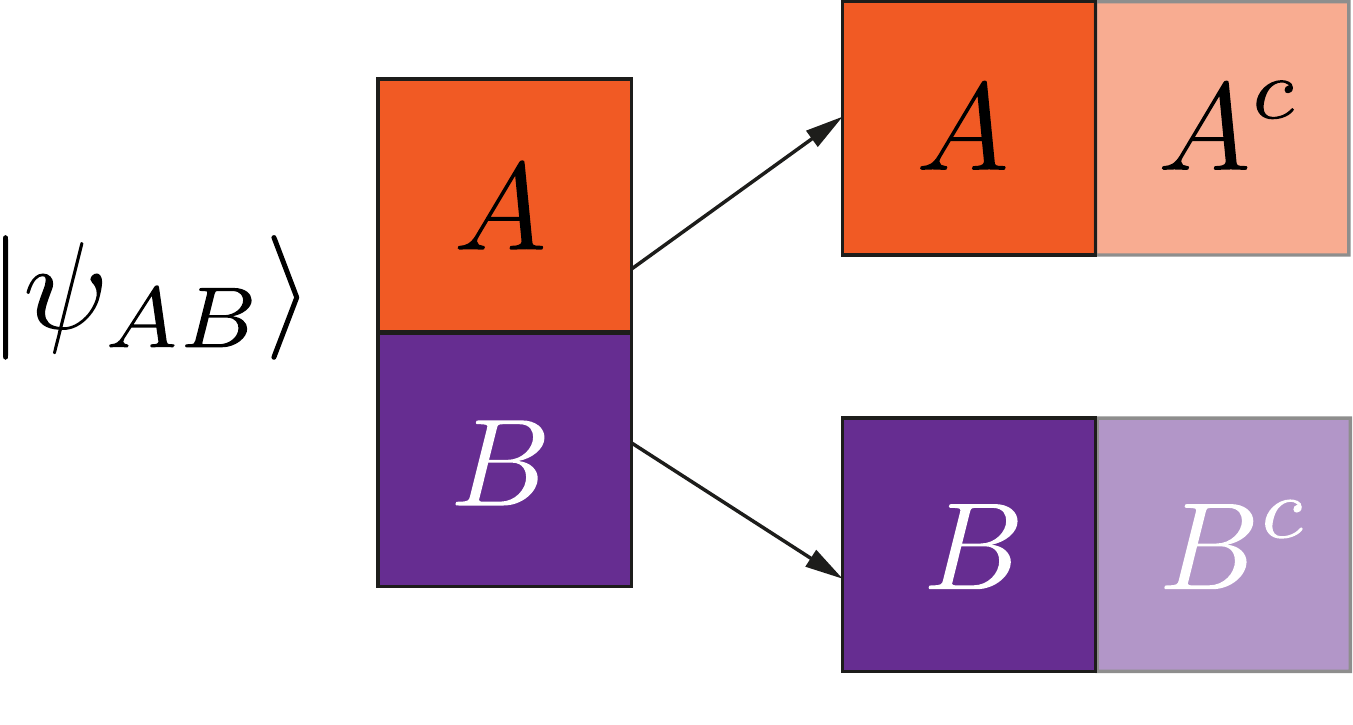}}
\caption{Illustration of mutual complexity. We start by a pure state on a system $AB$ which is then split into two mixes states on the systems $A$ and $B$. The sum of the complexities of purification of these mixed states using ancillary systems $A^c$ and $B^c$ is not necessarily equal to the complexity of the original pure state.}
\label{fig:MutualComplexity}
\end{figure}

When a mixed state $\rho_A$ is obtained from an original pure state $|\psi_{AB}\rangle$, it is often the case that the original state is not the optimal purification.
This is because in $|\psi_{AB}\rangle$
we work too hard to establish all the correlations between the $B$
degrees of freedom and mimic exactly those between 
$A$ and $B$.
To estimate how different are the correlations in the optimal purification from those in the original state we define the mutual complexity
\begin{equation}
    \Delta \mathcal{C}_{\text{mutual}} = \mathcal{C}(\rho_A)+\mathcal{C}(\rho_B)-\mathcal{C}(|\psi_{AB}\rangle)\,,
\end{equation}
see Fig. \ref{fig:MutualComplexity}. For example, when considering the process of forming the thermal state from tracing out half of the thermofield double state we obtain 
\begin{equation}
    \Delta \mathcal{C}_{\text{mutual}} = 2\mathcal{C}(\rho_{th})-\mathcal{C}(|TFD\rangle)\,.
\end{equation}
In particular in quantum field theory of a free scalar field, this quantity turns out to be finite (\ie  all the UV divergences cancel) and is proportional to the thermal entropy for the case of a massless scalar (the conformal limit). The mutual complexity in the diagonal basis in the various QFT examples studied in \cite{Caceres:2019pgf} was found to be subadditive, \ie it satisfies $\Delta \mathcal{C}^{\text{diag}}_{\text{mutual}}>0$.

Another interesting example of a mixed state of a free bosonic QFT is that of subregions of the vacuum state. We could as before, focus on the example of subregions of the vacuum state on the lattice for a free bosonic QFT. The authors of \cite{Caceres:2019pgf} have focused on a one dimensional spatial lattice with $N$ sites. The wavefunction of the vacuum state reads
\begin{equation}
    \Psi_0(\phi_k) \propto \prod_{k=0\ldots N-1} e^{-\frac{1}{2}\omega_k |\phi_k|^2}
\end{equation}
where $\omega_k$ and $\phi_k$ were defined in equations \eqref{FourierModesFS} and \eqref{FourierFreq} and we substitute $d=2$. Translating back this expression to position basis using equation \eqref{FourierModesFS} we obtain
\begin{equation}
    \Psi_0(\tilde\phi_a) \propto \prod_{a,b=0\ldots N-1} e^{-\frac{1}{2}M_{ab} \tilde \phi_a \tilde \phi_b}
\end{equation}
where 
\begin{equation}
    M_{ab} = \frac{1}{N} \sum_{k=0\ldots N-1}\omega_k e^{-\frac{2\pi i k}{N}(a-b)}\,.
\end{equation}
To obtain the subregions we divide our lattice in two subsets 
$A=\{x_0,\ldots x_j\}$ and $B=\{x_{j+1},\cdots x_{N-1}\}$ and trace out the region $B$ as follows
\begin{equation}
    \rho_A(x_A,x'_A) = \int dx_B \Psi_0(x_A,x_B) \Psi_0^*(x'_A,x_B).
\end{equation}
Similarly to what we did earlier with the single Harmonic oscillator it is possible to minimize the complexity over the essential purifications of this mixed state. In fact \cite{Caceres:2019pgf} used a simplifying assumption. They considered \emph{mode-by-mode purifications} which are introduced after bringing the density matrix to a diagonal form and then purifying each mode which is mixed separately. This is a subset of all possible purifications which provides a good approximation to the complexity of purification based on tests with small systems (purifying two by four).

The authors performed this task numerically and found that the original vacuum state is not always the optimal purification. This is similar to what happened before with the TFD and thermal states.

\begin{figure}[htbp]
\centerline{\includegraphics[scale=.58]{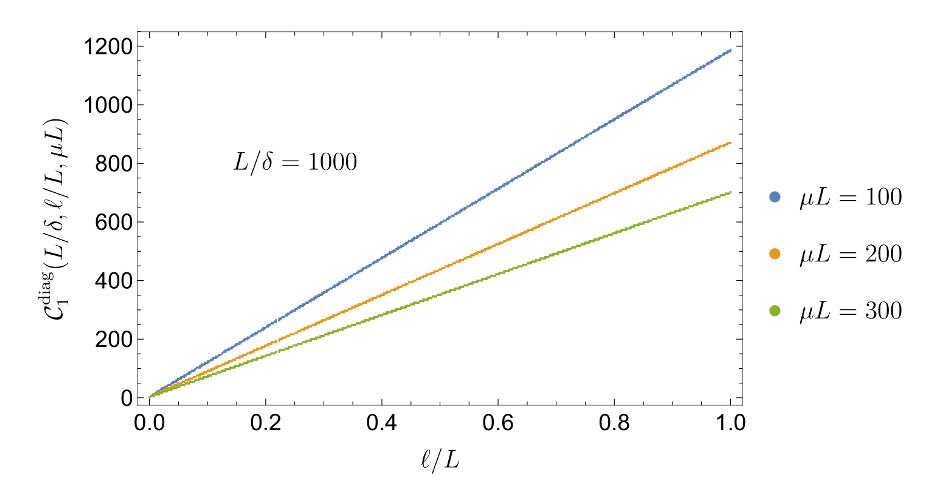}}
\caption{Complexity of subregions of the vacuum state as a function of the interval size. This figures makes it apparent that the leading contribution to complexity grows linearly with the subsystem size which is the aforementioned \emph{volume law}. Figure taken from \cite{Caceres:2019pgf}. Here the mass was fixed to be small $mL=0.01$ in order to mimic the results of a conformal field theory.}
\label{fig:SubVac1}
\end{figure}

\begin{figure}[htbp]
\centerline{\includegraphics[scale=.52]{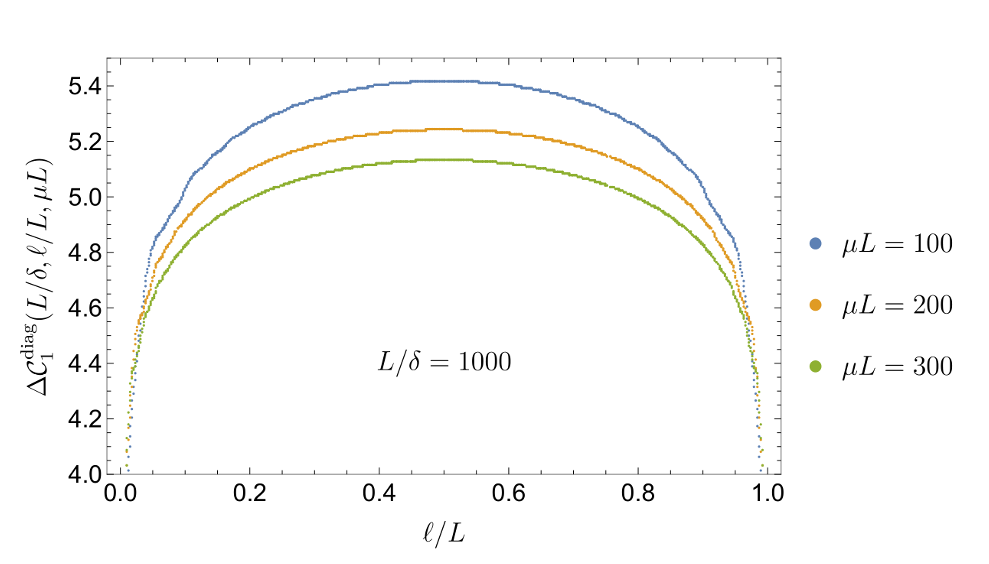}}
\caption{Mutual complexity of subregions of the vacuum as a function of the subregion size. This typical $\log(\sin(\#))$ behavior is reminiscent of the Calabrese-Cardy formula for the entanglement entropy. Figure taken from \cite{Caceres:2019pgf}. Here the mass was fixed to be small $mL=0.01$ in order to mimic the results of a conformal field theory.}
\label{fig:SubVac4}
\end{figure}

The results are presented in the plots. Fig. \ref{fig:SubVac1} presents the complexity as a function of the subregion size in the limit of small mass. The following expressions provides a good fit
\begin{equation}
\begin{split}
    \mathcal{C}_1^{UB,\text{diag}} =& \frac{\ell}{2\delta} \left|\log \mu\delta\right| + \frac{1}{2}f_1(\mu L)\log \left(\frac{L}{\pi \delta}\sin\frac{\pi \ell}{L}\right)\\
    &~~~~+\frac{\ell}{L}f_2(\mu L)+f_3(\mu L).
\end{split}
\end{equation}
Here, $\mu$ is the scale of the reference state, $L$ is the full system size, $\ell$ is the subregion size, $\delta$ is the cutoff and $f_1,f_2,f_3$ are functions of the reference state scale. These functions could not be determined very accurately because the numerical study examined only very few values of $\mu$. We see that the leading divergence is an area law and depends on the cutoff in a similar way to the leading divergences in the full vacuum complexity \eqref{divervac}. The subleading divergences are reminiscent of the entanglement entropy as we will see in a moment. The mutual complexity $\Delta\mathcal{C}_{\text{mutual}} = \mathcal{C}(\rho_A) +\mathcal{C}(\rho_B) - \mathcal{C} (|\psi_0\rangle)$ can also be evaluated and its dependence on the subregion size and cutoff can be fitted (see Fig. \ref{fig:SubVac4}) and one obtains in the limit of small mass 
\begin{equation}\label{Mutual-subregion}
    \Delta\mathcal{C}_{1,\text{mutual}}^{UB,\text{diag}} \approx f_1(\mu L) \left(\log \left(\frac{L}{\pi \delta}\sin\frac{\pi \ell}{L}\right)+f_4(\mu L)\right).
\end{equation}
Here, $f_4$ is yet another function of $\mu L$. Some proposed fits for $f_1$ and $f_4$ can be found in equation (7.10) of \cite{Caceres:2019pgf}. 
The above formula is very similar to the entanglement entropy formula by Calabrese and Cardy \cite{Calabrese:2004eu}-\cite{Calabrese:2005zw}. This hints at a deeper connection between the subleading divergences in complexity an the  entanglement entropy in non-dynamical situations.

\subsection{Complexity in CFT}
The approach of studying QFT state complexity restricted to Gaussian or nearly Gaussian states has its clear limitations. Many interesting physical systems are  strongly interacting. In particular, when making connection via holography between quantum information and black holes which is one of the prime motivations for studying QFT complexity, the relevant field theories are strongly interacting. These theories are however special in that they preserve a large spacetime symmetry group - the conformal symmetry. The abundance of symmetry is what helps make progress in this case. Therefore in this section we will focus on the question - can one utilize the conformal symmetry to define a complexity of states within conformal field theory.

This exploration began with the work of \cite{Caputa:2018kdj} who considered the geometric approach to complexity within 2d CFTs. In particular the authors focused on circuits in a unitary representation of the Virasoro algebra\footnote{Due to holomorphic factorization, each of the two copies of the Virasoro algebra could be considered separately.}
\begin{equation}
    [L_m,L_n] = (m-n) L_{m+n}+\frac{c}{12}m (m^2-1)\delta_{n+m,0}.
\end{equation}
The CFT was taken to live on a circle with angular coordinate $\theta\equiv \theta+2\pi$ and the corresponding stress tensor can be expressed as 
\begin{equation}
    T(\theta) = \sum_{n\in \mathbb{Z}}{ \left(L_{n} -\frac{c}{24}\delta_{n,0}\right) e^{-i n\theta}}.
\end{equation}

The circuits are constructed from the symmetry generators, 
\begin{equation}
\begin{split}
   \vspace{-10pt}U(\sigma) =&\, \overleftarrow{\mathcal{P}} \exp \int_0^\sigma d\sigma' \, Q(\sigma'),
    \\
    Q(\sigma) =&\, \int_0^{2\pi} \frac{d\theta}{2\pi} \epsilon(\sigma,\theta)T(\theta) 
=\sum_{n\in \mathbb{Z}}{\epsilon_n(\sigma) \left(L_{-n} -\frac{c}{24}\delta_{n,0}\right)},
\end{split}
\end{equation}
where the Fourier modes 
\begin{equation}
    \epsilon_n(\sigma) = \int_0^{2\pi} \frac{d\theta}{2\pi} \epsilon(\sigma, \theta)e^{i n\theta}
\end{equation}
serve as control functions along the circuit. They should satisfy  $\epsilon_n(\sigma)^* = -\epsilon_{-n}(\sigma)$ in order for the transformation to be unitary. In addition, in order to start our circuit at the identity we require $\epsilon_n(\sigma=0)=0$.

The Virasoro symmetry 
without its central extension\footnote{The central extension was later treated in \cite{Erdmenger:2020sup}.} is simply the group of diffeomorphisms of the circle $f(\theta)\in \text{Diff}(S^1)$. In particular, the function $\epsilon(\sigma,\theta)$ in the circuit above fixes the infinitesimal diffeomorphisms whose composition gives the total diffeomorphism function  $f(\sigma,\theta)$ at each point $\sigma$ along the circuit. Explicitly, 
$\epsilon(\sigma,f(\sigma,\theta)) = \del_\sigma f(\sigma,\theta)$.

The reference state serving as the starting point for the circuit is taken to be the chiral primary $|h\rangle$ satisfying
\begin{equation}
    L_0|h\rangle = h |h\rangle, \quad L_n|h\rangle =0 \text{ for } n>0.
\end{equation}

The authors of \cite{Caputa:2017yrh} considered two different cost functions along the circuit
\begin{equation}
\begin{split}\label{CMcosts}
    \mathcal{F}_1(\sigma) = & | \langle \psi(\sigma)| \del_\sigma \psi(\sigma) \rangle|\,,
\\
    \mathcal F_2(\sigma) = &\sqrt{ \langle \del_\sigma \psi(\sigma)| \del_\sigma \psi(\sigma) \rangle }\,,
\end{split}
\end{equation}
which become equivalent in the large central charge limit $\mathcal{F}_2 \simeq \mathcal{F}_1 (1+\mathcal{O}(1/c))$.
We should point out that the above $\mathcal{F}_1$ cost function is in fact different from the $F_1$ cost function in equation \eqref{NielsenYNorms}. The difference is reminiscent of exchanging the order of the absolute value in the complexity definition and the sum over circuit generators. The $\mathcal{F}_1$ cost function in equation \eqref{CMcosts}  generally has many null directions and therefore does not satisfy the mathematical definition of a norm, making it somewhat disadvantageous as a complexity measure.  Nevertheless, it has a nice geometric interpretation in terms of the coadjoint orbits of the Virasoro group and a connection to the Liouville action featuring in the path-integral approach to complexity, see section \ref{sec:PIC}. Another useful cost function is the Fubini-Study (FS) metric
\begin{equation}
\begin{split}\label{FSFSFS}
    \mathcal F_{FS}(\sigma) = &\sqrt{ \langle \del_\sigma \psi(\sigma)| \del_\sigma \psi(\sigma) \rangle - | \langle \psi(\sigma)| \del_\sigma \psi(\sigma) \rangle|^2}.
\end{split}
\end{equation}
This cost function has the advantage that it assigns zero contributions to circuits which only modify our state by an overall phase.

Using some algebraic manipulations based on the symmetry algebra it is possible to show that the $\mathcal{F}_1$ cost function is given by
\begin{equation}\label{F1virasoro}
    \mathcal F_1(\sigma) =  \int_0^{2\pi} \frac{d\theta}{2\pi} 
    \frac{\del_\sigma f(\sigma,\theta)}{\del_{\theta} f(\sigma,\theta)}
    \left(\frac{c}{24}-h +\frac{c}{12} \{f,\theta\}\right)
\end{equation}
where $\{f,\theta\} = \frac{f'''}{f'} - \frac{3}{2}\left(\frac{f''}{f'}\right)^2$ is the Schwarzian derivative.

It turns out that the complexity functional \eqref{F1virasoro} is related to the Polyakov action of induced gravity in two dimensions with a convenient choice of coordinates which means that induced 2d gravity governs the complexity of Virasoro circuits. Since the Polyakov and Liouville actions are related, this connects nicely to the path integral complexity proposal, see next subsection.

A similar computation for the Fubini-Study metric was carried in \cite{Flory:2020eot,Flory:2020dja} which leads to 
\begin{equation}
\begin{split}
    &\mathcal{F}_{FS}(\sigma)^2 = 
    \int_0^{2\pi} \frac{d\theta_1}{2\pi}\frac{d\theta_2}{2\pi}
    \frac{\del_\sigma f(\sigma,\theta_1)}{\del_{\theta_1} f(\sigma,\theta_1)}
    \frac{\del_\sigma f(\sigma,\theta_2)}{\del_{\theta_2} f(\sigma,\theta_2)} \times
    \\&~~~~~~\left[\frac{c}{32 \sin^4[(\theta_1-\theta_2)/2]}-
    \frac{h}{2 \sin^2[(\theta_1-\theta_2)/2]}\right]\,.
\end{split}
\end{equation}

From the above expressions for the cost functions we note that what was earlier a geodesic equation for the control functions $Y^I(\sigma)$, cf. equations \eqref{controlprob1}-\eqref{controlprob2}, now became an infinite dimensional geodesic problem with the index $I$ replaced by the continuous variable $\theta$. The geodesic equations for the control function take the form of integro-differential equations for the function $f(\sigma, \theta)$. 
The equations of motion are second order in $\sigma$ which allows to find circuits connecting two points in the Virasoro group. This makes the Fubini-Study norm a better suited complexity measure compared to the $\mathcal{F}_1$ cost function. The authors of \cite{Flory:2020dja,Flory:2020eot} used those equations of motion to find the complexity for going between the identity $f(\sigma=0,\theta)=\theta$ and a perturbation containing a single Fourier mode $f(\sigma=1, \theta) =\theta+\frac{\epsilon}{m}\sin(m\theta) $
with $\epsilon\ll 1$ and $m\in\mathbb{N}$. The sectional curvatures were found to be negative in most directions for physically relevant values of $h$ and $c$.

A similar approach can be employed to study the complexity of unitary circuits of the conformal algebra in higher dimensions \cite{Chagnet:2021uvi}. The conformal algebra consists of dilatations, translations, special conformal transformations and rotations - $D,P_\mu, K_\mu, L_{\mu\nu}$ respectively, satisfying the commutation relations\footnote{There is a small subtlety here: we use the Euclidean conformal algebra generators to construct unitary representations of the Lorentzian conformal algebra. This is easy to understand in terms of the following analogy: the Euclidean conformal generators play a similar role to $J_{\pm}$ generators in the quantum mechanical treatment of angular momentum.}
\begin{equation}\label{ECA}
\begin{split}
  & [D, P_\mu]  =  P_\mu~,  	~~~[D, K_\mu] = - K_\mu~, \quad
 \\
 & ~~[K_\mu, P_\nu]  =  2\left(\delta_{\mu\nu} D - L_{\mu\nu}\right)~,
 \end{split}
 \end{equation}
where the rotations have been omitted from the list (but they satisfy the usual commutation relations). The generators satisfy the following Hermiticity relations 
\begin{equation}\label{hermrel}
	\begin{split}
		D^\dagger  = D~,  \quad K_\mu^\dagger  = P_\mu~,  \quad L_{\mu\nu}^\dagger  = - L_{\mu\nu}~.\\
	\end{split}
\end{equation}
As the reference state we consider a scalar primary state $|\psi_R\rangle = |\Delta\rangle$ of scaling dimension $\Delta$ which satisfies
\begin{equation}
    D|\Delta\rangle = \Delta |\Delta\rangle, \quad
    K_\mu|\Delta\rangle = L_{\mu\nu} |\Delta\rangle=0.
\end{equation}

A general unitary circuit will pass through states $|\alpha(\sigma)\rangle = U(\sigma) |\Delta\rangle$ where the unitary $U(\sigma)$ is constructed as follows
\begin{equation}
    U(\sigma) =e^{i \alpha(\sigma)\cdot P} e^{i \gamma_D(\sigma) D} \left(\prod_{\mu < \nu} e^{i \lambda_{\mu\nu}(\sigma) L_{\mu\nu}}\right)  e^{i \beta(\sigma) \cdot K}
\end{equation}
where the various control functions $\alpha_\mu(\sigma)$, $\gamma_D(\sigma)$, $\lambda_{\mu\nu}(\sigma)$, $\beta(\sigma)$ have to satisfy some constraints to make sure that $U(\sigma)$ is unitary. For example, one of these constraints is $\text{Im}(\gamma_D) = -\frac{1}{2}\log(1 - 2 \, \alpha \cdot \alpha^* + \alpha^2 \alpha^{*2})$.
The $\mathcal{F}_1$ complexity cost function reads:
\footnotesize
\begin{equation}\label{F1HD}
\hspace{-5pt}	\frac{\mathcal{F}_1}{\Delta} =  \left|\dfrac{\dot \alpha \cdot \alpha^* -\dot \alpha^* \cdot \alpha+  \alpha^2 \,(\dot \alpha^* \cdot \alpha^*)-\alpha^{*2} (\dot \alpha \cdot \alpha)}{1 - 2 \, \alpha \cdot \alpha^* + \alpha^2 \alpha^{*2}} + i \text{Re}(\dot \gamma_D) \right| ,
\end{equation}
\normalsize
while the FS-metric is
\begin{equation}
	\label{FSHD}
	\begin{split}
	\frac{ds_{FS}^2}{d\sigma^2}=&2 \Delta \left[\dfrac{\dot \alpha \cdot \dot \alpha^{*} - 2|\dot \alpha \cdot \alpha|^2}{1 - 2 \, \alpha \cdot \alpha^* + \alpha^2 \alpha^{*2}} \right.
	\\
	&~~~~~\left.+ 2\dfrac{\left|\dot \alpha \cdot \alpha^* - \alpha^{*2} \, \alpha \cdot \dot \alpha \right|^2}{(1 - 2 \, \alpha \cdot \alpha^* + \alpha^2 \alpha^{*2})^2}\right] \, .
	\end{split}
\end{equation}
We see that the $\mathcal{F}_1$ complexity depends on the overall phase $\gamma_D$ of the state. In addition it is possible to show that the $\mathcal{F}_1$ cost function has many null-directions where the distance vanishes along  non-trivial circuits. Once again we see that these properties make the $\mathcal{F}_1$ cost function a less favorable measure of complexity. 
Upon restricting the two-dimensional cost functions \eqref{FSFSFS}-\eqref{F1virasoro} to diffeomorphisms corresponding to the global conformal group, one simply obtains the $d=2$ case of the higher dimensional cost functions \eqref{F1HD}-\eqref{FSHD}.

Minimizing the Fubini-Study cost, it can be demonstrated that the complexity of a target state $\alpha_T \equiv |\alpha(t=0)\rangle$ is simply
\begin{equation}
   \mathcal{C}_{FS} = \sqrt{\Delta \left[(\tanh^{-1}\Omega^S)^2+(\tanh^{-1}\Omega^A)^2\right]}
\end{equation}
where we can extract $\Omega^S$ and $\Omega^A$ from the combinations $\Omega^S\pm\Omega^A = \sqrt{2\alpha_T \cdot \alpha^*_T\pm 2 |\alpha_T^2|}$. We note that this result scales with $\sqrt{\Delta}$. 

In holography, the Fubini-Study line element has been related to the average of minimal and maximal distances between infinitesimally displaced timelike geodesics in the bulk (each representing the state at some point along the circuit), see \cite{Chagnet:2021uvi}. This connection was made by identifying the bulk symplectic form and the one associated to the FS metric in the phase space of the circuits. This suggests that a very natural connection can be made to holography by studying the relevant symplectic forms as was indeed suggested in \cite{Belin:2018bpg}. These ideas opens the path to an explicit holographic verification of the holographic complexity proposals. We come back to this point in the discussion section.

The above approach (both in 2d and in higher dimensions) considers only unitary circuits constructed from symmetry generators of the conformal groups and those circuits do not allow to move between any two states in the CFT Hilbert space but only between states in the same conformal family. The extension to a larger class of circuits remains unknown.

\subsection{Path-Integral Approach to Complexity}\label{sec:PIC}
A different approach to complexity is based on preparing the state using the Euclidean path integral. The authors of  \cite{Caputa:2017urj,Caputa:2017yrh} have proposed that the optimization over possible circuits preparing the state is equivalent to optimizing the metric on the space where the path integral is performed. 
Roughly speaking, we are to understand this metric as the density of gates in a discretized version of the path integral which in turn can be understood as a tensor network.\footnote{We discuss briefly tensor networks in section \ref{sec:TensorNetwork}.} The idea is that if some gates are not needed for the optimal circuit, they can be deleted, and this will change the effective geometry. The Euclidean time in the path integral is identified with the depth along the (non-unitary) circuit, and  it gives rise to an RG direction  $z=-(\tau-\delta)$ which captures the gradual introduction of entanglement into the state at different length scales; the state prepared at the final time is defined at a UV cutoff $\delta$. 

The simplest case is that of a two-dimensional CFT, because every metric can be brought to the form $ds^2 = e^{2\phi(z,x)} (dz^2+dx^2)$. In the UV we should have one gate for each cutoff-size region, so we should set $e^{2\phi(z=\delta,x)}=\frac{1}{\delta^2}$. 
The ground state wavefunction in the curved metric is proportional to the one with the flat metric due to conformal symmetry: 
\begin{equation}\label{WFsym}
    \Psi_{g_{ab} = e^{2\phi}\delta_{ab}} = e^{S_L[\phi]-S_L[0]} \Psi_{g_{ab} = \delta_{ab}}
\end{equation}
with a  proportionality factor given by the Liouville action 
\begin{equation}\label{Liouville}
    S_L[\phi] = \frac{c}{24\pi}\int_{-\infty}^\infty dx \int_\delta^\infty dz \left[(\del_x \phi)^2 + (\del_z \phi)^2 + \mu e^{2\phi} \right].
\end{equation}
The parameter $\mu$ can be rescaled by a shift of $\phi$, so it can be set to one. 
The circuit that prepares the state is thus effectively computing the Liouville action, and  
the optimization is equivalent to minimizing the prefactor $e^{S_L[\phi]}$ (see also \cite{Czech:2017ryf} who proposes another argument for the Liouville action in the language of tensor networks).
This leads to the following proposal for the complexity
\begin{equation}
    \mathcal{C}_\Psi = \min_{\phi} S_L[\phi(z,x)]\,.
\end{equation}
The conformal factor that minimizes the action, subject to the boundary condition described above, corresponds to the metric on the hyperbolic half-plane $ds^2 = (dz^2+dx^2)/z^2$, 
and it can be interpreted as the metric of a time slice of AdS$_3$. This leads to a complexity  $\mathcal{C}_\Psi = \frac{cL}{12\pi \delta}$ for the vacuum state of the CFT, which has the same structure of divergences which we saw earlier in the free field theory case (cf. equation \eqref{c1mutilde} for $d=2$). 

Using appropriate boundary conditions on the strip and on the cut plane, one can find the solutions corresponding to the TFD and to the mixed state for a subregion of the vacuum state, respectively. In all these cases, the evaluation of the Liouville action (supplemented by boundary terms) gives results that agree qualitatively with the free field theory results and with the CV and CA holographic conjectures which we describe below (\ie they have the same dependence on the cutoff, but different coefficients).

The generalization to higher dimensions is non-trivial, since the metric has more degrees of freedom than just the conformal factor. Restricting to the class of conformally flat metrics, one can write a natural generalization of the Liouville action:
\begin{equation}
    S_d \sim \int d^{d-1}x \, dz \left[e^{d\phi}+e^{(d-2)\phi}((\del_x \phi)^2+(\del_z \phi)^2)\right] \,.
\end{equation}
The optimization of this action gives again a constant-time slice of AdS$_d$ and a vacuum complexity that agrees with the free field theory results and with the holographic CV/CA results which we will describe in the next section. 
A different but also natural generalization of the Liouville action to higher dimensions would be an action that reproduces the conformal anomaly of the theory \cite{Riegert:1984kt,Levy:2018bdc}. Such action would have higher-derivative terms and would not be positive-definite, so its interpretation as complexity would be more problematic.

This framework allows to study also the complexity of a state created by the insertion of a primary operator. The Liouville equation is modified by a source term, and the corresponding geometry is the Poincar\'e disc with a conical defect. This agrees with the dictionary of AdS$_3$/CFT$_2$ to first order in $\Delta/c$, but an exact match seems to require quantizing the Liouville action; it is not clear how this could arise from the optimization problem (see however \cite{Boruch:2021hqs}). 

While for a CFT it is possible to perform the optimization varying only the background metric, 
for a generic QFT that has running couplings along the RG flow one expects to have to allow for variations of some parameters of the network. The case of a CFT perturbed by a relevant operator $\lambda \mathcal{O}$ was considered in \cite{Bhattacharyya:2018wym}. The condition \eqref{WFsym} that the wavefunction remains the same up to a prefactor is no longer a consequence of the symmetry but has to be enforced by choosing $\lambda(z)$ appropriately.  
The Liouville action is replaced by a functional $N[\phi,\lambda]$ which can be calculated order by order in an expansion in $\lambda$. The optimal geometry agrees with the backreaction of a scalar field on AdS$_3$.

\section{Complexity in Holography}
\label{sec:Holo1}

\subsection{Complexity Conjectures}
\label{sec:Complexity-conjectures}
In a series of papers starting in 2014 \cite{Susskind:2014rva,Susskind:2014moa,Brown1,Brown2,Stanford:2014jda}, Susskind and collaborators have argued that the notion of quantum complexity is crucial to understand the quantum and information-theoretic properties of black holes. A connection was in fact already suggested in \cite{Harlow:2013tf} in relation to the problem of decoding the information contained in the Hawking radiation. Susskind et al. made the connection much sharper by conjecturing, in the context of the AdS/CFT correspondence, a precise relation between the complexity of a state in the dual theory and the corresponding bulk geometry. 
The conjecture has two alternative forms: ``{\it Complexity=Volume}" (CV) and ``{\it Complexity=Action}" (CA).\footnote{An additional proposal which relates complexity to spacetime-volume was made in \cite{Couch:2016exn}, but we will not discuss it here.} In order to formulate them, let us 
denote by $\Sigma$ a surface at constant time on the AdS boundary, where the state is defined. 
CV postulates that the complexity of the state is equal to the volume of a maximal slice in the bulk $\mathcal{N}$ such that  $\partial \mathcal{N}=\Sigma$: 
\begin{equation}\label{CV-def}
    \mathcal{C}_V = \frac{\textrm{Vol}(\mathcal{N})}{G_N \ell_{CV}} \,, 
\end{equation}
where $\ell_{CV}$ is a length required to make the quantity dimensionless.\footnote{In the following we will take $\ell_{CV}$ to coincide with $\ell_{AdS}$, as in most of the literature.} 
CA postulates instead that the complexity is equal to the on-shell action of a Wheeler-DeWitt (WDW) patch, which is the domain of dependence of a Cauchy slice in the bulk anchored at the boundary on $\Sigma$:\footnote{In the following we will set $\hbar=1$.}
\begin{equation}\label{CA-def}
 \mathcal{C}_A = \frac{S_{WDW}}{\pi \hbar} \,.
\end{equation}

Let us see how these prescriptions work in the case of a two-sided eternal black hole in AdS, which is thought to be the holographic dual of the TFD state. The geometry has two asymptotic boundaries, where the two copies of the theory live, that are separated by a horizon, so the L and R theories are in an entangled state but do not interact with each other. 

\begin{figure}[htbp]
\centerline{\includegraphics[scale=.55]{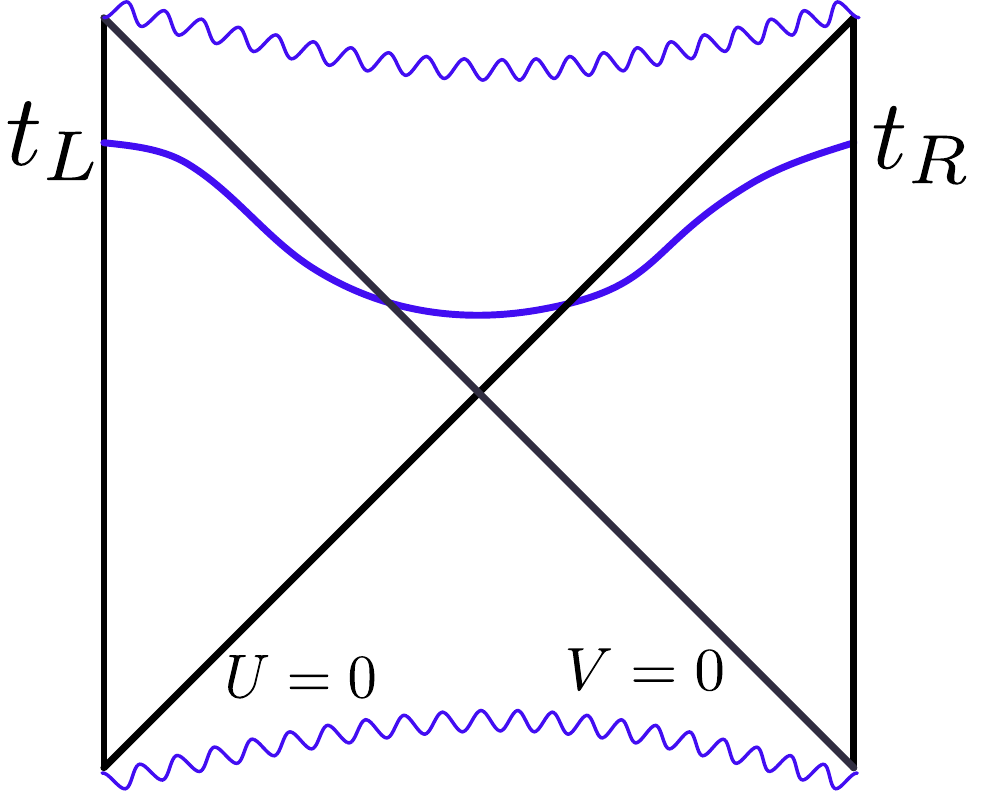}}
\caption{Penrose diagram of the two-sided black hole in AdS and the maximal-volume surface connecting two constant time slices on the opposite boundaries.}
\label{fig:PenroseCV}
\end{figure}

The metric of the Schwarzschild-AdS${}_{d+1}$ solution (with conformal boundary $\mathbb{R}\times S^{d-1}$) is 
\begin{equation}\label{AdS-BH}
\begin{split}
    ds^2 &= -f(r) dt^2 + \frac{dr^2}{f(r)} + r^2 d\Omega_{d-1}^2     \\
    & = - \frac{f(r) e^{-4 \pi T r_*}}{(2\pi T)^2} dU dV + r^2 d \Omega_{d-1}^2 \,, \\
    f(r) & = 1 + \frac{r^2}{\ell_{AdS}^2} -\frac{\mu}{r^{d-2}}
\end{split}
\end{equation}
where $\mu$ is proportional to the mass of the black hole: 
\begin{equation}
    M = \frac{(d-1)\omega_{d-1}}{16 \pi G_N} \mu \,.
\end{equation}
We have denoted by $\omega_{d-1}$ the area of the sphere $S^{d-1}$. 
The mass determines also the Hawking temperature via $f(r_h)=0$, $f'(r_h)= 4 \pi T$, where $r_h$ is the horizon radius. The entropy of the black hole is given by  $S=\omega_{d-1} r_h^{d-1}/(4G_N)$.
In the second line of \eqref{AdS-BH}, the metric is expressed in terms of the Kruskal coordinates $U$ and $V$ that cover the maximal analytical extension of the spacetime. The full spacetime can be divided in four regions, depending on the signs of $U$ and $V$ (see Fig. \ref{fig:PenroseCV}). 
The relation with the $t_R,r$ coordinates defined on the right quadrant is 
\begin{equation}
    U = -e^{-2 \pi T (t_R-r_*)} \,, V = e^{2 \pi T (t_R+r_*)}
\end{equation}
where $r_*$ is the tortoise coordinate defined by $d r_* = dr/f(r)$. We can see that the original coordinates only cover the region $U<0,V>0$. 
The metric has an isometry $U \to e^{-a} U, V \to e^{a} V$ which is just time translation $t_R \to t_R+a$, but on the left boundary it translates time in the opposite direction: $t_L \to t_L -a$. We chose the time coordinates $t_{L},t_{R}$ to run in the same direction on both sides of the Penrose diagram. The isometry reflects the invariance of the TFD state under the evolution generated by $H_L-H_R$.
In Kruskal coordinates, the boundaries are located at $UV=-1$, the horizon is the union of the lines $U=0$ and $V=0$, and the black hole singularity is at a constant value of $UV >0$. 

\paragraph{CV conjecture}: 
Let us now consider a bulk hypersurface connecting constant time slices at $t_L, t_R$. We can use the isometry to set $t_L = t_R\equiv \frac{t}{2}$. 

Describing the surface by an embedding $t(r)$, its volume is calculated as\footnote{Here, we are being slightly cavalier in our treatment since of course the time coordinate is singular when crossing the horizon. The proper treatment would be to convert those expressions to coordinates that interpolate  smoothly across the horizon (for example $V$ and $r$, see \cite{Carmi:2017jqz}).}
\begin{equation}\label{volume-slice}
    \textrm{Vol}(\mathcal{N})= \omega_{d-1} \, \int dr \, r^{d-1} \sqrt{-f(r) t'(r)^2+ \frac{1}{f(r)}} \,.
\end{equation}
We can integrate the equation for extremizing \eqref{volume-slice} using the existence of an integral of motion $\gamma$:\footnote{Note that this integral blows up at the horizon. However, when writing it we mean that one should compute it using the principal value prescription.}
\begin{equation}
    t(r) = \int_{r_0}^\infty \frac{dr}{f \sqrt{1+\gamma^{-2}f r^{2d-2}}} \,, \quad t(\infty) = t_R \,.
\end{equation}
The vanishing of the denominator gives the turning point $r_0$ of the surface: $\gamma^2 = |f(r_0)| r_0^{2d-2}$ behind the horizon. So $\gamma \leq \gamma_{max} =\textrm{max} ( \sqrt{|f|} r^{d-1})$, and as $\gamma \to \gamma_{max}$ the integral diverges logarithmically, so $t_R \to \infty$. Using the integral of motion, the volume can be rewritten as 
\begin{equation}\label{CVintegral}
    \textrm{Vol}(\mathcal{N})= 2 \, \omega_{d-1} \, \int_{r_0} dr \frac{r^{2d-2}}{\sqrt{\gamma^2+f(r) r^{2d-2}}} \,.  
\end{equation}
Comparing the last two equations, we see that the integrals for the time and the volume have the same logarithmic divergence at the lower integration limit, \ie the region when $r \approx r_0$, so we can estimate
\begin{equation}\label{vol-linear-growth}
\textrm{Vol}(\mathcal{N}) \sim  \omega_{d-1} \gamma_{max} \, t \quad \textrm{as} \,\, t \to \infty \,.
\end{equation}
The maximal volume then grows linearly in time, and this can be attributed to the growth of the region behind the horizon, the ER bridge that connects the L and R theories. For a black hole with a large mass one finds 
\begin{equation}
   \gamma_{max} \sim \frac{\mu \ell_{AdS}}{2} \,, \quad \frac{d \mathcal{C}_V}{d t} \sim \frac{8 \pi}{d-1} M \sim \frac{8 \pi}{d} TS \,.
\end{equation}
So the volume grows at a rate proportional to the total energy. 
The volume has also a divergence from the upper integration limit $r \to
\infty$. This is the typical UV divergence coming from the AdS
boundary, and as usual we regulate it with a radial cutoff $r_{\text{max}} = \ell_{AdS}^2 / \delta$.
We find that the leading divergent term is\footnote{The factor of $2$ in this equation comes from having two asymptotic boundaries of the eternal black hole.}
\begin{equation}\label{vol-divergenceold}
    \textrm{Vol}(\mathcal{N})_{div} \sim \frac{2}{d-1} \, \omega_{d-1}  \frac{\ell_{AdS}^{~2d-1}}{\delta^{d - 1}} \,.
\end{equation}
This leads to a complexity
\begin{equation}\label{vol-divergence}
    \mC_{V,div} \sim  \frac{\tilde c}{d-1} \,  \frac{\text{Vol}}{\delta^{d - 1}} \,,
\end{equation}
where $\tilde c=\ell_{AdS}^{d-1}/G_N$ is proportional to the central charge of the theory \cite{Buchel:2009sk} and $\text{Vol}=2\omega_{d-1}\ell_{AdS}^{d-1}$ is the total spatial volume of the two boundary time slices.  
Notice that this term is time-independent; this is easy to understand, since
when $r$ is large we can neglect the $\gamma^2 $ term in the denominator of equation \eqref{CVintegral}.
Moreover it is also state-independent: different states correspond to
asymptotically-AdS geometries with the same metric at leading order and
corrections  of relative order $1 / r^{d}$. Therefore the difference of the volume in two different states is finite, and can be regularized by a state-independent subtraction. 
This state-independent subtraction can be done by focusing on the \emph{complexity of formation}  which we defined in equation \eqref{formationdef} where we subtracted from the complexity of the TFD state at $t=0$ that of two copies of the vacuum state (here empty AdS), this yields in the high temperature limit in $d>2$ \cite{Formation}
\begin{equation}\label{formationcv}
    \Delta \mC_V = 4\sqrt{\pi} \frac{(d-2)\Gamma\left(1+\frac{1}{d}\right)}{(d-1)\Gamma\left(\frac{1}{2}+\frac{1}{d}\right)} S+\ldots,
\end{equation}
where the dots indicate corrections away from high temperatures. Note that the complexity of formation is proportional to the entropy, just like what we found in the free field theory in equation \eqref{formationfinalfree}, although with a different coefficient. In $d=2$ the coefficient of the entropy in this expression vanishes and we are left with a constant complexity of formation.
In particular, if we compare the complexity of the BTZ black hole to that of the Neveu-Schwarz vacuum in the boundary theory we obtain 
$\Delta \mC_V =8\pi c/3$ where $c=3\ell_{AdS}/(2G_N)$ is the central charge, whereas comparing to the Ramond vacuum instead yields $\Delta \mC_V =0$.
In all these examples the complexity of formation is non-negative. This property was proven in general in asymptotically AdS spaces in $d=3$ and in some symmetric spaces in other dimensions in \cite{Engelhardt:2021mju}.

When the boundary geometry is not flat, the subtraction contains \eqref{vol-divergence} as the leading term, but also additional subleading divergences that we will not discuss here; their  structure was analyzed in \cite{Carmi:2016wjl,Reynolds:2016rvl}.

\paragraph{CA conjecture}: As stated before, we need to find the
domain of dependence of a Cauchy slice in the bulk, ending on the boundary at
$t_L = t_R$. This is the part of the bulk that can be unambiguously
reconstructed if one knows only the initial conditions on the slice. It is
easy to see that the  WDW patch consists of points
that are  spacelike-separated from all boundary points in $\Sigma$. The boundary of the WDW patch
is then obtained by considering the innermost null geodesics starting from the boundary
at the given time. In Kruskal coordinates, denoting the coordinate of the
boundary time slices as $(U_L, V_L )$, $(U_R, V_R)$, with $U_L V_L=U_R V_R=-1$, $U_L/V_L=V_R/U_R$, these geodesics are the surfaces $U = U_L$, $U = U_R$, $V = V_L$, $V = V_R$. 

\begin{figure}[htbp]
\centerline{\includegraphics[scale=.6]{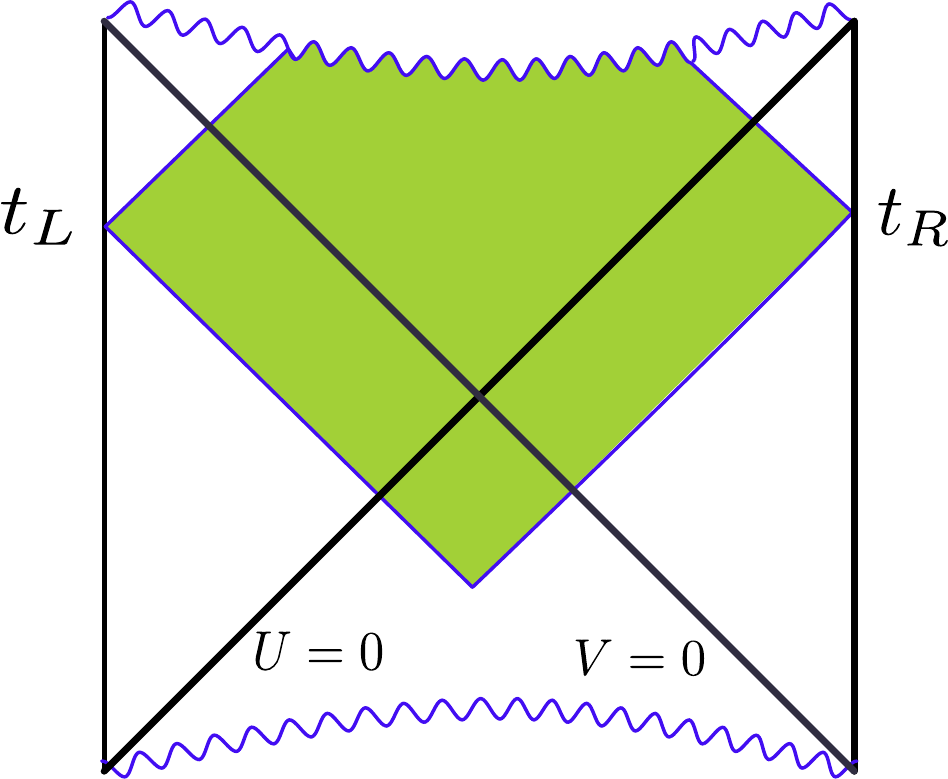}}
\caption{The Wheeler-DeWitt patch used in the computation of CA.}
\label{fig:Penrose-WDW}
\end{figure}

If we are considering a solution to Einstein gravity with a cosmological constant, then naively, the on-shell action will be proportional to the spacetime volume of the WDW patch. However the spacetime region we consider has boundaries, and it is well-known that in the presence of boundaries the Einstein-Hilbert action has to be supplemented by additional boundary terms. For spacelike or timelike boundaries these are  the Gibbons-Hawking boundary terms .  However, these terms are not well-defined on null surfaces, due to the fact that the induced metric is degenerate. Furthermore, the boundaries of the WDW patch are not smooth. They consist of multiple components that intersect along codimension-two corners. The complete action appropriate in this situation was found in \cite{Lehner:2016vdi} (see also \cite{Parattu:2015gga,Carmi:2016wjl,Neiman:2012fx}) and can be written as a sum of terms $S=\sum_j S_j$ associated to regions of codimension $j$. The terms are\footnote{Note that the sign of the null-boundary terms is flipped compared to the appendix of \cite{Lehner:2016vdi}v1 and \cite{Carmi:2016wjl}v4 where there was a sign mistake, see footnote 11 in \cite{Vaidya1}.} 
\begin{equation}\label{Action-bdy}
    \begin{split}
         16 \pi G_N S_0 & =  \int_M d^{d+1} x \sqrt{-g} (R-2 \Lambda) \,, \\
         16 \pi G_N S_1 & = 2 \epsilon_K \int_{B_\pm} d^d x \sqrt{|h|} K \\
         & + 2 \int_{B_0} d^{d-1}\theta d\lambda \sqrt{\gamma} \left(\epsilon_\kappa \kappa - \Theta 
         \log (\ell_{ct}| \Theta|) \right)\,, \\
         16 \pi G_N S_2 &= 2 \epsilon_a \int_J d^{d-1}\theta \sqrt{\sigma} a \,. 
    \end{split}
\end{equation}
Here $B_\pm$, $B_0$ are the spacelike $(+)$, timelike $(-)$ or null $(0)$ components of the boundary, $K$ is the trace of the extrinsic curvature, $(\theta^a, \lambda)$ are coordinates on $B_0$ such that $\lambda$ is a parameter on the null generators of the surface, increasing towards the future; $\kappa$ is defined by $k^\mu \nabla_\mu k_\nu = \kappa k_\nu$ for the vector field  normal to the surface $k^\mu \partial_\mu = \partial_\lambda$, $\Theta = \partial_\lambda \log \sqrt{\gamma}$ is the trace of the second fundamental form, which gives the expansion rate of the congruence of null generators. $J$ denotes the joints, or corners, arising from the intersection of two boundary components. There can be different types of joints: for $J=B_\pm \cap B_0$, $a=\log |n\cdot k|$, and for $J=B_0 \cap B_0'$, $a= \log |\frac{k\cdot k'}{2} |$. The normal vectors have to be taken pointing outwards from the region $M$ for timelike surfaces and be future-oriented for spacelike and null surfaces. The factors $\epsilon_K, \epsilon_\kappa, \epsilon_a$ are signs:
$\epsilon_K = 1$ for a timelike boundary while for a spacelike boundary 
$\epsilon_K= 1 (-1)$ if the region $M$ lies in the future (past) of the boundary component; 
$\epsilon_\kappa= 1 (-1)$ if the region $M$ lies in the future (past) of the boundary component; and $\epsilon_a=-1$ if the volume of interest lies to the future (past) of the null segment and the joint lies to the future (past) of the segment, otherwise $\epsilon_a=1$, see appendix C of \cite{Lehner:2016vdi}.\footnote{Other types of joints not involving null surfaces can of course exist but we will not need them, see \cite{Carmi:2016wjl} for a full discussion.}   
The boundary term on the null boundaries is given in \eqref{Action-bdy} using a particular parametrization, but one can show that it is reparametrization-invariant, thanks to the term involving $\Theta$.\footnote{The use of this counterterm was first advocated in  \cite{Parattu:2015gga}.} Notice that this term requires the introduction of a length scale $\ell_{ct}$ on top of the AdS scale.

With all these ingredients at hand, we can compute the action of the WDW patch. It is UV divergent, and there are different ways to regulate it: we can compute the action of the WDW patch restricted to the part of the bulk within the cutoff, or alternatively we can compute the WDW patch in the cutoff space, with null geodesics starting from the cutoff surface $UV=-1+4 \pi T \delta$. The two regularization schemes lead to the same result for the leading divergence \cite{Reynolds:2016rvl,Carmi:2016wjl,Caceres:2019pgf}\footnote{One must take care to include the additional counterterms at the boundary of AdS that are used for holographic renormalization \cite{Skenderis:2002wp} when the WDW patch has a boundary at the cutoff surface, see footnote 79 of \cite{Caceres:2019pgf}.}

\begin{equation}\label{action-divergence}
S_{div} \sim \frac{2}{4\pi G_N} \frac{\ell_{AdS}^{2d-2}}{\delta^{d-1}} \omega_{d-1} \log \left(  (d-1)  \frac{\ell_{ct}}{\ell_{AdS}}
\right) \,.
\end{equation}
This gives a divergence in the complexity 
\begin{equation}
    \mathcal{C}_{A,div} \sim \frac{\tilde c}{4\pi^2} \frac{\text{Vol}}{\delta^{d-1}} \log \left(  (d-1)  \frac{\ell_{ct}}{\ell_{AdS}}
\right),
\end{equation}
where, as before, $\tilde c=\ell_{AdS}^{d-1}/G_N$ is proportional to the central charge of the theory and $\text{Vol}$ is the total spatial volume of the two boundary time slices. 
This has the same structure as \eqref{vol-divergence}: it is extensive in the field theory volume and diverges as $\delta^{1-d}$; the prefactor is different, but in  both cases it depends on an arbitrary length scale (recall that in CV the scale enters in the prescription \eqref{CV-def}). 

As for the time dependence, one can see that thanks to the time-translation isometry, the action of the part of the WDW patch outside the horizon is time-independent. For late times, the part behind the past horizon becomes vanishingly small, so the only contribution comes from the part within the future horizon.\footnote{There may be subtleties in this statement, as discussed in \cite{Brown2,Lehner:2016vdi}.} 
The patch extends to the singularity. However, the relevant contribution to the action is finite due to the fact that the sphere shrinks there and there is no need to regularize the singularity. The computation done in \cite{Brown2} gives 
\begin{equation}\label{action-growth}
    \frac{dS}{dt} = 2 M \,.
\end{equation}
It is interesting to note that this result is independent of the counterterm scale $\ell_{ct}$.

As before, it is interesting to consider the \emph{complexity of formation} \eqref{formationdef} where we subtracted from the complexity of the TFD state at $t=0$ that of two copies of empty AdS. This yields at high temperatures in $d>2$ \cite{Formation}
\begin{equation}\label{formationca}
    \Delta \mC_A =  \frac{(d-2)}{d\pi} \cot\left(\frac{\pi}{d}\right) \, S+\ldots,
\end{equation}
where the dots indicate corrections away from high temperatures. Once again we find the proportionality of the complexity of formation to the entropy.  In $d=2$ the coefficient of the entropy in this expression vanishes and we are left with a constant complexity of formation. In particular, if we compare the complexity of the BTZ black hole to that of the Neveu-Schwarz vacuum in the boundary theory we obtain $\Delta \mC_A =- c/3$ where $c=3\ell_{AdS}/(2G_N)$ is the central charge whereas comparing to the Ramond vacuum instead yields $\Delta \mC_A =0$.

\subsection{Comparison between CV and CA} 
\label{sec:Comparison}

The first thing to notice is that both the CV and the CA results contain some ambiguities. CV requires a length scale for dimensional reasons; CA appears at first to be more canonically defined, but as we have seen, the presence of null  boundary terms naturally  reintroduces an additional scale. 
Moreover, the action could be modified by additional boundary terms. For example when dealing with charged black holes it turns out that the complexity can depend strongly on the boundary conditions one imposes on the  associated Maxwell field  \cite{Goto:2018iay}. 

Comparing with the results of the previous sections, we see that both the volume of maximal slices and the action of the WDW patch show the same behavior as  the complexity in the free-field theory examples from section \ref{sec:Complexity-free-QFT}. First, the UV divergent part obeys a volume law, and depends on the cutoff as $\delta^{1-d}$, the same as the free-field theory result for $\mathcal{C}_1^{UB}$ in equation \eqref{c1mutilde}. If we consider instead the free-field result for $\mathcal{C}_2$ in equation \eqref{divervac}, we see that it has a different power law and cannot be matched the holographic result. Comparing to our holographic results in equations \eqref{vol-divergence} and \eqref{action-divergence} we are led to identify 
\begin{equation}
    |\tilde\mu| \propto \ell_{AdS}/\ell_{CV} \propto \log (\ell_{ct}(d-1)/\ell_{AdS})
\end{equation}
where here we introduced back the length scale $\ell_{CV}$ involved in the definition of the CV proposal. We see that in fact the choice we could make in the field theory side for the scale of the reference state is  naturally identified with the freedom which we have in the CV and CA proposals.

Second, 
the linear growth in time matches the expectation 
from the circuit model \eqref{circuit-growth}. 
Recall that using the relation with the Lyapunov exponent under the assumption of maximal chaos, the circuit time $n$ is related to the physical time as $n \propto  T t$, and the number of qubits $N$ is proportional to the entropy of the system. With these identifications, the rate of growth of the complexity for a black hole is expected to be proportional to $TS$ at late times. 
This expectation is borne out both by CV and CA. It is worth noting that the linear growth of complexity for a very long time is not reproduced in the free field theory model in section \eqref{sec:TFDcompandtdep}. Indeed, in such a simple theory the dynamical properties of complexity are expected to differ significantly from those of  chaotic systems.

The result for CA in equation \eqref{action-growth} may look more satisfactory, giving a rate of linear growth exactly equal to the mass, while for CV there is a proportionality factor that depends on the dimension. However, given the uncertainty in the identification of time, and the fact that the definition of complexity itself does not fix the normalization, we should be skeptical about the significance of the precise prefactor. 
Nevertheless, the holographic prescription fixes a particular normalization, and 
one may still be tempted to conjecture that \eqref{action-growth} is a universal result for holographic models. This turns out not to be the case: for charged and rotating black holes the rate is a non-trivial function of the charge and angular momentum, and does not coincide with $M$. Initially \cite{Brown1,Brown2} speculated that $M$ might give an absolute upper bound on the rate of growth of complexity, based on an analogy with the Lloyd's bound on the rate of computation \cite{LLoyd} (which in turn is based on the orthogonality bounds discussed in section \ref{sec:Primer}). This turns out to be false as well: 
 it was shown in \cite{Carmi:2017jqz} that at late times the limiting value of the rate of change in complexity using CA is approached from above, thus violating the supposed bound by an amount that can be made arbitrarily large. Another counterexample was given in \cite{Swingle:2017zcd}: in the case of hyperscaling-violating solutions of Einstein-Maxwell dilaton theory the growth was found to be enhanced compared to the CFT case: $dS/dt= 2E (1+\frac{z-1}{d-\theta})$ where $E$ is the energy (equal to $M$ in the $z=1$ case). This however would still be compatible with a putative bound given by $2TS$.  In fact a counterexample was given already in the initial paper \cite{Brown1,Brown2}: the bound is violated for large charged black holes\footnote{By this we mean, charged black holes whose horizon radius is much larger than the AdS scale.} and this violation is most pronounced close to extremality, but in general such black holes are unstable to the emission of light charged particles. 
 Recently a version of the holographic Lloyd's bound was proven for the case of CV: it was shown \cite{Engelhardt:2021mju} that under certain energy conditions, in asymptotically-AdS spacetimes in $d\geq3$, the rate of growth of $\mathcal{C}_V$ is bounded by $\frac{8\pi M}{d-1} f(M)$,  where $f(M)$ is a function equal to 1 for $M\leq \hat M$ with $\hat M$ a mass scale near the Hawking-Page transition, and $f(M)=1+2(M/\hat M)^{1/(d-2)}$  for $M>\hat M$.
We will comment further on the  bounds on the rate of computation in the discussion section. 

Finally, let us note that the complexity of formation in holography using the CV
\eqref{formationcv} and CA \eqref{formationca} proposals was found to be proportional to the entropy in $d>2$. We observed a similar behavior in free field theory where the mass was set to zero \eqref{formationfinalfree}. While the dependence of the proportionality coefficient on the dimension was different in all these cases, as we already mentioned earlier, this coefficient is somewhat arbitrary in the prescriptions for evaluating complexity.

\subsection{Tensor Network Model}
\label{sec:TensorNetwork}

A different perspective on the growth of the complexity can be gained by considering a tensor network model. This gives another argument for the linear growth of complexity with a prefactor proportional to the temperature times the entropy of the system \cite{Brown2}. Tensor networks have been used as a computational tool to provide an efficient representation of states (\eg of a spin system) that are less entangled than a typical state. Typically one is interested in the ground state of a local Hamiltonian, which has area-law entanglement entropy (with logarithmic corrections for a gapless system)  whereas the typical state has a volume law entanglement entropy. 
We cannot give a full account of the topic in this review, the reader can find more details, \eg  in the recent review  \cite{Jahn:2021uqr}.

\begin{figure}[htbp]
\centerline{\includegraphics[scale=.8]{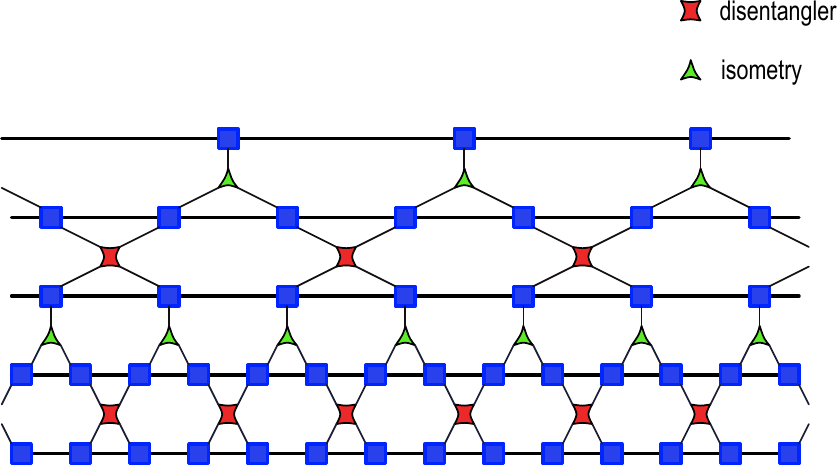}}
\caption{Illustration of a MERA circuit, implementing the RG flow from the bottom (UV) to the top (IR).}
\label{fig:TensorNetwork}
\end{figure}

It has been proposed that Tensor networks can provide a discretized picture of AdS/CFT, 
in particular using the MERA (Multi-Entanglement Renormalization Ansatz) tensor networks which are especially designed for constructing ground states of critical systems \cite{Swingle:2009bg}. In a MERA network, the ground state state of a critical system is produced by iterating two types of operations, as illustrated in Fig. \ref{fig:TensorNetwork}. One operation is the disentangler, which introduces entanglement between the pair of qubits that it acts on; the other is the isometry, which makes a coarse-graining of the degrees of freedom. The effect of the two operations is that entanglement is introduced in the state at increasingly larger length scales.

Schematically, one starts from an unentangled state at a UV scale $\Lambda$, for a system of length $L$. One layer of the circuit acts on the state with an operator $V$ and  gives the wave function at the coarse-grained scale $\psi(2 L, \frac{\Lambda}{2}) = V \psi(L, \Lambda)$. 

\begin{figure}[htbp]
\centerline{\includegraphics[scale=1.15]{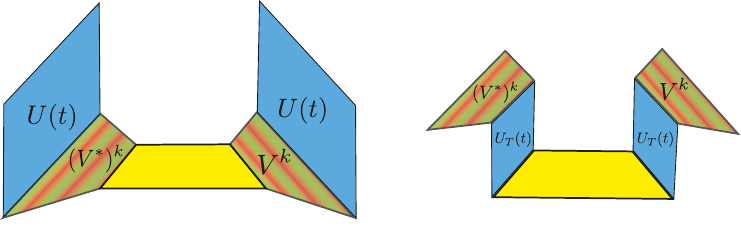}}
\caption{Illustration of a tensor network constructing the TFD state and its time evolution.}
\label{fig:TensorNetwork2}
\end{figure}

The thermofield double state at temperature $T$ has entanglement at length scales smaller than $1/T$ on each side while points at larger distances are unentangled. Therefore the circuit that builds two copies of the ground state also builds to a good approximation the finite-temperature TFD at short length scales. At the scale of the temperature the state is given by 
\begin{equation}
\begin{split}
    &\ket{TFD(L,T)} = V^k \otimes (V^*)^k \ket{TFD(L/2^k, \Lambda)} \,,\\
    &\quad  \,.
\end{split}
\end{equation} 
where $k = \log_2 \frac{\Lambda}{T}$ is the number of gates in layers in the circuit. The operation of $V$ in the circuit constructing the TFD is depicted in red/green in Fig. \ref{fig:TensorNetwork2}.

Now, if we consider the evolution of the TFD state in time, we should attach a unitary time evolution operator to the UV part of the circuit. This evolution is described in blue in Fig. \ref{fig:TensorNetwork2}. Naively then, we could expect that the complexity grows as $N \Lambda \, t$, (with $N=(L\Lambda)^{d-1}$ the number of UV degrees of freedom) since the Hamiltonian acts on all the UV degrees of freedom. 
However, it turns out that this is not the most efficient way to prepare the TFD state at finite time.
In fact, by swapping the action of the $V$ operators with the time evolution operators we can convince ourselves that it is more efficient in the complexity sense (it requires fewer operations) to act with an effective Hamiltonian on the IR degrees of freedom, see the right panel of Fig. \ref{fig:TensorNetwork2}. Since we are describing a critical system, we can use the fact that  $H(L) V^k = V^k 2^{-k\Delta} H(L/2^k)$,  namely that the Hamiltonian is a scaling operator, with dimension $\Delta=1$. Then we can act on the IR state with a renormalized Hamiltonian;  this is much more efficient since the number of sites on which we need to act is reduced by a factor of $2^k$ after $k$ steps. At the scale $T$ the number of sites is $LT$, and so the expected growth rate of complexity is reduced to $T(LT)^{d-1} \sim T S(T)$. In this way we recover the same prefactor in the rate of growth of complexity that arose from the epidemic model. 

We hasten to add that the argument is very heuristic, and the precise correspondence of tensor networks with holography is far from being completely established. 

\section{Additional Tests of the Holographic Conjectures }
\label{sec:Holo2}

\subsection{Shock Waves}

A particularly important support for the complexity conjectures can be obtained by studying their  behavior under a perturbation of the system, and comparing it to the predictions from the circuit model in section \ref{sec:fast-scramblers}. In \cite{Stanford:2014jda} the authors considered the evolution of the TFD state after the application of a precursor: 
\begin{equation}
    U_L(t_L) U_R(t_R) W_L(t_w) \ket{TFD} \,.
\end{equation}
Here $W$ is a local CFT operator of energy $E \ll M$ -- more precisely, $E=\mathcal{O}(1)$, while $M = \mathcal{O}(N^2)$. 
The operator acts on the boundary at a time $t_w$, and creates an excitation which propagates in the bulk along a null line. 
As the excitation moves towards the horizon its energy gets more and more blue-shifted, so its backreaction cannot be ignored, even though the initial energy of the excitation is small. The backreaction is described by a shock wave \cite{Shenker:2013pqa}. 
For simplicity, we consider the case of AdS$_3$, and we take an excitation created by an operator which is  smeared uniformly along the circle at the boundary, sent from the left at some very early time.\footnote{The case of localized shocks was analyzed in \cite{Roberts:2014isa}, see also \cite{Sfetsos:1994xa}; the higher-dimensional generalization was considered in \cite{Stanford:2014jda,Vaidya1,Vaidya2}.} 
In this case the perturbed metric can be written in Kruskal coordinates as
\begin{equation}
    ds^2 = - A(r) (2 dU dV - 2 h  \delta(U) dU^2) + r^2 d\phi^2
\end{equation}
where $A(r)= f(r)e^{-4 \pi T r_*}/(8\pi^2 T^2)$ can be read by comparing to the unperturbed metric in equation \eqref{AdS-BH}. The perturbation can be interpreted as a shift in the $V$ coordinate across the horizon: $V \to V- h \theta(U)$.
The bulk stress energy tensor is localized on the shock wave; we can write it as 
\begin{equation}
T_{UU} =\frac{E}{16 \pi G_N M} e^{2 \pi T |t_w|} \delta(U) \,,
\end{equation}
where $E$ is the energy of the excitation inserted at $t_w=0$. Solving Einstein's equations gives 
\begin{equation}\label{shocksol}
    h \sim \, e^{2\pi T(|t_w|-t_*)} \,, \quad t_* = \frac{1}{2\pi T} \log \frac{M}{E} \,,
\end{equation}
where $t_*$ is the scrambling time. The solution is valid in the limit where $E \to 0, |t_w| \to \infty$, with $h$ fixed. In this limit the shock wave propagates along the horizon. 

\begin{figure}[htbp]
\centerline{\includegraphics[scale=.5]{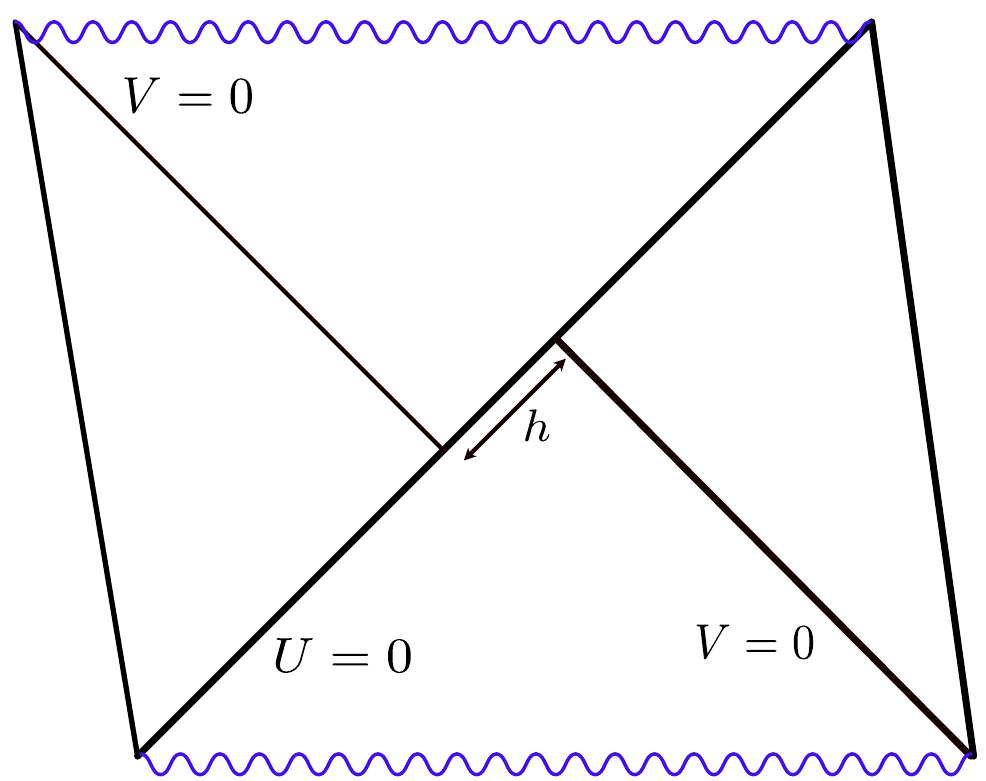}}
\caption{Penrose diagram of the AdS$_{3}$ black hole geometry perturbed by a shock wave.}
\label{fig:Shockwave}
\end{figure}

Due to the shift in $V$, the maximal slices are displaced when they cross the horizon. The modification of the volume can be computed analytically in the 3d case, and the corresponding complexity is given, up to an additive constant which is UV divergent but time-independent, by the formula \cite{Susskind:2018pmk,Shenker:2013pqa} 
\begin{equation}\label{BTZ-geodesics}
   \mathcal{C}_V \sim S \log \left[ \cosh \left(\pi T (t_L+t_R)\right) +\mathfrak{c}\, h \, e^{\pi T( t_L-t_R)} \right]\,,
\end{equation}
where $S$ is the entropy, $\mathfrak{c}$ is some order one constant and $h$ is given in \eqref{shocksol}. 
Setting $t_L=t_R=0$, the formula has the same  dependence on $|t_w|$ and the scrambling time $t_*$ as the result of the epidemic model \eqref{epidemic-comp}: it grows exponentially with $|t_w|$ for $|t_w| \ll t_*$, and linearly for $|t_w| \gg t_*$. Also as a function of $t_L$, $t_R$ at fixed $h$ we can see different regimes. Setting for instance $t_L=-t_R$, we have exponential growth in $t_L$ for $t_L \ll t_* - |t_w|$ 
followed by a linear growth at late times. 

The formula \eqref{BTZ-geodesics} is actually a good approximation also for the complexity in shockwave backgrounds in higher-dimensional AdS black holes, because one can argue, with a reasoning similar to the one that led to equation \eqref{vol-linear-growth}, that the main contribution comes from  a region where $r$ is almost constant, and therefore the volume of the angular directions only contributes an overall factor but does not change the shape of the maximal surface.

More explicitly, we can evaluate the leading late-time result as follows \cite{Stanford:2014jda}: one finds that the volume of a maximal surface connecting the left boundary at $t_L$ to the horizon at $(U=0,V_R)$ 
is given by 
\begin{equation}
\textrm{Vol}(t_L,V_R) \sim  \frac{\omega_{d-1} \gamma_{max}}{2 \pi T} \log(V_R e^{2\pi T t_L}) \,.
\end{equation}
The remaining part of the surface goes from $(U=0,V_R-h)$ to the boundary at $t_R$. Minimizing the sum of the two contributions over $V_R$ gives $V_R = h/2$, and 
\begin{equation}\label{switchback}
\begin{split}
     \textrm{Vol} & =\frac{\omega_{d-1} \gamma_{max}}{2 \pi T}\left( \log \left(\frac{h}{2} e^{2\pi T t_L}\right)+\log \left(\frac{h}{2} e^{-2 \pi T t_R} \right) \right) \\
     & = \omega_{d-1} \gamma_{max} (t_L-t_R +2 |t_w| - 2 t_*)+\mathcal{O}(1)\,.
\end{split}
\end{equation}
In this derivation we assumed $t_L > t_w$, $t_R < - t_w$. 
The argument can be extended to more complicated insertions of the form $U_L(t_L)W_L(t_1)\ldots W_L(t_n)U_R(t_R)$. We describe here the results of  \cite{Stanford:2014jda}, that constructed the geometries corresponding to multiple shock waves 
created by the insertion of operators on the left side at different times, building up on the work of  \cite{Shenker:2013yza}. Since the times $t_1, \ldots t_n$ do not have to be ordered, one has to  distinguish the operator insertions that are time-ordered from those that are not. The former give rise to shock waves that propagate in the same direction, and only give small perturbations to the geometry. The latter create shock waves propagating in opposite directions and have a larger effect. The geometry corresponding to multiple shock waves can be constructed patching together portions of AdS along the horizon with shifts $(V,U) \to (V,U)\pm 2e^{-2\pi T (t_*\pm t_i)}$, where the coordinate being shifted, as well as the sign in the exponent, depends on the direction of the shock wave. One finds, in agreement with the expectation from the circuit model,  that the complexity grows linearly with the time difference between insertions, and with the offset from the switchback coming from points where the time contour folds; the generalization of equation \eqref{switchback} to multiple time insertions then reads 
\begin{equation}
    \textrm{Vol} \sim |t_L - t_1| + |t_1 - t_2| + \ldots |t_R + t_n| - 2 n_s t_* \,.
\end{equation}
This result is valid only in the limit when all the time differences between the different shocks and between the shocks and the boundary times are very large compared to $t_*$; the exact formula, just as for a single shock, will also exhibit different regimes where the volume grows exponentially. It was shown in \cite{Brown2} that the same behavior is obtained also using the CA prescription, although with more cumbersome calculations, especially in the case of multiple shock waves.

In the limit $E \to 0$ we have considered, the energy of the shock is negligible and it does not change the mass of the black hole. The case of a finite-energy shock was considered in \cite{Vaidya2}. 
In that case, with a single shockwave, the complexity grows linearly at late times at a rate proportional to the final mass of the black hole (after it has absorbed the shock), whereas at early times there is a linear growth with a slope proportional to the energy of the shockwave, and a relatively sharp transition between the two regions.\footnote{These results are valid for the CV proposal only in the limit of high temperatures or for planar black holes and for the CA proposal at any temperature and horizon geometry.}  

\begin{figure}[htbp]
\centerline{\includegraphics[scale=.4]{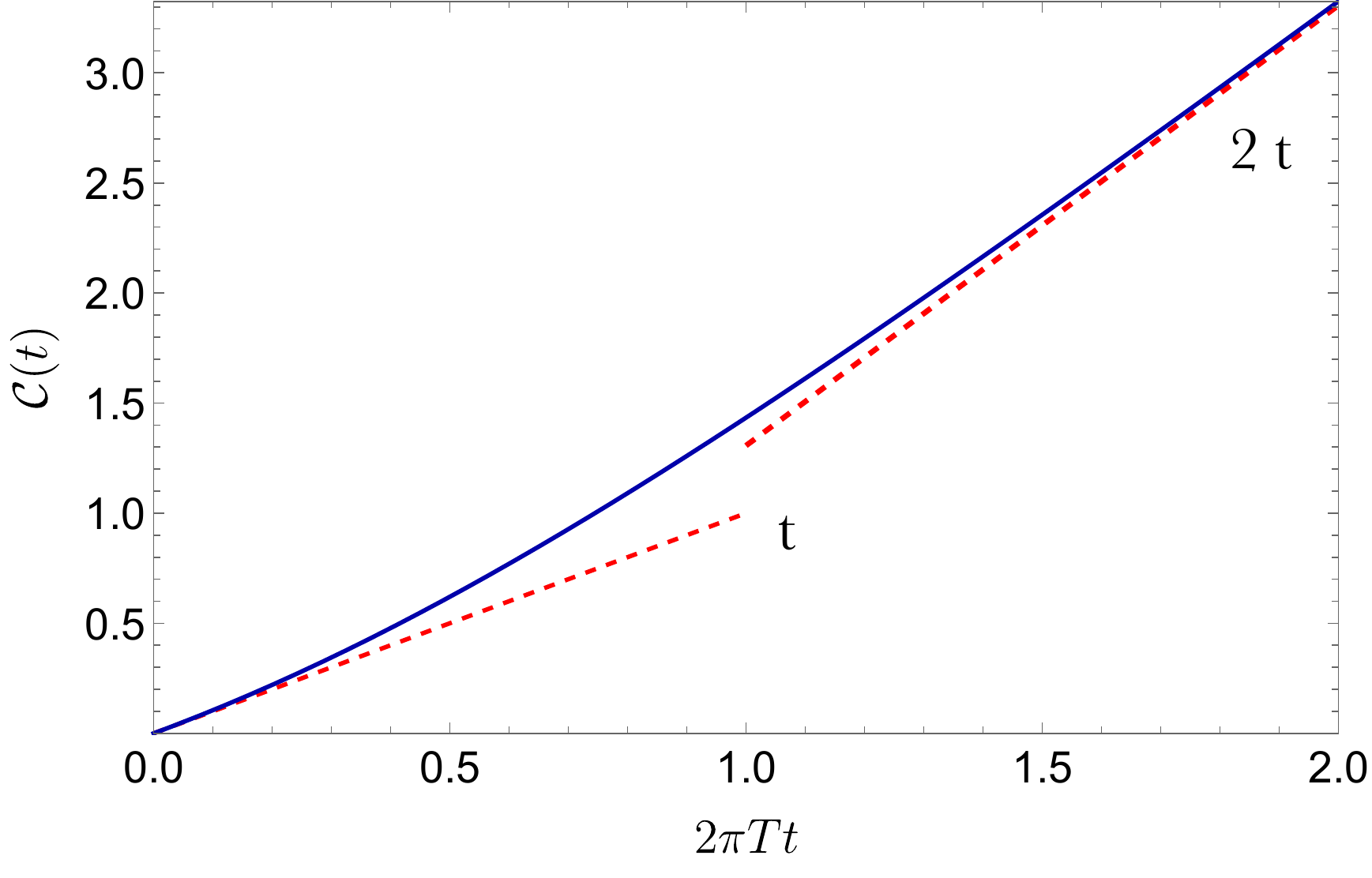}}
\caption{Complexity growth calculated in the epidemic model with an insertion of a heavy operator of size $N_0=N/2$.}
\label{fig:HeavyShockComplexity}
\end{figure}

As observed in \cite{Vaidya2}, the AdS$_3$ result \eqref{BTZ-geodesics} is in agreement with the epidemic model of section \ref{sec:fast-scramblers} and agrees also with the holographic result for light shocks, but does not account for the early-time growth of finite-energy shocks. The epidemic model we used assumed that the perturbation is generated by a simple operator. In fact, it is easy enough to account for the insertion of a heavy operator. We simply have to modify the initial conditions for the number of infected sites. The solution is given by $s(n_0+n)$, with $s(n)$ the number of infected sites at the $n$-th step as in \eqref{episolinv} and $s(n_0) = N_0$ is the size of the operator serving as the initial perturbation. We want to consider the case when the initial size is a finite fraction of the total size, $N_0= \alpha N$. From equation \eqref{episol} we find
\begin{equation}
    n_0 = \frac{1}{k-1} \log \left( \frac{1-(1-\alpha)^{k-1}}{(1-\alpha)^{k-1}} \right) + n_*.
\end{equation}
The time needed for the infection to spread to the whole system is now $n_*- n_0$, and we see that it is much shorter than the scrambling time, since it does not scale with $N$.\footnote{At first glance it is not obvious that $n_0<n_*$. To explain this, we should be slightly more careful in defining the scrambling time. Since the system is only fully infected asymptotically as $n\rightarrow \infty$, it is natural to define the scrambling time as the time at which the perturbation spread throughout half of the system $\tilde n_* = \frac{1}{k-1}\log\left(\frac{N}{s_0(k-1)} \left(2^{k-1}-1\right)\right)$. With this definition,  we can show that $n_0<\tilde n_*$ when $\alpha<1/2$.}
The early and late time behavior of the complexity can be obtained from the corresponding limits of \eqref{episolinv}, taken without the assumption $s_0 \ll N$. One finds 
\begin{equation}
\begin{split}
    \mathcal{C}(n) & \sim N_0 n \,, \quad n \ll 1 \,, \\
    & \sim N n \,, \quad n \gg 1 \,.
\end{split}
\end{equation}
We see that the behavior is the same as for a finite-energy shock: there is an early-time linear regime with rate controlled by the size of the perturbation, and a later-time linear regime  with rate given by the size of the system,\footnote{Note however that in the case of a gravitational shock the perturbation adds energy to the system, so the final size (\ie energy) is the sum of the initial size and the perturbation, but in the circuit model the perturbation does not increase the size of the system.} as illustrated in Fig. \ref{fig:HeavyShockComplexity}. The timescale of the transition between the two asymptotic regimes (called the delay time in \cite{Vaidya2}) is not controlled by the scrambling time but is of order $t_d \sim 1/T$. 
In the case $k=2$ we can give a formula for the full evolution of the complexity: 
\begin{equation}
    \mathcal{C}(n) = N \log \left(  
    \frac{N-1+e^{n_0+n}}{N-1+e^{n_0}}
    \right).
\end{equation}

\subsection{Subregions}

We have considered in the previous sections the complexity of pure states defined by a full holographic geometry. 
We can also consider mixed states associated to a subregion of the boundary. The information about the density matrix $\rho_A$ of this mixed
state is encoded holographically in the entanglement wedge, \ie the bulk
domain of dependence of the part of the constant time slice contained between
the boundary region and the corresponding RT surface \cite{Czech:2012bh,Wall:2012uf,Headrick:2014cta}.  We recall that the RT surface computes holographically the entanglement entropy of a region on the boundary, and is the minimal surface in the bulk anchored on the boundary of the entangling region, for a review see \cite{rangamani2017holographic}. 
It is natural to extend the complexity conjectures to the case of subregions. The extension of CV was first suggested in \cite{Alishahiha:2015rta} for the case of static geometries; they proposed
to take the volume of the maximal bulk slice bounded by $A$ and by the RT surface. 
In the case of time-dependent geometries the prescription proposed in \cite{Carmi:2016wjl} makes use of the HRT surface \cite{Hubeny:2007xt} which is the appropriate covariant generalization of the RT surface.

The extension of CA, also proposed in \cite{Carmi:2016wjl}, is to take the
action of the region formed by the intersection of the entanglement wedge of $A$ with the WDW patch of any boundary constant-time slice that contains $A$ (one can show that the prescription is independent of the choice 
of the slice).  

The case of a subregion given by a ball $B$ of radius $R$ in the vacuum (\ie pure AdS) was considered in\cite{Alishahiha:2015rta}. Using the CV proposal, one finds for the leading divergence
\begin{equation}
   \mathcal{C}_{V,div} =\frac{\tilde c}{(d-1)} \frac{\text{Vol}(B)}{\delta^{d-1}}.
\end{equation}
This has a volume law, just like the complexity of the full system. In the case of a BTZ black hole, for a segment of length $x$, one has 
\begin{equation} \label{subregion-CV}
    \mathcal{C}_V = \frac{2 c}{3}\left( \frac{x}{\delta} - \pi \right) \,,
\end{equation}
with $c$ the central charge of the dual theory. This result was generalized to multiple segments in \cite{Abt:2017pmf}, who found  
\begin{equation}
     \mathcal{C}_V = \frac{2 c}{3}\left( \frac{x_{tot}}{\delta} - \pi(2 \chi - \frac{m}{2}) \right) \,,
\end{equation}
where $\chi$ is the Euler characteristic of the extremal surface, and $m$ the number of joints between the boundary segments and the RT surface. 
Notice that the finite term is topological, and surprisingly there is no dependence on the temperature of the black hole. This is the case also in global AdS$_3$, but not for higher dimensions \cite{Alishahiha:2015rta,Ben-Ami:2016qex}.

Using the CA prescription for the same situation of a segment in planar AdS$_3$ gives \cite{Auzzi:2019vyh} 
\begin{equation}
    \mathcal{C}_A = \frac{x}{\delta} \frac{c}{6 \pi^2} \log \left( \frac{\ell_{ct}}{\ell_{AdS}}\right) - \frac{c}{3\pi^2}\log  \left( \frac{2 \ell_{ct}}{\ell_{AdS}}\right) \log \left( \frac{x}{\delta} \right) + \frac{c}{24} \,,
\end{equation}
and for the planar BTZ black hole  
\begin{equation}
    \mathcal{C}_A = \frac{x}{\delta} \frac{c}{6 \pi^2} \log \left( \frac{\ell_{ct}}{\ell_{AdS}}\right) - \log  \left( \frac{2 \ell_{ct}}{\ell_{AdS}}\right) \frac{S_{EE}(x)}{\pi^2} + \frac{c}{24} \,,
\end{equation}
where 
\begin{equation}
    S_{EE}(x) = \frac{c}{3} \log \left( \frac{1}{\pi T \delta} \sinh (\pi T x) \right) 
\end{equation}
is the entanglement entropy of the segment. 
In comparison with \eqref{subregion-CV}, CA has a subleading logarithmic divergence that persists also in the limit of zero temperature. Notice that the entanglement entropy appears in this formula in the same way as in the field theory result for the mutual complexity \eqref{Mutual-subregion}, although in order to compare the two we should take the limit $L \to \infty$ in the latter. 
However, the relation between complexity and entanglement becomes more intricate for the case  of multiple segments, see for example the case of two-segments in holography \cite{Auzzi:2019vyh} and in field theory \cite{Camargo:2020yfv} and it  certainly does not hold in dynamical situations since the time dependence of the two quantities is drastically different, as we have already seen.

It was observed in \cite{Agon:2018zso} that while CV subregion complexity is additive in a pure state (\ie $\Delta \mathcal{C}_{V}(\rho_A, \rho_{A^c})=0$, where $A^c$ is the region complementary to $A$), and is in general superadditive, $\Delta \mathcal{C}_{V} \leq 0$,  for CA complexity one cannot make a general statement: it can be subadditive or superadditive, and it may change behavior depending on the value of the counterterm scale  $\ell_{ct}$. However if $\ell_{ct}$  is selected such that the leading divergence in the complexity is positive, then the CA complexity is found to be superadditive  $\Delta \mathcal{C}_A<0$ \cite{Agon:2018zso,Caceres:2019pgf}. This contrasts with the field theory results of 
section  \ref{sec:Complexity-Mixed-States} where the complexity was found to be subadditive $\Delta \mathcal{C}^{\text{diag}} >0$ in the diagonal basis. In the physical basis on the other hand, the complexity was found to be superadditive in several cases \cite{Caceres:2019pgf}.

\subsection{Defects and Boundaries}

Another interesting situation to consider is the presence of boundaries or defects in the field theory. Defects in a CFT that preserve part of the conformal symmetry have been investigated extensively, including their holographic realizations. The simplest model to consider is the thin-brane model, where the defect extends in the AdS bulk as a brane \cite{Azeyanagi:2007qj} (different models were considered in  \cite{Auzzi:2021nrj,Baiguera:2021cba}).  The action is the Einstein-Hilbert action coupled to the action of the brane:
\begin{equation}
\begin{split}
    S = &\frac{1}{16\pi G_N} \int d^3 x \sqrt{-g}\left(R + \frac{2}{\ell_{AdS}^2}\right)
    \\
    &- \frac{T}{8 \pi G_N} \int d^2x \sqrt{-h}\,.
\end{split}
\end{equation}
The gravity solution is obtained by gluing two patches of vacuum AdS$_3$ along the brane, in the way specified by the Israel-Stewart matching conditions \cite{Israel:1966rt}. In this model there are three parameters: the central charges of the theories joined by the defect, $c_{L,R}$, and the tension of the brane $T$. The dependence of the complexity of the vacuum on the tension was studied in \cite{Chapman:2018bqj} for the case of a 2d CFT, with $c_L = c_R =c$.  
When the theory is put on a circle of length $L$, with two defects at the diametrically opposed points $x=0, x= L/2$, one finds 
\begin{equation}\label{defect-cpxty}
\begin{split}
     \mathcal{C}_V & = \frac{4 c}{3} \left( \frac{\pi L}{\delta} + 2 \log\left( \frac{2 L}{\delta} \right) \sinh(2 y^*)
    \right) \,, \\
    \mathcal{C}_A & = \frac{c}{3 \pi} \left( 
  \frac{L}{\delta} \log \left( \frac{e \ell_{ct}}{\ell_{AdS}}\right)+ \frac{\pi}{2} \right) \,,
\end{split}
\end{equation}
where $y^*$ is related to the tension via  $ T \ell_{AdS} = 2 \tanh y^*$. Remarkably there is no dependence on $y^*$ in the CA result, which is completely unaffected by the presence of the defect. One may be tempted to take this surprising result as evidence against the CA conjecture. However it turns out that this is consistent with the result obtained in a simple model of a conformal defect for a free scalar in 2d. This defect is also characterized by a single parameter that determines the matching condition: 
\begin{equation}
    \begin{pmatrix} \partial_x \phi_- \\ \partial_t \phi_- \end{pmatrix} = \begin{pmatrix} \lambda & 0 \\ 0 &\lambda^{-1} \end{pmatrix} \begin{pmatrix} \partial_x \phi_+ \\ \partial_t \phi_+ \end{pmatrix}
\end{equation}
where $\phi_{\pm}$ is the value of the field at the two sides of the defect. 
When one defect is placed at $x=0$ and the opposite defect (which has $\lambda$ replaced with $ \lambda^{-1}$) at $x=L/2$, the spectrum  of the theory is not affected by the defect and therefore the vacuum complexity is unaffected as well, see equation \eqref{c12QFT}. 

This calculation was also extended to the case of a subregion symmetrical across the defect. Just as in the case without defect, the CA subregion has a logarithmic divergence depending on $\ell_{ct}$, but still independent of the defect's parameter. 

Instead of a defect, one can consider the case where the CFT has a boundary. The holographic description of a BCFT with the thin-brane model was proposed in \cite{Takayanagi:2011zk}, and using this proposal the complexity was considered for a CFT of dimension $d$, with the boundary on a hyperplane, in \cite{Sato:2019kik}. 
They found that in $d>2$ CV and CA have qualitatively similar behavior. In $d=2$, similarly to \eqref{defect-cpxty}, CV has a logarithmic divergence which is absent in CA, but CA has also a finite contribution which is tension-dependent.  One should notice however that there is an ambiguity coming from the joints at the boundary: the null normals to the boundary and the WDW patch are orthogonal, so the prescription \eqref{Action-bdy} is not well-defined in this case. 

The same result in $d=2$ was found also in \cite{Braccia:2019xxi}), who in addition also computed the vacuum complexity of a finite harmonic chain with Dirichlet boundary conditions. 
The computation is similar to the one in \ref{sec:QuantumComputationalComplexity}, but now the boundary condition breaks the translational invariance, so the zero mode is lifted and one can take the massless limit; this can be more directly compared to the holographic result, and once again it was found that the $\mathcal{C}_1$ complexity is in qualitative agreement with the CV proposal.

\begin{figure}[htbp]
\centerline{\includegraphics[scale=1]{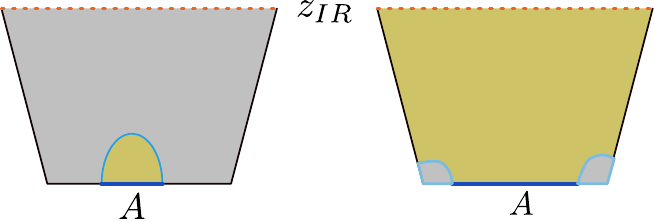}}
\caption{Illustration of the entanglement and complexity phase transition in a system with two boundaries, as a function of the size of the boundary region $A$. The region inside the RT surface is colored in yellow. Note that in the right figure this region extends to the IR cutoff and an IR regulator is needed to give a finite result. This effect is due to working within the Poincar\'e patch and is not present when considering global AdS.}
\label{fig:PhaseTransition}
\end{figure}

It is interesting to observe that in the case of a subregion in a BCFT, the holographic complexity exhibits a phase transition ``inherited'' from the entanglement entropy \cite{Braccia:2019xxi}. Depending on the ratio of the subregion length and the distance from the boundary, the transition is determined by the minimum area of two possible configurations of the RT surface: one where the surface is the same as it would be without boundary, and the other where the surface ends on the brane in the bulk. At the transition point, the two surfaces have the same area, so the entanglement entropy is continuous, but the complexity changes discontinuously, see Fig. \ref{fig:PhaseTransition}. 
This type of transitions in the entanglement entropy were used extensively for studying the formation of islands in the context of the Page curve of black hole evaporation \cite{Penington:2019npb,Almheiri:2019psf,Penington:2019kki,Almheiri:2019qdq,Almheiri:2019hni,Rozali:2019day,Chen:2019uhq,Chen:2020hmv,Chen:2020uac,Hernandez:2020nem,Schneiderbauer:2019anh,Schneiderbauer:2020isp}. It would be interesting to see if the phase transition of the complexity can give additional insights into this problem.
A similar discontinuity appears also without defect  or boundary, in the case of a subregion consisting of two disconnected segments \cite{Auzzi:2019vyh}.

\section{Summary and Outlook}
\label{sec:SumDisc}

In this introductory, review we started by  presenting the most basic ideas related to quantum complexity in relation to quantum computing, as one measure of the difficulty of solving a problem with a quantum algorithm. We have established some generic properties that can be deduced with simple counting arguments on the space of operators.  We have introduced the geometric approach of Nielsen, which replaces gate complexity with a notion of continuous complexity. This has many advantages, not least that it is in many cases more amenable to explicit computations. We have illustrated the method on examples of increasing system size (\ie dimension of the Hilbert space): first a single qubit, then a harmonic oscillator, and finally a free QFT. In the last two cases, the complexity is computable for the class of Gaussian states (or equivalently, operators that are generated by gates quadratic in the oscillators).  We presented a partial further extension to the case of a CFT, in which case the states that can be considered are  those that belong to a single conformal family, \ie are descendants of a single primary state. We also presented the additional problems that arise when considering mixed states, mostly using one particular definition of complexity, namely the complexity of purification. 

We then moved to the holographic complexity conjectures. We showed, working with the example of the eternal two-sided black hole dual to the TFD state, that both CV and CA reproduce qualitatively the features expected for complexity: the divergence structure matches the free-field theory result, and the behavior in time matches the growth expected for a chaotic, fast-scrambling system. We showed that a crucial property of complexity, the switchback effect, is present in simple holographic models where the perturbation of the system is represented by a shock wave. Finally we presented the extension of the conjectures to the case of subregions of the boundary theory, and an application to the thin-brane holographic models of CFTs with defects and boundaries. 

While CV and CA give qualitatively similar answers in most cases, we showed that for subregions and defects/boundaries there were significant differences. This raises an important question: which one, if any, of the two conjectures is the correct one? In fact, complexity is not a single observable, but a family of them. The holographic definitions have some ambiguities, but much fewer than the QFT definition which depends on the choice of a cost function, a basis of gates, penalty factors etc.\footnote{The authors of \cite{Fu:2018kcp} have argued, based on the analysis of multi-boundary solutions in AdS, that the holographic complexity is not compatible with local gates; see also the recent work \cite{Susskind:2021esx} on the effect of scaling the number of legs in the gates with the number of degrees of freedom.} 
It could be that the two holographic conjectures correspond each to a specific choice of these parameters, and all the other choices do not have a natural bulk interpretation, or at least we have not found it yet. If it is true, it would be extremely interesting to understand which complexity is naturally singled out by holography and why. 
We definitely do not have a ``smoking gun" signature comparable to other precision tests of the AdS/CFT correspondence, which require supersymmetry or integrability in order to interpolate between weak and strong coupling. It has not been explored whether supersymmetry and/or integrability play a role in the complexity story.

The tensor network description of holography could shed some light on this question, but it needs to be understood better, particularly for what concerns the dynamical aspects. Another approach is to attack the problem from the other end, as it were, namely to develop further the techniques for studying complexity in QFT. Since holographic theories are strongly coupled, it is essential to develop tools to go beyond Gaussian states and free theories. 
For the moment, only a few attempts have been made using perturbation theory. 
As we explained, the computations are manageable only when one can exploit a symmetry of the system; for this reason it seems promising to consider CFTs, but for the moment it is not known how to compute the relative complexity of two states that do not belong to the same conformal family. 
As we have seen, in free theories the complexity can be found in terms of the spectrum of the theory. Presumably in a CFT there will be some dependence on the OPE coefficients as well. It would be interesting to understand this dependence, and to determine whether some part of complexity has universality properties.  

Penalty factors are a crucial ingredient  of the complexity geometry. As we have seen in the single qubit case, but is true more generally, their effect is to create negative sectional curvature, which in turn is associated to diverging geodesics and chaotic behavior (notice however that in the case of coherent states we found a section with the geometry of hyperbolic space even without any penalty factors). It is therefore important to try and understand how the complexity in QFT is affected by penalty factors (see \cite{Akal:2019ynl,Auzzi:2020idm} for some work in this direction).  This would also help in understanding better the relation between complexity and chaos \cite{Balasubramanian:2019wgd}.

An important open question is whether there are universal bounds on the growth rate of complexity.
As we have seen in section \ref{sec:Comparison}, 
in many cases CA saturates a bound inspired by the Lloyd's bound, which yields a maximum computation rate proportional to the energy of the system. However, on one hand, one can find holographic counterexamples where the bound is violated, and on the other hand, the Lloyd's bound, seen as a bound on computational speed, requires some assumptions on how the computation is performed; in particular, it assumes that the operations performed by the gates map a state into an orthogonal state. This assumption is not satisfied by the ``simple" gates, namely gates that are close to the identity, which are the type of gates used in the definition of continuous complexity. It was argued in \cite{Cottrell:2017ayj} that the holographic results imply that a black hole is modeled by simple gates, if one assumes a serial circuit. They introduce two time scales: the time $\tau_{comp}$ required to perform an operation, and the time $\tau_{coh}$ which characterizes the spread of the wavefunction, and can be related to the density of states for a system with many degrees of freedom using a saddle-point approximation . For holographic systems $\tau_{coh} \gg \tau_{comp}$, implying that the gates are simple. 
However, it seems more reasonable that a circuit modeling a black hole will be parallel, namely many gates can act simultaneously on different qubits (generically we expect as many as $S/2$). The analysis in this case becomes more subtle. This is a question that certainly warrants further investigation. 

Apart from the question of the bounds, the fact that the complexity grows linearly in time is in itself highly significant, and it has important implications for quantum computability. As discussed in \cite{Susskind:2018fmx}, if we assume that black holes behave as universal quantum circuits, then their linear growth of complexity for an exponentially long time implies that there exist problems that can be solved by a classical computer with polynomial space and \emph{arbitrary time} (\ie they are in the complexity class PSPACE) but which cannot be solved by a quantum computer in polynomial time. Of course, in order to reach this conclusion it is not enough to argue that the growth is generically linear, but one has to prove it. This has been done recently in  \cite{Haferkamp:2021uxo}, for the case of \emph{ random circuits} built from two-qubit gates, where each gate is drawn randomly according to the Haar measure on $SU(4)$. The proof is  basically a refinement of the counting argument, and it shows that the complexity is bounded below by a linear function of time, with probability 1.  It is believed that this kind of circuit should be a good model for chaotic quantum dynamics generated by a time-independent Hamiltonian. The result was proven for the exact gate complexity, while it is not yet proven for approximate or continuous complexity.

We have mentioned in the introduction that one of the most important questions concerning the quantum information properties of gravity is the difficulty of decoding the Hawking radiation emitted by a black hole. The holographic conjectures we have presented addresses a different, albeit not unrelated, problem, namely the difficulty of distinguishing different states of a black hole. The fact that the holographic duality relates a quantity of the boundary theory that is difficult to compute (in the colloquial sense of the word) with one in the bulk that is easy to compute does not come as a surprise to people who are familiar with the correspondence. However, in the quantum information-theoretic setting we attribute a precise meaning to the difficulty, and we can wonder, as \cite{Bouland:2019pvu} did, whether this property of the correspondence violates the extended Church-Turing thesis, which postulates that any physical process can be efficiently simulated on a quantum computer. Even though the volume, or the action, of the wormhole is not exactly a physical observable, nevertheless one can argue that it is a quantity that can be easily extracted from a coarse knowledge of the metric. Therefore a quantity of high complexity can be efficiently determined by evolving in the bulk; this suggests that the conversion of bulk quantities into boundary quantities, namely the holographic dictionary, must be extremely complex. As pointed out by \cite{Susskind:2020kti}, this Gedanken experiment requires that the bulk observer has access to the black hole interior, so the horizon will play a role in keeping the Church-Turing thesis valid, under the condition  that one only considers the space accessible to outside observers. 

Considering the problem of decoding Hawking radiation, one encounters a different puzzle, observed in \cite{Brown:2019rox} where a possible solution was also proposed. Suppose a black hole is let to radiate for a not too long time.\footnote{To be more precise, the time should be much shorter than the exponential time $t\sim e^S$ at which the complexity saturates, but sufficiently long that there is substantial entanglement between the interior modes and the radiation; for instance, one can take a time of the order of the Page time.} 
According to the ER=EPR conjecture  \cite{Maldacena:2013xja}, there is a wormhole that connects the interior to the radiation, but the volume grows linearly with time and according to CV the complexity of the state is only polynomial in the entropy at this time, in contrast with the result that the distillation of the information from the radiation is exponentially hard. The solution proposed in \cite{Brown:2019rox} is that the difficulty of the distillation task is in fact measured by a different quantity, since one is not allowed to use all possible gates but only those that act on the radiation without acting on the interior. A different holographic conjecture was proposed for this {\it restricted complexity}, which involves the area of the maximum cross-section of the wormhole and of the minimal surface in the throat that connects it to the asymptotic region. This shows that there are probably different notions of complexity that can be useful for answering different questions about the quantum information-theoretic aspects of gravity, and there is still much to be understood. 

Another important question concerns the implications of complexity for many-body systems.
In order to characterize properties such as scrambling, chaos, and thermalization, extensive use has been made mostly of two type of observables: low-point correlation functions (especially out-of-time-order correlators), and entanglement entropy.  
Quantum computational complexity captures properties of the quantum state of a system that are more refined than those visible through these observables. This is why it is sensitive to the evolution of the microstates in the ensemble corresponding to a black hole. It is likely that it can also be used to give new insights into the mechanisms underlying the approach to equilibrium and thermalization, and possibly detect new types of phase transitions (see \eg \cite{doi:10.1146/annurev-conmatphys-031214-014726,Liu:2019aji,Camilo:2020gdf}).

We should finally point out again that we did not aim at writing a comprehensive review of the subject, therefore we left out  many topics that we felt were too advanced for an introduction, such as the thermodynamics and resource theory aspects of complexity \cite{Brown:2017jil,Susskind:2018pmk,Bernamonti:2019zyy,Bernamonti:2020bcf}, the relation with bulk dynamics (in the sense of reconstructing the Einstein equations in the bulk from the complexity of the boundary) \cite{Czech:2017ryf}, alternative conjectures, most notably the one in \cite{Couch:2016exn} (sometimes called CV~2.0), complexity in de Sitter space \cite{Chapman:2021eyy,Susskind:2021esx,Reynolds:2017lwq,Geng:2019ruz}, the evolution of complexity after a quench \cite{Alves:2018qfv,Camargo:2018eof}, the relation between complexity and chaos \cite{Balasubramanian:2019wgd,Balasubramanian:2021mxo}, other notions of complexity such as the \emph{ operator complexity} \cite{Barbon:2019wsy,Rabinovici:2020ryf}. We hope that our readers will be encouraged to delve further into this fascinating subject and contribute to its development.

\section*{Acknowledgments}
We gratefully acknowledge discussions with Adam Chapman, Juan Hernandez and Shan-Ming Ruan. We are also grateful to our many collaborators in papers on the subject of complexity in the past few years. The work of SC is supported by the Israel Science Foundation (grant No. 1417/21). SC acknowledges the support of Carole and Marcus Weinstein through the BGU Presidential Faculty Recruitment Fund. 

\bibliographystyle{unsrt}
%%%% problematic command, needs to be removed at some point. Orders references.
%\bibliographystyle{plain}
\bibliography{references}   % name your BibTeX data base
 
\end{document}